\documentclass[%
  reprint,
  showpacs,preprintnumbers,
  amsmath,amssymb,
  aps,
  prb,
longbibliography
]{revtex4-1}

\usepackage{graphicx}
\usepackage{dcolumn}
\usepackage{bm}
\usepackage{multirow}
\usepackage{braket}
\usepackage{color}
\usepackage{ulem}

\usepackage{times}
\usepackage{booktabs}

\begin{document}

\preprint{APS/123-QED}

\title{
Noncoplanar multiple-$Q$ spin textures by itinerant frustration:\\ Effects of single-ion anisotropy and bond-dependent anisotropy
}

\author{Satoru Hayami and Yukitoshi Motome}
\affiliation{
Department of Applied Physics, The University of Tokyo, Tokyo 113-8656, Japan
}
 
\begin{abstract}
We theoretically investigate multiple-$Q$ spin textures, which are composed of superpositions of spin density waves with different wave numbers, for an effective spin model of centrosymmetric itinerant magnets. 
Our focus is on the interplay between biquadratic interactions arising from the spin-charge coupling and magnetic anisotropy caused by the spin-orbit coupling. 
Taking into account two types of the magnetic anisotropy, single-ion anisotropy and bond-dependent anisotropy, we elucidate magnetic phase diagrams for an archetypal triangular-lattice system in the absence and presence of an external magnetic field. 
In the case of the single-ion anisotropy, we find a plethora of multiple-$Q$ instabilities depending on the strength and the sign of the anisotropy (easy plane or easy axis), including a noncoplanar triple-$Q$ state regarded as a skyrmion crystal with topological number of two, and coplanar and noncoplanar double-$Q$ states. 
In an external magnetic field, we find that another noncoplanar triple-$Q$ state, a skyrmion crystal with topological number of one, is stabilized by the in-plane (out-of-plane) magnetic field under the easy-plane (easy-axis) anisotropy. 
A part of the results, especially for the relatively large biquadratic interaction, qualitatively reproduce those in the Kondo lattice model which explicitly includes itinerant electrons [S. Hayami and Y. Motome, Phys. Rev. B {\bf 99}, 094420 (2019)].
We also examine the stability of the field-induced skyrmion crystal by rotating the field direction. 
As a biproduct, we show that a triple-$Q$ state with nonzero chirality appears in the presence of the biquadratic interaction and the easy-axis anisotropy. 
Meanwhile, we find that the bond-dependent anisotropy also stabilizes both types of skyrmion crystals. 
We show that, however, for the skyrmion crystal with topological number of one, Bloch- and N\'eel-type skyrmion crystals are selectively realized depending on the sign of the bond-dependent anisotropy, since this anisotropy selects a particular set of the helicity and vorticity. 
Moreover, we find yet another multiple-$Q$ states with nonzero spin scalar chirality, including two types of meron crystals with the skyrmion numbers of one and two. 
The systematic investigation of multiple-$Q$ instabilities in triangular itinerant magnets will provide a reference to complex magnetic textures in centrosymmetric magnetic metals. 
\end{abstract}
\maketitle

\section{Introduction}
\label{sec:Introduction}

Superpositions of spin density waves, which are termed as multiple-$Q$ magnetic states, have attracted much interest in various fields of condensed matter physics~\cite{Bak_PhysRevLett.40.800,Shapiro_PhysRevLett.43.1748,bak1980theory,batista2016frustration}. 
Different ways of taking a linear combination lead to different types of spin textures. 
One of the fundamental examples is found in a superposition of collinear states in the axial next-nearest-neighbor Ising model, which exhibits peculiar temperature dependence of spatial spin modulations called the devil's staircase~\cite{Elliott_PhysRev.124.346,Fisher_PhysRevLett.44.1502,Bak_PhysRevLett.49.249,selke1988annni,ShibataPhysRevB.52.10232,Fobes_PhysRevB.96.174413}. 
Another interesting example is represented by a superposition of spiral states, which results in noncollinear and noncoplanar spin textures, such as magnetic vortices~\cite{Bak_PhysRevLett.40.800,Momoi_PhysRevLett.79.2081,Kamiya_PhysRevX.4.011023,Wang_PhysRevLett.115.107201} and skyrmion crystals~\cite{Bogdanov89,Bogdanov94,rossler2006spontaneous,Muhlbauer_2009skyrmion,yu2010real,nagaosa2013topological}. 
Such superpositions of spirals are intriguing, as they often carry nonzero vector chirality, $\bm{S}_i \times \bm{S}_j$, and/or scalar chirality, $\bm{S}_i \cdot (\bm{S}_j \times \bm{S}_k)$, which are sources of an emergent electromagnetic field for electrons through the spin Berry phase mechanism~\cite{berry1984quantal,Loss_PhysRevB.45.13544,Xiao_RevModPhys.82.1959}. 
Indeed, the chirality degrees of freedom in the multiple-$Q$ states generate interesting phenomena, such as the topological Hall effect~\cite{Loss_PhysRevB.45.13544,Ye_PhysRevLett.83.3737,Ohgushi_PhysRevB.62.R6065,tatara2002chirality}, the spin Hall effect~\cite{Katsura_PhysRevLett.95.057205,Zhang_PhysRevLett.113.196602,zhang2018spin}, and nonreciprocal transport~\cite{ishizuka2020anomalous, Hayami_PhysRevB.101.220403,hayami2020phase,Hayami_PhysRevB.102.144441}. 

Such noncollinear and noncoplanar multiple-$Q$ states are ubiquitously found in a wide range of materials. 
From the viewpoint of the microscopic mechanism, however, there are several different origins depending on the systems. 
We here discuss three of them in the following. 
The first one is (i) the relativistic spin-orbit coupling in the absence of spatial inversion symmetry in the lattice structure. 
It induces an effective antisymmetric exchange interaction called the Dzyaloshinskii-Moriya (DM) interaction~\cite{dzyaloshinsky1958thermodynamic,moriya1960anisotropic}, which favors a twist in the spin texture. 
For instance, the interplay among the ferromagnetic interaction, the DM interaction, and an external magnetic field stabilizes a triple-$Q$ spiral density wave termed as the skyrmion crystal~\cite{bak1980theory,rossler2006spontaneous,Yi_PhysRevB.80.054416,Mochizuki_PhysRevLett.108.017601,Gungordu_PhysRevB.93.064428}. 
Since the discovery of skyrmion crystals in B20 compounds~\cite{Muhlbauer_2009skyrmion,yu2010real}, a number of candidates in this category have been studied intensively, and various types of skyrmions have been explored, such as the Bloch-type skyrmion~\cite{Muhlbauer_2009skyrmion,yu2010real,seki2012observation}, N\'eel-type skyrmion~\cite{kezsmarki_neel-type_2015,Kurumaji_PhysRevLett.119.237201}, antiskyrmion~\cite{koshibae2016theory,nayak2017discovery,hoffmann2017antiskyrmions}, and bi-skyrmion crystals~\cite{yu2014biskyrmion,lee2016synthesizing}.  
Recently, multiple-spin chiral interactions, which can be regarded as higher-order extensions of the DM interaction, have been studied to understand the peculiar noncoplanar magnetism at surfaces and interfaces~\cite{brinker2019chiral,Laszloffy_PhysRevB.99.184430,grytsiuk2020topological,Brinker_PhysRevResearch.2.033240,Mankovsky_PhysRevB.101.174401}. 

The second mechanism is based on (ii) competing interactions between the magnetic moments. 
For example, geometrical frustration arising from nonbipartite lattice structures leads to noncollinear and noncoplanar multiple-$Q$ states, combined with, e.g., the effect of further-neighbor interactions~\cite{Okubo_PhysRevB.84.144432,Okubo_PhysRevLett.108.017206,Rosales_PhysRevB.87.104402}, quantum fluctuations~\cite{Kamiya_PhysRevX.4.011023,Wang_PhysRevLett.115.107201,Marmorini2014,Ueda_PhysRevA.93.021606}, and disorder by impurities~\cite{Maryasin_PhysRevLett.111.247201,maryasin2015collective,Hayami_PhysRevB.94.174420}. 
Bond-dependent exchange anisotropy, e.g., of compass and Kitaev type, can also induce magnetic vortices and skyrmion crystals~\cite{Michael_PhysRevB.91.155135,Lukas_PhysRevLett.117.277202,Rousochatzakis2016,yao2016topological,Maksimov_PhysRevX.9.021017,amoroso2020spontaneous}. 
Frustration rooted in the competing exchange interactions and the magnetic anisotropy also gives rise to a plethora of multiple-$Q$ states~\cite{Rousochatzakis2016,leonov2015multiply,Lin_PhysRevB.93.064430,Hayami_PhysRevB.93.184413,Lin_PhysRevLett.120.077202,Binz_PhysRevLett.96.207202,Binz_PhysRevB.74.214408,Park_PhysRevB.83.184406,zhang2017skyrmion}. 
Note that the magnetic anisotropy in this mechanism originates from the spin-orbit coupling in centrosymmetric systems, in contrast to (i). 
In addition, multiple-spin interactions beyond the bilinear exchange interaction provide another way to induce the multiple-$Q$ states through the frustration~\cite{Kurz_PhysRevLett.86.1106,heinze2011spontaneous,Yoshida_PhysRevLett.108.087205,paul2020role,grytsiuk2020topological,Mankovsky_PhysRevB.101.174401}. 

The third mechanism is (iii) itinerant nature of electrons. 
The kinetic motion of electrons can induce effective magnetic interactions through the coupling between spin and charge degrees of freedom. 
The typical example is the Ruderman-Kittel-Kasuya-Yosida (RKKY) interaction appearing when the spin-charge coupling is much smaller than the bandwidth~\cite{Ruderman,Kasuya,Yosida1957}. 
The RKKY interaction is long-ranged and favors a single-$Q$ spiral state whose wave number is set by the Fermi surface. 
On the other hand, when the Fermi surface has a structure so that the bare susceptibility exhibits multiple peaks in momentum space, the instability toward the single-$Q$ spiral state occurs at the multiple wave numbers simultaneously. 
This is another type of frustration distinguished from that in the mechanism (ii), which we call 
{\it itinerant frustration}. 
In this case, higher-order contributions from the spin-charge coupling lift the degeneracy. 
Among many contributions, an effective positive biquadratic interaction in momentum space plays an important role~\cite{Akagi_JPSJ.79.083711,Akagi_PhysRevLett.108.096401,Hayami_PhysRevB.90.060402,Ozawa_doi:10.7566/JPSJ.85.103703,Hayami_PhysRevB.94.024424,Hayami_PhysRevB.95.224424,lounis2020multiple,hayami2020multiple} in stabilizing multiple-$Q$ states, such as the triple-$Q$ states in hexagonal crystal systems~\cite{Martin_PhysRevLett.101.156402,Akagi_JPSJ.79.083711,Kato_PhysRevLett.105.266405,Barros_PhysRevB.88.235101,Ozawa_PhysRevLett.118.147205,Venderbos_PhysRevLett.108.126405,Venderbos_PhysRevB.93.115108,Barros_PhysRevB.90.245119,Ghosh_PhysRevB.93.024401}, the double-$Q$ states in tetragonal crystal systems~\cite{Solenov_PhysRevLett.108.096403,hayami_PhysRevB.91.075104,Ozawa_doi:10.7566/JPSJ.85.103703,Hayami_doi:10.7566/JPSJ.89.103702}, and the triple-$Q$ states in cubic crystal systems~\cite{Chern_PhysRevLett.105.226403,Hayami_PhysRevB.89.085124}.  

More recently, further interesting situations have been studied by considering the interplay between the mechanisms (i)-(iii) mentioned above. 
For instance, a synergetic effect between (i) the antisymmetric exchange interactions by the spin-orbit coupling and (iii) the multiple spin interactions by the spin-charge coupling results in more exotic multiple-$Q$ states, such as the triple-$Q$ and quartet-$Q$ hedgehog crystals~\cite{Okumura_PhysRevB.101.144416,okumura2020tracing} and  sextuple-$Q$ states~\cite{Okada_PhysRevB.98.224406}. 
Competition between (ii) the single-ion anisotropy and (iii) the spin-charge coupling induces a triple-$Q$ skyrmion crystal under the magnetic field~\cite{Wang_PhysRevLett.124.207201,Hayami_PhysRevB.99.094420,Su_PhysRevResearch.2.013160}. 
Moreover, a Bloch-type skyrmion crystal is realized even in a Rashba-type metal by taking into account (i), (ii), and (iii)~\cite{hayami2018multiple,Hayami_PhysRevLett.121.137202}. 

These series of studies to investigate when and how the multiple-$Q$ states appear are important to understand the microscopic origins of the multiple-$Q$ states found in materials. 
Recently, unconventional multiple-$Q$ states have been found in $d$- and $f$-electron systems, such as the vortices in MnSc$_2$S$_4$~\cite{Gao2016Spiral,gao2020fractional}, CeAuSb$_2$~\cite{Marcus_PhysRevLett.120.097201,Seo_PhysRevX.10.011035}, and Y$_3$Co$_8$Sn$_4$~\cite{takagi2018multiple}, the skyrmions in SrFeO$_3$~\cite{Ishiwata_PhysRevB.84.054427,Ishiwata_PhysRevB.101.134406,Rogge_PhysRevMaterials.3.084404}, Co-Zn-Mn alloys~\cite{karube2018disordered}, EuPtSi~\cite{kakihana2017giant,kaneko2018unique,tabata2019magnetic}, Gd$_2$PdSi$_3$~\cite{kurumaji2019skyrmion,Hirschberger_PhysRevLett.125.076602,Hirschberger_PhysRevB.101.220401,Nomoto_PhysRevLett.125.117204,moody2020charge}, Gd$_3$Ru$_4$Al$_{12}$~\cite{hirschberger2019skyrmion,Hirschberger_10.1088/1367-2630/abdef9}, and GdRu$_2$Si$_2$~\cite{khanh2020nanometric,Yasui2020imaging}, and the hedgehogs in MnSi$_{1-x}$Ge$_{x}$~\cite{tanigaki2015real,kanazawa2017noncentrosymmetric,fujishiro2019topological,Kanazawa_PhysRevLett.125.137202}. 
Furthermore, there remain several unidentified multiple-$Q$ states distinguished from the above states, especially in centrosymmetric materials~\cite{Ishiwata_PhysRevB.101.134406,kurumaji2019skyrmion,Hirschberger_PhysRevLett.125.076602,Hirschberger_PhysRevB.101.220401,hirschberger2019skyrmion,khanh2020nanometric}. 
Due to the crystal symmetry and the short period of the magnetic textures, their mechanisms might be accounted for by (ii) and (iii), although their origins are still under debate. 

To understand the microscopic origins and encourage further experimental exploration of exotic multiple-$Q$ states, in this paper, we push forward the theoretical study in a more systematic way on the interplay between (ii) the magnetic anisotropy caused by the spin-orbit coupling in centrosymmetric systems and (iii) the multiple-spin interactions arising from the spin-charge coupling. 
Taking an archetypal hexagonal model, we study how the itinerant frustration is relieved by their interplay and what types of the multiple-$Q$ states are generated. 
By introducing two types of magnetic anisotropy, single-ion anisotropy and bond-dependent anisotropy, to the effective bilinear-biquadratic model for itinerant magnets, we elaborate magnetic phase diagrams in a wide parameter range of the biquadratic interaction, the magnetic anisotropy, and the magnetic field in a systematic way. 
We uncover a variety of multiple-$Q$ states, including those which have never been reported. 
Our results provide deeper understanding of the multiple-$Q$ states emergent from the synergy between the spin-charge coupling and the spin-orbit coupling in centrosymmetric systems. 

The rest of the paper is organized as follows. 
We start by showing a brief summary of the main results in this paper in Sec.~\ref{sec:Brief Summary of main results}. 
In Sec.~\ref{sec:Model and Method}, we present an effective bilinear-biquadratic spin model on a triangular lattice including the two types of magnetic anisotropy, and outline the numerical method. 
In Sec.~\ref{sec:Single-ion Anisotropy}, we discuss the effect of the single-ion anisotropy. 
We obtain the magnetic phase diagram including three multiple-$Q$ states by changing the single-ion anisotropy and the biquadratic interaction in the absence of the magnetic field in Sec.~\ref{sec:At zero field_SIA}. 
Then, in Secs.~\ref{sec:Field along the $z$ direction_SIA}-\ref{sec:Field rotation in the $zx$ plane_SIA}, we show a further variety of multiple-$Q$ instabilities in the magnetic fields applied in different directions. 
In Sec.~\ref{sec:Bond-exchange Anisotropy}, we discuss the effect of the bond-dependent anisotropy. 
We find five multiple-$Q$ states at zero field and more in the field. 
Section~\ref{sec:Summary} is devoted to the concluding remarks.

\section{Brief Summary of main results}
\label{sec:Brief Summary of main results}

\begin{figure*}[htb!]
\begin{center}
\includegraphics[width=1.0\hsize]{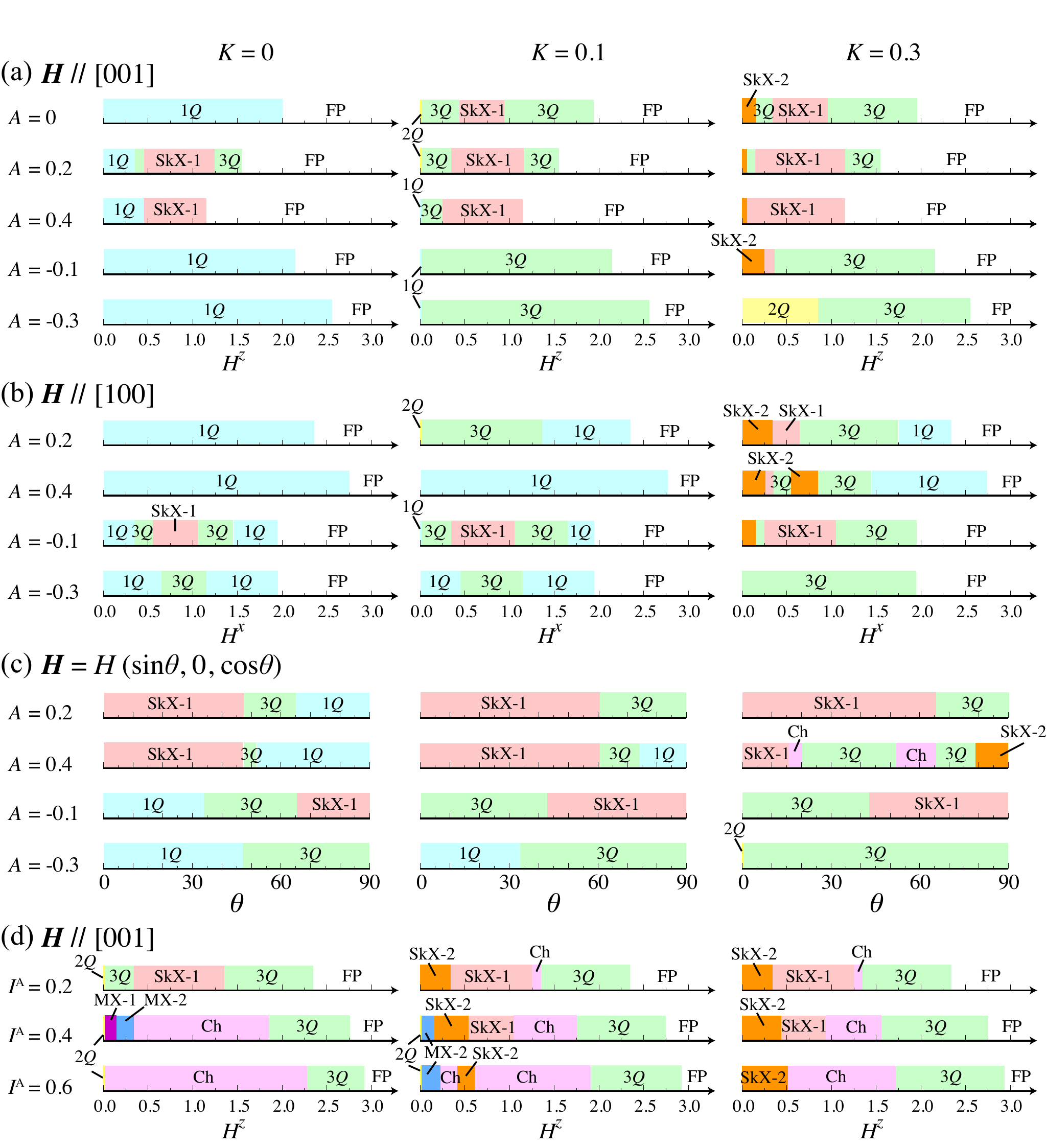} 
\caption{
\label{fig:Summary}
Schematics of the magnetic phase diagram of the model in Eq.~(\ref{eq:Ham}) in the presence of (a)-(c) the single-ion anisotropy $A$ and (d) the bond-dependent anisotropy $I^{\rm A}$. 
$K$ stands for the coupling constant for the biquadratic exchange interaction: 
The left, middle, and right panels are the results at $K=0$, $0.1$, and $0.3$, respectively. 
(a) and (d) are for the [001] magnetic field, (b) is for the [100] field, and (c) is for the field in the $xz$ plane with $H=0.8$. 
$1Q$, $2Q$, $3Q$, SkX-1, SkX-2, MX-1, MX-2, Ch, and FP stand for the single-$Q$ state, double-$Q$ state, triple-$Q$ state, $n_{\rm sk}=1$ skyrmion crystal, $n_{\rm sk}=2$ skyrmion crystal, $n_{\rm sk}=1$ meron crystal, $n_{\rm sk}=2$ meron crystal, multiple-$Q$ states with 
nonzero uniform scalar chirality, and the fully-polarized state, respectively. 
The detailed magnetic and chirality structures in (a)-(c) are presented in Sec.~\ref{sec:Single-ion Anisotropy} and those in (d) in Sec.~\ref{sec:Bond-exchange Anisotropy}. 
}
\end{center}
\end{figure*}

Before starting the detailed discussions, we summarize the main results of this paper, i.e., 
where multiple-$Q$ states appear in the phase diagram through the interplay among the biquadratic interaction, the magnetic anisotropy, and the magnetic field. 
We investigate two types of the magnetic anisotropy, the single-ion anisotropy and the bond-dependent anisotropy, with and without the biquadratic interaction and the magnetic field. 
The schematic phase diagrams are shown for typical parameter sets in Fig.~\ref{fig:Summary}, where $K$ represents the biquadratic interaction, positive (negative) $A$ represents the easy-axis (easy-plane) single-ion anisotropy, $I^{\rm A}$ represents the bond-dependent anisotropy, and $\bm{H}=(H^x,H^y,H^z)$ is an external magnetic field. See Sec.~\ref{sec:Model and Method} for the details of the model and parameters. 

For the isotropic case ($A=I^{\rm A}=0$), we find a triple-$Q$ skyrmion crystal in a magnetic field for small $K=0.1$ and another one evolved from zero field for large $K=0.3$, as displayed in the top row of Fig.~\ref{fig:Summary}(a).   
The former is characterized by the topological number of one ($n_{\rm sk}=1$ skyrmion crystal), while the latter has topological number of two ($n_{\rm sk}=2$ skyrmion crystal); see Secs.~\ref{sec:At zero field_SIA} and \ref{sec:Isotropic case_Hz}. 

The stability of the $n_{\rm sk}=1$ and $n_{\rm sk}=2$ skyrmion crystals against the single-ion anisotropy $A$ is discussed in Secs.~\ref{sec:Field along the $z$ direction_SIA}-\ref{sec:Field rotation in the $zx$ plane_SIA}. 
We show that the $n_{\rm sk}=2$ skyrmion crystal remains stable 
against both small easy-axis and easy-plane anisotropy as shown in Figs.~\ref{fig:Summary}(a) and \ref{fig:Summary}(b), qualitatively similar to the result obtained for the Kondo lattice model~\cite{Hayami_PhysRevB.99.094420}. 
We find two types of modulations of the $n_{\rm sk}=2$ skyrmion crystal by systematically changing $A$ and $H^z$: One is characterized by a superposition of the magnetic vortices in the $xy$ spin component and the sinusoidal wave in the $z$ spin component, and the other is characterized by a superposition of the magnetic vortices in both $xy$ and $z$ spin components; see Sec.~\ref{sec:Field along the $z$ direction_SIA}. 
Besides, we find another $n_{\rm sk}=2$ skyrmion crystal in the magnetic field along the $H^x$ direction for the easy-axis anisotropy, as shown in the case with $A=0.4$ in Fig.~\ref{fig:Summary}(b), whose spin texture is characterized by dominant double-$Q$ modulations in the $z$ spin component and a subdominant modulation in the $xy$ spin components; see Sec.~\ref{sec:With easy-axis anisotropy_Hx} for the details. 
Meanwhile, we find that the $n_{\rm sk}=1$ skyrmion crystal remains stable in the magnetic field along the $z$ direction for the easy-axis anisotropy and along both $x$ and $z$ directions for the easy-plane anisotropy, as shown in Figs.~\ref{fig:Summary}(a) and \ref{fig:Summary}(b); see Secs.~\ref{With easy-axis anisotropy_Hz}, \ref{sec:With easy-plane anisotropy_Hz}, and \ref{sec:With easy-plane anisotropy_Hx}.
We examine the systematic evolution of the $n_{\rm sk}=2$ and $n_{\rm sk}=1$ skyrmion crystals by rotating the magnetic field in the $xz$ plane, as shown in Fig.~\ref{fig:Summary}(c); see Sec.~\ref{sec:Field rotation in the $zx$ plane_SIA}. 
The $n_{\rm sk}=2$ skyrmion crystal found for the field along the $x$ direction under the easy-axis anisotropy is rapidly destabilized by rotating the field to the $z$ direction. 
On the other hand, the $n_{\rm sk}=1$ skyrmion crystal is stable in a wide range of fields and anisotropy, as shown in the second row of Fig.~\ref{fig:Summary}(c). 
The range of the field angle where the $n_{\rm sk}=1$ skyrmion crystal is stabilized tends to be wider for larger $K$ for both easy-axis and easy-plane anisotropy; see Secs.~\ref{sec:With easy-axis anisotropy_Hrot} and \ref{With easy-plane anisotropy_Hrot}. 
Moreover, we find a triple-$Q$ state with nonzero scalar chirality in the magnetic field in the $xz$ plane under the easy-axis anisotropy, as shown in the second row of Fig.~\ref{fig:Summary}(c). 
We also show that there appear triple-$Q$ states, which are topologically trivial, around the skyrmion crystals, as shown in Figs.~\ref{fig:Summary}(a)-\ref{fig:Summary}(c). 
Also, in the large $K$ and small $A$ region, we find a double-$Q$ state, whose spin configuration is coplanar at zero field, as shown in the right bottom row of Fig.~\ref{fig:Summary}(a); see Sec.~\ref{sec:With easy-plane anisotropy_Hz}. 

The effect of the bond-dependent anisotropy $I^{\rm A}$ is discussed in Sec.~\ref{sec:Bond-exchange Anisotropy}. 
We find that the $n_{\rm sk}=2$ skyrmion crystal is also stabilized at zero field by introducing the bond-dependent anisotropy, as shown in Fig.~\ref{fig:Summary}(d). 
Interestingly, this state has a spontaneous ferromagnetic moment along the $z$ direction, in contrast to the case of the single-ion anisotropy.
Accordingly, the sign of the scalar chirality is selected to be opposite to that of the $z$ component of the ferromagnetic moment. 
In other words, it lifts the degeneracy between the skyrmion and antiskyrmion~\cite{amoroso2020spontaneous}; see Sec.~\ref{sec:At zero field_bond}. 
When the magnetic field is applied along the $z$ direction, we obtain two types of the $n_{\rm sk}=1$ skyrmion crystals: The one is characterized by a periodic array of the uniaxially-elongated skyrmions and the other shows a periodic array of the isotropic ones.
Besides the skyrmion crystals, we show that the bond-dependent anisotropy induces a $n_{\rm sk}=1$ meron crystal, including one meron and three antimerons in the magnetic unit cell. 
Furthermore, we find a $n_{\rm sk}=2$ meron crystal in the large $I^{\rm A}$ and small $K$ region, which includes four merons in the magnetic unit cell. 
We also obtain multiple-$Q$ states with nonzero uniform scalar chirality other than the skyrmion and meron crystals under the magnetic field as shown in Fig.~\ref{fig:Summary}(d), which have not been found in the case of the single-ion anisotropy. 
See Sec.~\ref{sec:Field along the $z$ direction} for all the details. 
Meanwhile, we could not find the instability toward the skyrmion and meron crystals against the in-plane magnetic field (not shown).

\section{Model and Method}
\label{sec:Model and Method}
We introduce an effective spin model for itinerant magnets with the magnetic anisotropy in Sec.~\ref{sec:Model}. 
We outline the method of numerical simulations and measured physical quantities in Sec.~\ref{sec:Numerical calculations}. 

\subsection{Model}
\label{sec:Model}
When an itinerant electron system consists of itinerant electrons and localized spins coupled via the exchange interaction, like in the Kondo lattice model, one can derive an effective spin model for the localized spins by tracing out the itinerant electron degree of freedom.
The model includes the exchange interactions in momentum space and two types of magnetic anisotropy in general. 
We consider such a model whose Hamiltonian is explicitly given by 
\begin{align}
\label{eq:Ham}
\mathcal{H}&= \mathcal{H}^{\rm 
BBQ}+\mathcal{H}^{\rm SIA}+\mathcal{H}^{\rm BA}+\mathcal{H}^{\rm Z},  
\end{align}
where
\begin{align}
\label{eq:Ham_JK}
\mathcal{H}^{\rm BBQ}&= 2\sum_\nu
\left[ -J\bm{S}_{\bm{Q}_{\nu}} \cdot \bm{S}_{-\bm{Q}_{\nu}}
+\frac{K}{N} (\bm{S}_{\bm{Q}_{\nu}} \cdot \bm{S}_{-\bm{Q}_{\nu}})^2 \right], \\
\label{eq:Ham_SIA}
\mathcal{H}^{\rm SIA}&= -A \sum_{i} (S^z_i)^2, \\
\label{eq:Ham_B}
\mathcal{H}^{\rm BA}&= 2\sum_\nu
\left[ -J \sum_{\alpha\beta}I^{\alpha \beta}_{\bm{Q}_{\nu}} S^\alpha_{\bm{Q}_{\nu}} S^\beta_{-\bm{Q}_{\nu}} \right. \nonumber \\
&\ \ \ \ \ \ \ \ \ \ \left.+\frac{K}{N} \left(\sum_{\alpha\beta}I^{\alpha \beta}_{\bm{Q}_{\nu}} S^\alpha_{\bm{Q}_{\nu}} S^\beta_{-\bm{Q}_{\nu}}\right)^2 \right], \\
\label{eq:Ham_Z}
\mathcal{H}^{\rm Z}&= - \sum_{i} \bm{H}\cdot \bm{S}_i. 
\end{align}

The first term $\mathcal{H}^{\rm BBQ}$ represents the bilinear-biquadratic interactions in momentum space, which was originally derived from the perturbation expansion with respect to the spin-charge coupling in the Kondo lattice model~\cite{Hayami_PhysRevB.95.224424}; 
$J$ and $K$ are the positive coupling constants for the isotropic bilinear and biquadratic exchange interactions, which are obtained by the second- and fourth-order perturbation analyses in terms of the spin-charge coupling in the Kondo lattice model, respectively. 
Although the coupling constants can be derived from the perturbation theory, we regard them as phenomenological parameters in order to cover the whole magnetic phase diagram in the model in Eq.~(\ref{eq:Ham}), as in the previous study~\cite{Hayami_PhysRevB.99.094420}. 
Both interactions in Eq.~(\ref{eq:Ham_JK}) are defined in momentum space for a particular set of the wave numbers $\bm{Q}_\nu$; $\bm{S}_{\bm{Q}_\nu} = (1/\sqrt{N})\sum_{i}\bm{S}_i e^{-i \bm{Q}_\nu \cdot \bm{r}_i}$ is the Fourier component of the spin $\bm{S}_i = (S_i^x, S_i^y, S_i^z)$ at site $i$, where $N$ is the number of spins. 
In the present study, we consider the triangular lattice in the $xy$ plane ($x$ is taken along the bond direction), and assume that $\bm{Q}_\nu$ originate from the six peaks of the bare susceptibility dictated by the Fermi surface in the presence of sixfold rotational symmetry of the lattice. 
Specifically, we choose a set of $\bm{Q}_\nu$ as $\bm{Q}_1=(\pi/3,0,0)$, $\bm{Q}_2=(-\pi/6,\sqrt{3}\pi/6,0)$, and $\bm{Q}_3=(-\pi/6,-\sqrt{3}\pi/6,0)$ in the following calculations (the lattice constant is taken to be unity). 
The other contributions with different $\bm{q}$ dependences (including $\bm{q}=\bm{0}$ component) are ignored by assuming distinct peak structures of the bare susceptibility~\cite{Hayami_PhysRevB.95.224424}.
Hereafter, we set $J=1$ as the energy unit. 

The second and third terms in Eq.~(\ref{eq:Ham}) represent the magnetic anisotropy that we focus on in the present study. 
The second term $\mathcal{H}^{\rm SIA}$ in Eq.~(\ref{eq:Ham_SIA}) represents the local single-ion anisotropy. 
The positive (negative) $A$ represents the easy-axis (-plane) anisotropy. 
The effect of the single-ion anisotropy on the instability toward multiple-$Q$ magnetic orderings has been investigated for chiral~\cite{Butenko_PhysRevB.82.052403,Wilson_PhysRevB.89.094411,Lin_PhysRevB.91.224407,leonov2016properties,Leonov_PhysRevB.96.014423}, frustrated~\cite{leonov2015multiply,Lin_PhysRevB.93.064430,Hayami_PhysRevB.93.184413}, and itinerant magnets~\cite{Hayami_PhysRevB.99.094420,Wang_PhysRevLett.124.207201,Su_PhysRevResearch.2.013160}, although the analysis including itinerant electrons explicitly has not been performed extensively due to the huge computational cost. 

The third term $\mathcal{H}^{\rm BA}$ in Eq.~(\ref{eq:Ham_B}) represents the anisotropic exchange interaction dependent on the bond direction.
Due to the sixfold rotational symmetry and mirror symmetry of the triangular lattice, the anisotropic tensor $I^{\alpha \beta}_{\bm{Q}_{\nu}}$ satisfies the relation, $
-I^{xx}_{\bm{Q}_{1}}=I^{yy}_{\bm{Q}_{1}}=2I^{xx}_{\bm{Q}_{2}}=-2I^{yy}_{\bm{Q}_{2}}=2I^{xy}_{\bm{Q}_{2}}/\sqrt{3}=2I^{yx}_{\bm{Q}_{2}}/\sqrt{3}=2I^{xx}_{\bm{Q}_{3}}=-2I^{yy}_{\bm{Q}_{3}}=-2I^{xy}_{\bm{Q}_{3}}/\sqrt{3}=-2I^{yx}_{\bm{Q}_{3}}/\sqrt{3} \equiv I^{\rm A}$ and otherwise zero. 
This type of interaction specifies the spiral plane according to the sign of $I^{\rm A}$: 
A positive (negative) $I^{\rm A}$ favors the proper-screw (cycloidal) spiral state. 
This term originates from the relativistic spin-orbit coupling irrespective of inversion symmetry~\cite{shibuya2016magnetic,Hayami_PhysRevLett.121.137202,takagi2018multiple}, in contrast to the antisymmetric Dzyaloshinskii-Moriya interaction in the absence of inversion symmetry. 
Similar interactions have been discussed in terms of the short-ranged bond-dependent interaction in magnetic insulators, such as the compass and Kitaev interactions~\cite{Shekhtman_PhysRevB.47.174,jackeli2009mott,Li_PhysRevB.94.035107,Maksimov_PhysRevX.9.021017,Motome_2020}. 

The last term $\mathcal{H}^{\rm Z}$ in Eq.~(\ref{eq:Ham_Z}) represents the Zeeman coupling to an external magnetic field. 
In the presence of the single-ion anisotropy ($A\neq 0$), we apply the magnetic field in the $z$ and $x$ directions, i.e., the [001] and [100] directions, and also rotate it in the $xz$ plane 
(note that the [100] and [010] fields are equivalent when $I^{\rm A}=0$). 
Meanwhile, in the presence of the bond-dependent anisotropy ($I^{\rm A}\neq 0$), 
we apply the magnetic field along the [100], [010], and [001] directions, but show the results only for the most interesting [001] case (we do not find any chiral spin textures in the [100] and [010] cases).

\subsection{Numerical calculations}
\label{sec:Numerical calculations}

We study the magnetic phase diagram of the model in Eq.~(\ref{eq:Ham}) by using simulated annealing from high temperature. 
Our simulations are carried out with the standard Metropolis local updates in real space. 
We present the results for the system with $N=96^2$ spins. 
In each simulation, we first perform the simulated annealing to find the low-energy configuration by gradually reducing the temperature with the rate $T_{n+1}=\alpha T_{n}$, where $T_n$ is the temperature in the $n$th step.
We set the initial temperature $T_0=0.1$-$1.0$ and take the coefficient $\alpha=0.99995$-$0.99999$. 
The final temperature is typically taken at $T=0.01$ for zero field and $T=0.0001$ for nonzero field (we need lower temperature for nonzero fields to resolve keen competition between different phases). 
The target temperatures are reached by spending totally $10^5$-$10^6$ Monte Carlo sweeps. 
At the final temperature, we perform $10^5$-$10^6$ Monte Carlo sweeps for measurements after $10^5$-$10^6$ steps for thermalization. 
We also start the simulations from the spin structures obtained at low temperatures to determine the phase boundaries between different magnetic states.

We identify the magnetic phase for each state obtained by the simulated annealing by calculating the spin and scalar chirality configurations. 
The spin structure factor is defined as 
\begin{align}
S^{\alpha \alpha}_s(\bm{q}) = \frac1N \sum_{j,l} \langle S_j^{\alpha} S_l^{\alpha} \rangle e^{i \bm{q} \cdot (\bm{r}_j-\bm{r}_l)},  
\end{align}
where $\bm{r}_j$ is the position vector at site $j$. 
As the magnetic interaction in the $\bm{Q}_\nu$ channel tends to stabilize the magnetic order with wave number $\bm{Q}_\nu$, we focus on the magnetic moment with the $\bm{Q}_\nu$ component, which is given by 
\begin{align}
\label{eq:mq}
m^\alpha_{\bm{Q}_\nu} = \sqrt{\frac{S^{\alpha \alpha}_s(\bm{Q}_\nu)}{N}}.
\end{align}
In the case of the single-ion anisotropy, we measure the in-plane component $(m^{xy}_{\bm{Q}_\nu})^2=(m^{x}_{\bm{Q}_\nu})^2+(m^{y}_{\bm{Q}_\nu})^2$, while in the case of the bond-dependent anisotropy, we measure the in-plane components of the magnetic moments, $m^{\bm{Q}_{\parallel} }_{\bm{Q}_\nu}$ and $m^{\bm{Q}_{\perp}}_{\bm{Q}_\nu}$, which are parallel and perpendicular to the $\bm{Q}_\nu$ direction, respectively; we take $(m^{\bm{Q}_{\parallel} }_{\bm{Q}_\nu},m^{\bm{Q}_{\perp}}_{\bm{Q}_\nu}, m^{z}_{\bm{Q}_\nu})$ to form the orthogonal coordinates. 
We also calculate the uniform component of the magnetization $m^\alpha_0$. 

Meanwhile, the chirality structure factor is defined as 
\begin{align}
\label{eq:chiralstructurefactor}
S_{\chi}(\bm{q})= \frac{1}{N}\sum_{\mu}\sum_{\bm{R},\bm{R}' \in \mu} \langle \chi_{\bm{R}}
\chi_{\bm{R}'}\rangle e^{i \bm{q}\cdot (\bm{R}-\bm{R}')}, 
\end{align}
where $\bm{R}$ and $\bm{R}'$ represent the position vectors at the centers of triangles, 
and $\mu=(u, d)$ represent upward and downward triangles, respectively; 
$\chi_{\bm{R}}= \bm{S}_j \cdot (\bm{S}_k \times \bm{S}_l)$ is the local spin chirality at $\bm{R}$, where $j,k,l$ are the sites on the triangle at $\bm{R}$ in the counterclockwise order. 
The scalar chirality with the $\bm{Q}_\nu$ component 
is defined as 
\begin{align}
\label{eq:chi_q}
&\chi_{\bm{Q}_\nu}=\sqrt{\frac{S_{\chi}(\bm{Q}_\nu)}{N}}. 
\end{align}
The uniform component is given by $\chi_{0}$. 
Note that, in this definition, a staggered arrangement of $\chi_{\bm{R}}$ also gives a nonzero $\chi_{0}$; we distinguish uniform and staggered ones by real-space pictures.

\section{Single-ion Anisotropy}
\label{sec:Single-ion Anisotropy}
In this section, we investigate the effect of the single-ion anisotropy $A$ for the Hamiltonian $\mathcal{H}= \mathcal{H}^{\rm BBQ}+\mathcal{H}^{\rm SIA}+\mathcal{H}^{\rm Z}$ (i.e., $\mathcal{H}^{\rm BA} = 0$). 
The magnetic phase diagram at zero magnetic field is presented in Sec.~\ref{sec:At zero field_SIA}. 
Then, the field-induced magnetic orders are discussed in Secs.~\ref{sec:Field along the $z$ direction_SIA}-\ref{sec:Field rotation in the $zx$ plane_SIA} for different field directions: the field along the $z$ direction in Sec.~\ref{sec:Field along the $z$ direction_SIA}, the field along the $x$ direction in Sec.~\ref{sec:Field along the $x$ direction_SIA}, and the field rotated in the $xz$ plane in Sec.~\ref{sec:Field rotation in the $zx$ plane_SIA}.

\subsection{At zero field}
\label{sec:At zero field_SIA}

\begin{figure}[htb!]
\begin{center}
\includegraphics[width=1.0 \hsize]{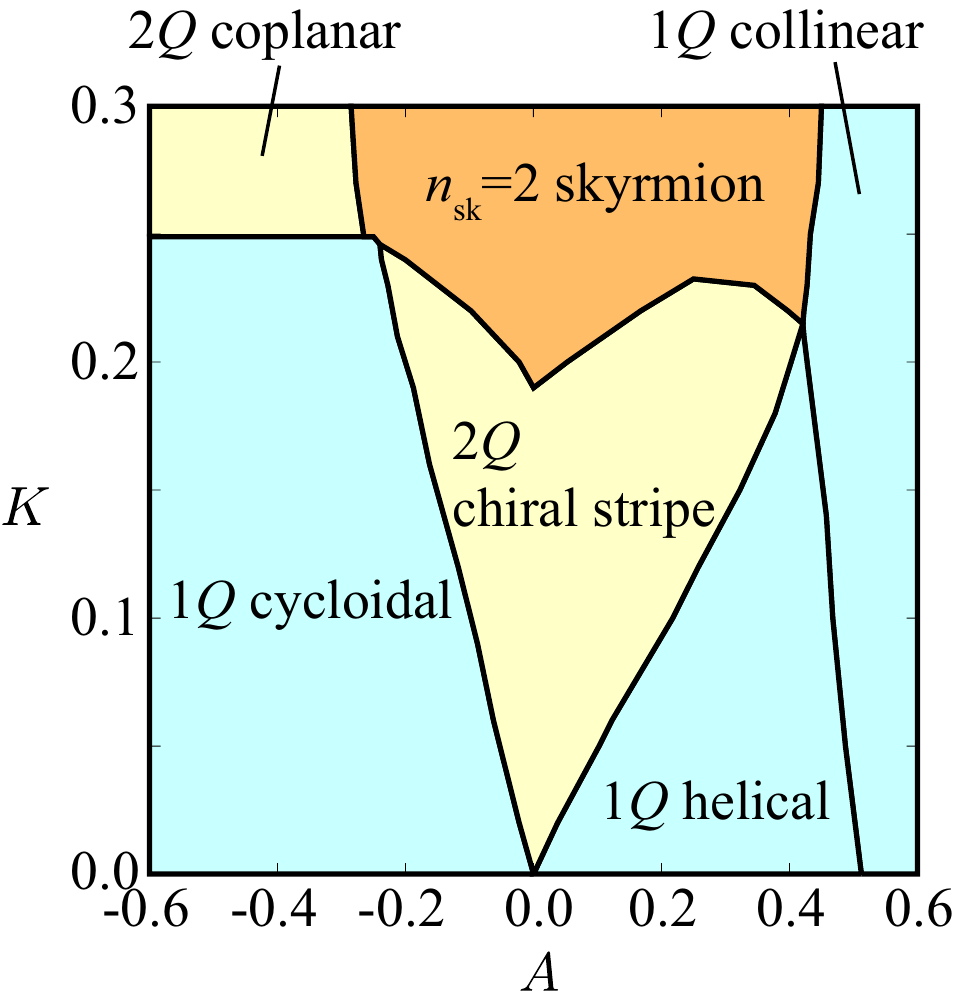} 
\caption{
\label{fig:souzu_A-K}
Magnetic phase diagram of the model in Eq.~(\ref{eq:Ham}) with $\mathcal{H}^{\rm BA}=\mathcal{H}^{\rm Z}=0$ obtained by the simulated annealing down to $T=0.01$.  
$A>0$ ($A<0$) represents the easy-axis 
(plane) anisotropy. 
See the text for details. 
}
\end{center}
\end{figure}

\begin{figure}[htb!]
\begin{center}
\includegraphics[width=1.0 \hsize]{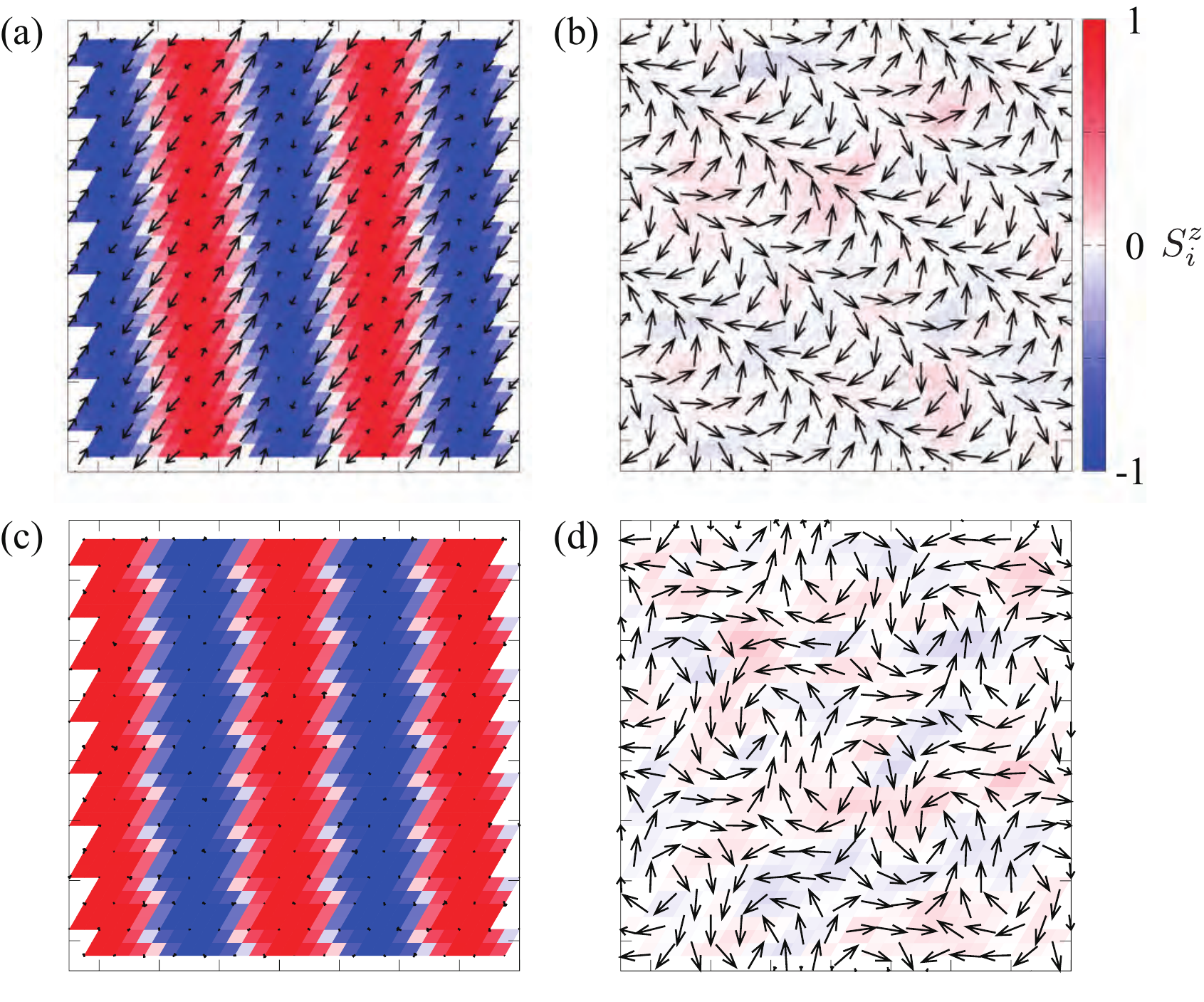} 
\caption{
\label{fig:Spin_zerofield_SIA}
Real-space spin configurations of (a) the single-$Q$ (1$Q$) helical state at $A=0.2$ and $K=0$, (b) the 1$Q$ cycloidal state at $A=-0.4$ and $K=0$, (c) the 1$Q$ collinear state at $A=0.6$ and $K=0$, and (d) the double-$Q$ (2$Q$) coplanar state at $A=-0.4$ and $K=0.3$.
The contour shows the $z$ component of the spin moment~\cite{comment_contour}, and the arrows represent the $xy$ components. 
}
\end{center}
\end{figure}

\begin{figure}[htb!]
\begin{center}
\includegraphics[width=1.0 \hsize]{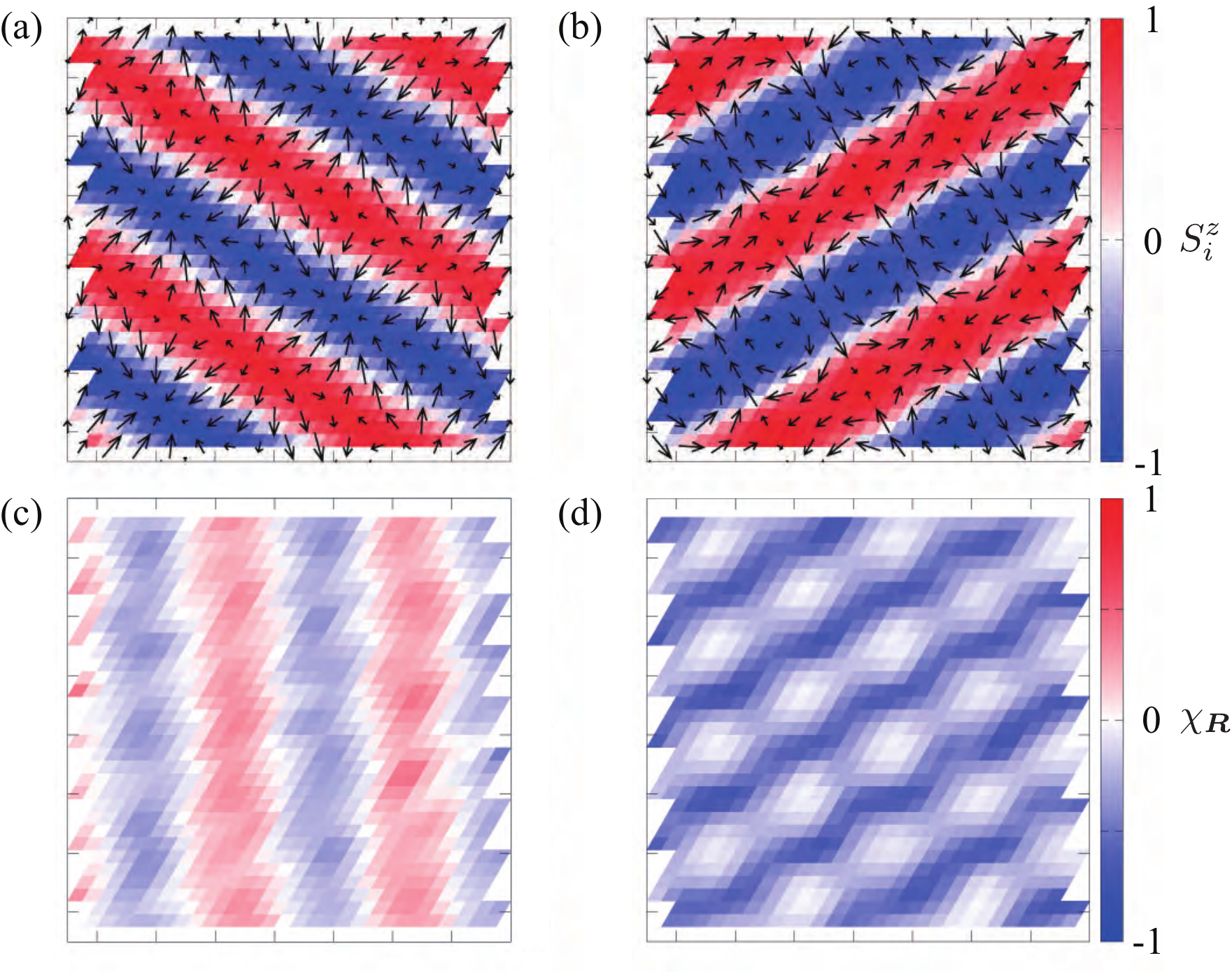} 
\caption{
\label{fig:Spin_zerofield_SIA_chiral}
Real-space spin configurations of (a) the 2$Q$ chiral stripe state at $A=0.1$ and $K=0.1$ and (b) the $n_{\rm sk}=2$ skyrmion crystal at $A=0.2$ and $K=0.3$. 
The contour shows the $z$ component of the spin moment, and the arrows represent the $xy$ components. 
(c) and (d) display the real-space chirality configurations corresponding to (a) and (b), respectively. 
}
\end{center}
\end{figure}

\begin{figure*}[htb!]
\begin{center}
\includegraphics[width=1.0 \hsize]{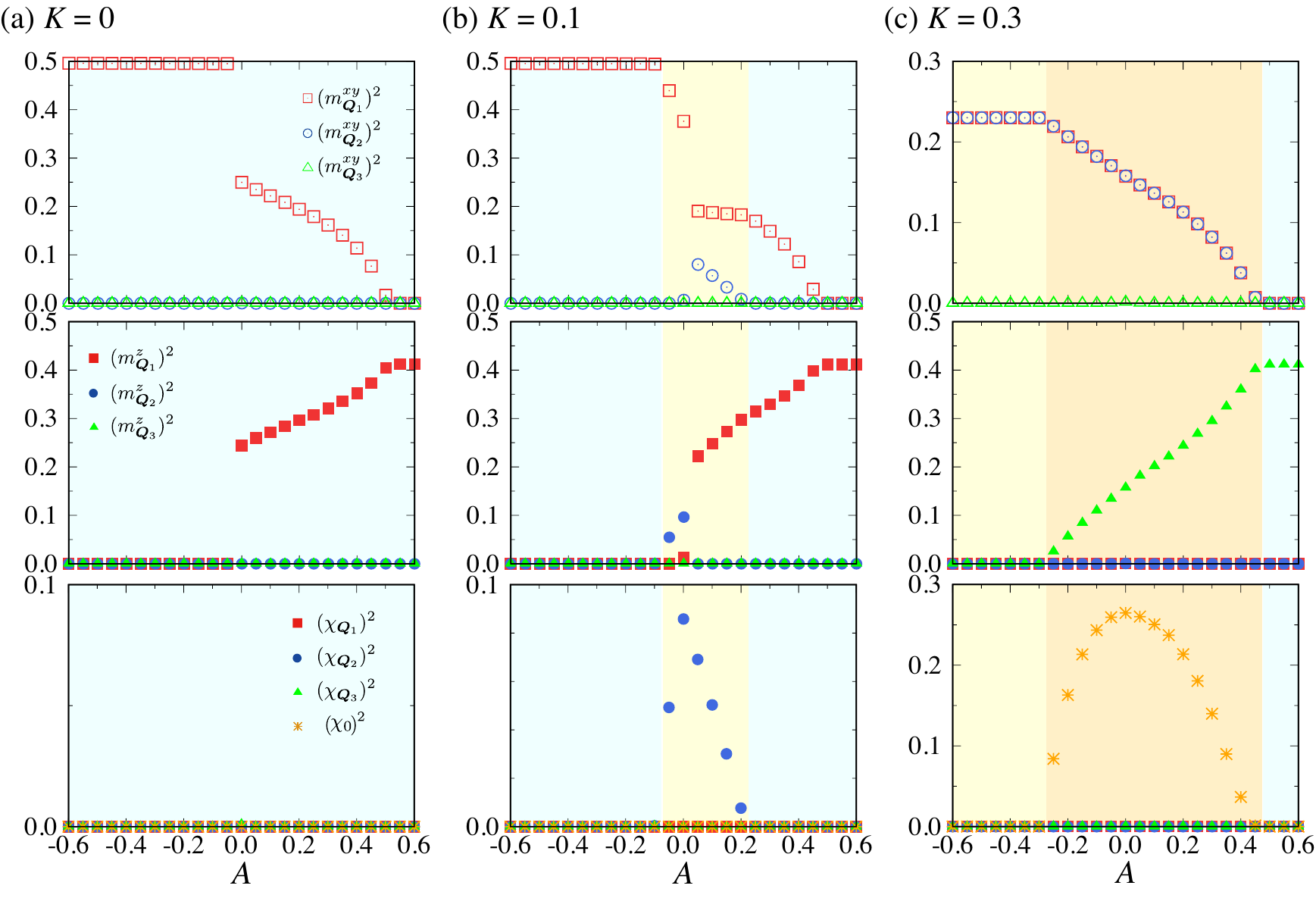} 
\caption{
\label{fig:Adep_H=0}
$A$ dependence of (first row) $(m^{xy}_{\bm{Q}_\nu})^2$, (second row) $(m^{z}_{\bm{Q}_\nu})^2$, and (third row) $ (\chi_{\bm{Q}_\nu})^2$ and $(\chi_0)^2$ for (a) $K=0$, (b) $K=0.1$, and (c) $K=0.3$. 
See also Fig.~\ref{fig:souzu_A-K}. 
}
\end{center}
\end{figure*}

First, we present the magnetic phase diagram for the model in Eq.~(\ref{eq:Ham}) with $\mathcal{H}^{\rm BA} = \mathcal{H}^{\rm Z} = 0$ obtained by the simulated annealing in Fig.~\ref{fig:souzu_A-K}. 
The result includes six phases, whose real-space configurations of spin and chirality are shown in Figs.~\ref{fig:Spin_zerofield_SIA} and \ref{fig:Spin_zerofield_SIA_chiral}. 
Each magnetic phase is characterized by the magnetic moments with the $\bm{Q}_\nu$ components, $\bm{m}_{\bm{Q}_\nu}$, the spin scalar chirality with the $\bm{Q}_\nu$ components, $\chi_{\bm{Q}_\nu}$, and the uniform component $(\chi_0)^2$.  
$A$ dependences of these quantities at $K=0$, $0.1$, and $0.3$ are shown in Fig.~\ref{fig:Adep_H=0}. 
Due to the sixfold rotational symmetry, $\bm{Q}_1$, $\bm{Q}_2$, and $\bm{Q}_3$ are symmetry-related; e.g., the single-$Q$ state with $\bm{m}_{\bm{Q}_1} \neq 0$ is equivalent with that with $\bm{m}_{\bm{Q}_2} \neq 0$ or $\bm{m}_{\bm{Q}_3} \neq 0$. 
Thus, three types of the single-$Q$ states are energetically degenerate, and hence, they are obtained randomly in the simulated annealing starting from different initial configurations. 
Similar degeneracy occurs also for other multiple-$Q$ states.
In the following, we show the results in each ordered state by appropriately sorting $(\bm{m}_{\bm{Q}_\nu})^2$ and $ (\chi_{\bm{Q}_\nu})^2$ for better readability. 

At $K=0$ where the model is reduced to the simple bilinear model with the single-ion anisotropy, the single-$Q$ state is stabilized for all $A$, although the spiral plane depends on the sign of $A$; 
the spins rotate  in the $xy$ plane for $A<0$, while in the $xz$ (or $yz$) plane for $A>0$, as shown in Figs.~\ref{fig:Spin_zerofield_SIA}(a) and \ref{fig:Spin_zerofield_SIA}(b), respectively.
This is represented by nonzero $(m^{xy}_{\bm{Q}_1})^2$ for $A<0$ and nonzero $(m^{z}_{\bm{Q}_1})^2$ for $A>0$, as shown in Fig.~\ref{fig:Adep_H=0}(a). 
The former is an in-plane cycloidal spiral, while the latter is an out-of-plane cycloidal or proper-screw spiral. 
We call the former the single-$Q$ cycloidal state and the latter the single-$Q$ helical state. 
In the single-$Q$ cycloidal state, $(m^z_{\bm{Q}_1})^2$ is zero and $(m^{xy}_{\bm{Q}_1})^2$ does not depend on $A$, while in the single-$Q$ helical state, both $(m^{xy}_{\bm{Q}_1})^2$ and $(m^{z}_{\bm{Q}_1})^2$ are nonzero and their ratio changes as $A$.  
This indicates that the spiral plane in the cycloidal and helical states are circular and elliptical, respectively.  
While increasing positive $A$, $(m^{xy}_{\bm{Q}_1})^2$ decreases and $(m^{z}_{\bm{Q}_1})^2$ increases to gain the energy from the single-ion anisotropy. 
While further increasing $A$, $(m^{xy}_{\bm{Q}_1})^2$ vanishes and the $1Q$ collinear state with $(m^z_{\bm{Q}_1})^2 \neq 0$ is realized for $A \gtrsim 0.55$. 
The real-space spin texture in the $1Q$ collinear state is shown in Fig.~\ref{fig:Spin_zerofield_SIA}(c). 

By introducing the biquadratic interaction $K$, the double-$Q$ state is stabilized in the small $|A|$ region, as shown in Fig.~\ref{fig:souzu_A-K}. 
The spin and chirality components are shown in the case of $K=0.1$ in Fig.~\ref{fig:Adep_H=0}(b). 
This double-$Q$ state is composed of two helices, as indicated by the nonzero $(\bm{m}_{\bm{Q}_1})^2$ and $(\bm{m}_{\bm{Q}_2})^2$ with different intensities, $(\bm{m}_{\bm{Q}_1})^2 > (\bm{m}_{\bm{Q}_2})^2$. 
At the same time, this state shows nonzero $(\chi_{\bm{Q}_2})^2$. 
The spin and chirality configurations obtained by the simulation are presented in Figs.~\ref{fig:Spin_zerofield_SIA_chiral}(a) and \ref{fig:Spin_zerofield_SIA_chiral}(c), respectively. 
This type of the double-$Q$ state has been found in the itinerant electron systems without the single-ion anisotropy, such as the Kondo lattice model with the weak spin-charge coupling~\cite{Ozawa_doi:10.7566/JPSJ.85.103703} and the $d$-$p$ model with the strong Hund's-rule coupling~\cite{yambe2020double}, where it is called the double-$Q$ chiral stripe state~\cite{Ozawa_doi:10.7566/JPSJ.85.103703}.  
In the limit of $A \to 0$, 
the real-space spin configuration is given by~\cite{Ozawa_doi:10.7566/JPSJ.85.103703} 
\begin{align}
\label{eq:doubleQ}
\bm{S}_i =
\left(
    \begin{array}{c}
     \sqrt{1-b^2+b^2 \cos \bm{Q}_2 \cdot \bm{r}_i}   \cos\bm{Q}_1 \cdot \bm{r}_i  \\
     \sqrt{1-b^2+b^2 \cos \bm{Q}_2 \cdot \bm{r}_i}  \sin \bm{Q}_1 \cdot \bm{r}_i  \\
      b \sin \bm{Q}_2 \cdot \bm{r}_i 
          \end{array}
  \right)^{\rm T}, 
\end{align}
which is approximately regarded as a superposition of the dominant spiral wave with $\bm{Q}_1$ in the $xy$ plane and the sinusoidal wave with $\bm{Q}_2$ along the $z$ direction. 
$b$ represents the amplitude of the latter component. 
In the case of $A=0$, the spiral plane is arbitrary; the energy is unchanged for any global spin rotation. 
A nonzero $A$ fixes the spiral plane. 
For $A>0$, the $\bm{Q}_1$ spiral is laid on the $xz$ (or $yz$) plane and becomes elliptical, and the 
sinusoidal $\bm{Q}_2$ component runs along the $y$ (or $x$) direction. 
On the other hand, for $A<0$, the double-$Q$ chiral stripe consists of the dominant spiral in the $xy$ plane and the additional sinusoidal wave along the $z$ direction, as shown in Fig.~\ref{fig:Adep_H=0}(b). 
By increasing $|A|$, the double-$Q$ chiral stripe continuously turns into the single-$Q$ cycloidal state for $A<0$ and the single-$Q$ helical state for $A>0$, which are connected to those at $K=0$. 
Note that the former approximately corresponds to $b \to 0$ in Eq.~(\ref{eq:doubleQ}). 
The region of the double-$Q$ chiral stripe state is extended by increasing $K$, as shown in Fig.~\ref{fig:souzu_A-K}. 

For larger $K$, two different multiple-$Q$ phases appear: the $n_{\rm sk}=2$ skyrmion crystal 
for $-0.3 \lesssim A \lesssim 0.5$ and the double-$Q$ coplanar state for $A \lesssim -0.3$, as shown in Fig.~\ref{fig:souzu_A-K}.  
The $n_{\rm sk}=2$ skyrmion crystal is a triple-$Q$ magnetic state by a superposition of three sinusoidal waves orthogonal to each other, $\bm{m}_{\bm{Q}_1} \perp \bm{m}_{\bm{Q}_2} \perp \bm{m}_{\bm{Q}_3}$~\cite{Ozawa_PhysRevLett.118.147205,Hayami_PhysRevB.95.224424}. 
The typical spin configuration is shown in Fig.~\ref{fig:Spin_zerofield_SIA_chiral}(b). 
While $(\bm{m}_{\bm{Q}_1})^2=(\bm{m}_{\bm{Q}_2})^2=(\bm{m}_{\bm{Q}_3})^2$ at $A=0$, the intensities at $\bm{Q}_\eta$ for the $xy$ component become larger (smaller) than those for the $z$ component for $A<0$ ($A>0$), as shown in Fig.~\ref{fig:Adep_H=0}(c). 
The $xy$ component always shows the double-$Q$ structure with equal intensities, while the $z$ component is single-$Q$. 
This magnetic structure has a noncoplanar spin configuration, leading to nonzero scalar chirality, as shown in Figs.~\ref{fig:Spin_zerofield_SIA_chiral}(d) and \ref{fig:Adep_H=0}(c), which gives rise to the topological Hall effect. 

By increasing $A$, the $xy$ spin component vanishes as shown in the top panel of Fig.~\ref{fig:Adep_H=0}(c), and then the $n_{\rm sk}=2$ skyrmion crystal turns into the $1Q$ collinear state continued from the smaller $K$ region. 
Meanwhile, when decreasing $A$, the $z$ spin component vanishes as shown in the middle panel of Fig.~\ref{fig:Adep_H=0}(c), and the double-$Q$ coplanar state with $(m_{{\bf Q}_1})^2=(m_{{\bf Q}_2})^2>0$ and $(m_{{\bf Q}_3})^2=0$ is realized whose spin texture is shown in Fig.~\ref{fig:Spin_zerofield_SIA}(d). 

A similar phase sequence of the double-$Q$ coplanar, $n_{\rm sk}=2$ skyrmion crystal, and single-$Q$ collinear states while changing the single-ion anisotropy was obtained also for the original Kondo lattice model~\cite{Hayami_PhysRevB.99.094420}. 
Thus, our effective spin model in Eq.~(\ref{eq:Ham}) can capture the instability toward multiple-$Q$ states in itinerant magnets qualitatively in the large $K$ region, as demonstrated for the isotropic case~\cite{Hayami_PhysRevB.95.224424}. 
However, by closely comparing the results, we find at least two differences between the two models. 
One is the nature of the phase transitions: In the effective spin model, the transitions from the $n_{\rm sk}=2$ skyrmion crystal to the single-$Q$ collinear and double-$Q$ coplanar states appear to be of second order with continuous changes of the magnetic moments $m_{\bf{Q}_\eta}$ and the uniform scalar chirality $(\chi_0)^2$, while the results in the Kondo lattice model indicate the first-order transitions with clear jumps in these quantities. 
The other difference is that a noncoplanar double-$Q$ phase appears in a narrow region between the $n_{\rm sk}=2$ skyrmion crystal and the double-$Q$ coplanar state in the Kondo lattice model. 
These differences might be attributed to some factors which are omitted in the derivation of the effective spin model from the Kondo lattice model, such as the interactions at wave numbers other than ${\bf Q}_\eta$ and other types of magnetic interactions dropped off in the perturbation processes in itinerant magnets. 
Nevertheless, our result indicates that the effective spin model is useful to investigate the multiple-$Q$ instability in the Kondo lattice model, since it provides us with an overall picture of the emergent multiple-$Q$ phases, by a considerably smaller computational cost than that by the direct numerical simulation of the Kondo lattice model.

\begin{figure*}[htb!]
\begin{center}
\includegraphics[width=1.0 \hsize]{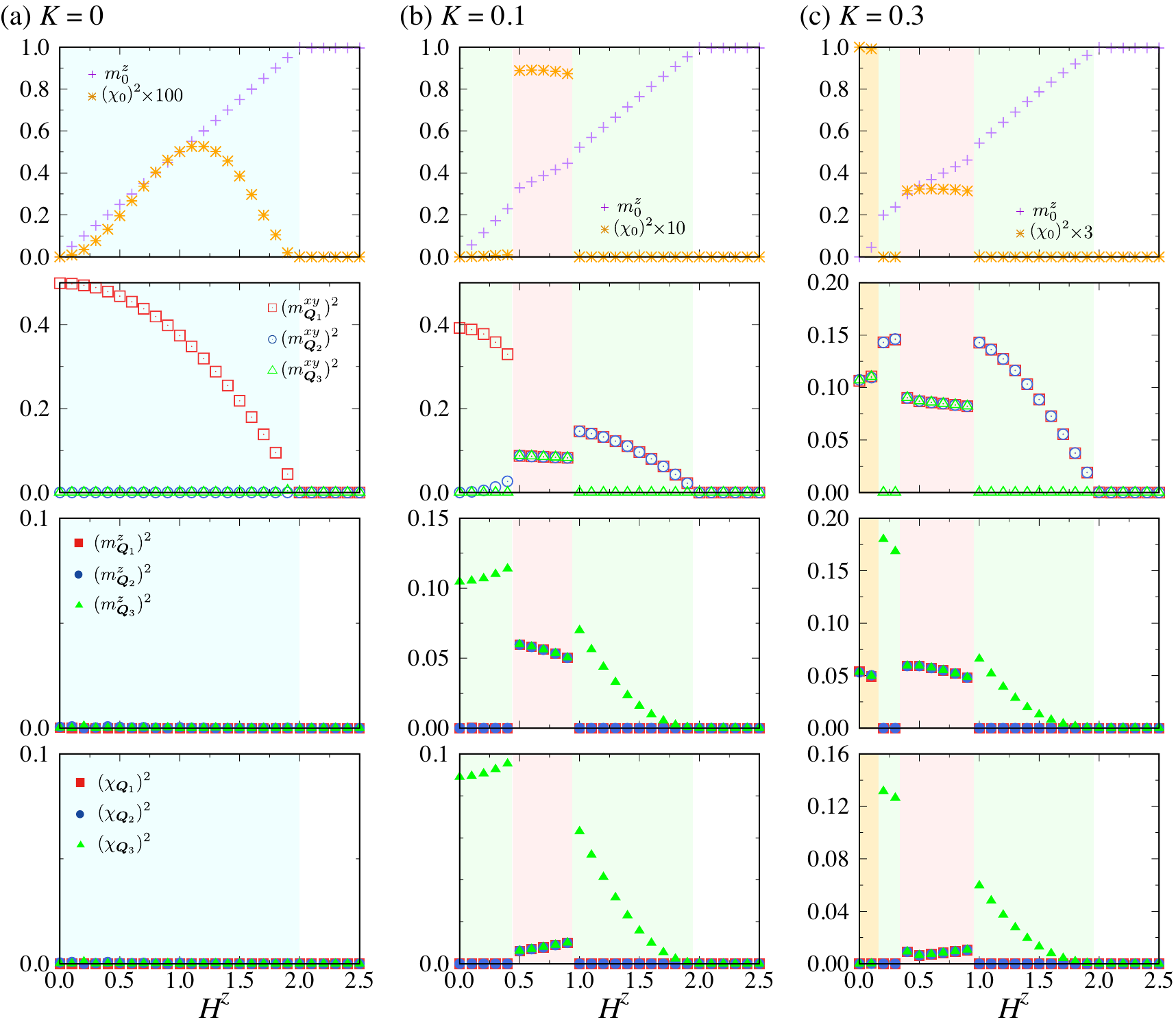} 
\caption{
\label{fig:SIA_Hz_A=0}
$H^z$ dependence of (first row) 
$m^z_0$ and 
$(\chi_0)^2$, (second row) $(m^{xy}_{\bm{Q}_\nu})^2$, (third row) $(m^{z}_{\bm{Q}_\nu})^2$, and (fourth row) $ (\chi_{\bm{Q}_\nu})^2$ for (a) $K=0$, (b) $K=0.1$, and (c) $K=0.3$ at $A=0$. 
}
\end{center}
\end{figure*}

\begin{figure}[htb!]
\begin{center}
\includegraphics[width=1.0 \hsize]{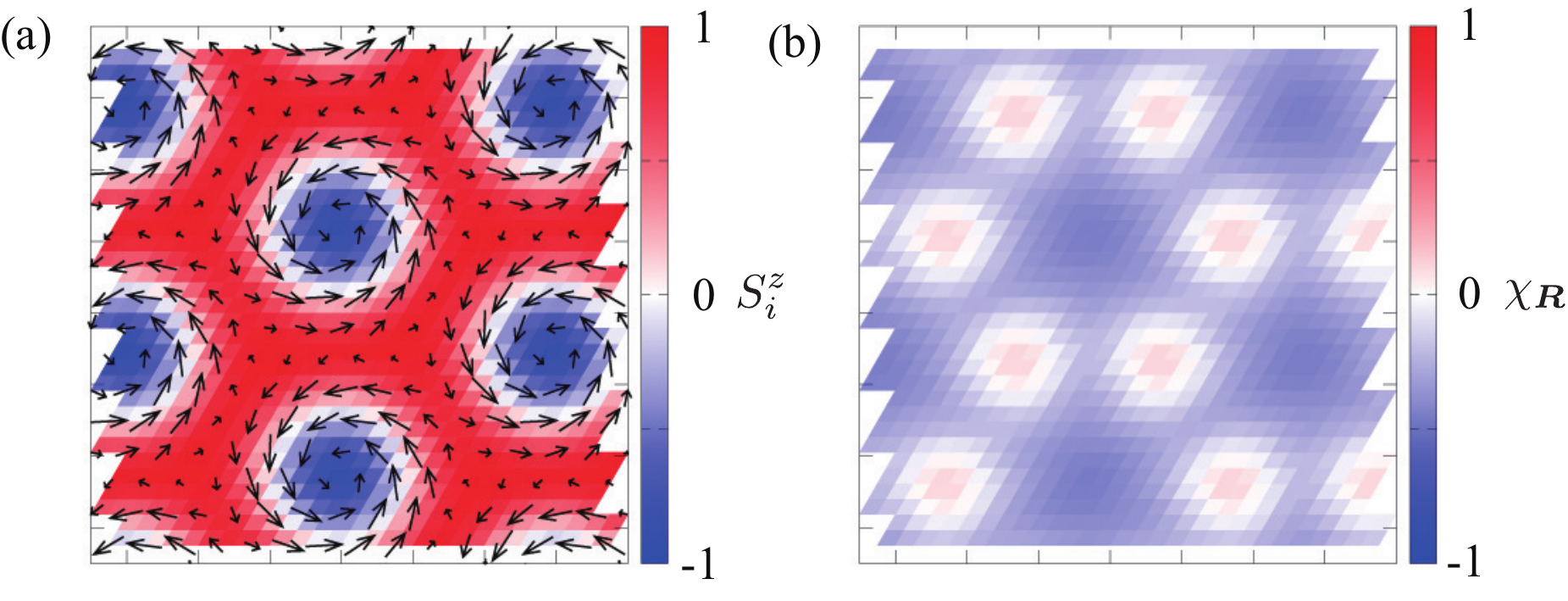} 
\caption{
\label{fig:Spin_nsk1SkX}
(a) Real-space spin configuration of the $n_{\rm sk}=1$ skyrmion crystal at $K=0.1$ and $H=0.6$. 
The contour shows the $z$ component of the spin moment, and the arrows represent the $xy$ components. 
(b) Real-space chirality configuration corresponding to (a). 
}
\end{center}
\end{figure}

\subsection{Field along the $z$ direction}
\label{sec:Field along the $z$ direction_SIA}
Next, we examine the effect of the magnetic field along the $z$ direction. 
We take $\bm{H}=(0,0,H^z)$ in the Zeeman Hamiltonian $\mathcal{H}^{\rm Z}$ in Eq.~(\ref{eq:Ham_Z}). 
We show the results for the isotropic case at $A=0$ in Sec.~\ref{sec:Isotropic case_Hz}, and the effects of the easy-axis and easy-plane anisotropy in Secs.~\ref{With easy-axis anisotropy_Hz} and \ref{sec:With easy-plane anisotropy_Hz}, respectively. 
We discuss the results in this section in Sec.~\ref{sec:Summary of this section_1}. 

\subsubsection{Isotropic case}
\label{sec:Isotropic case_Hz}

We first consider the situation in the absence of the single-ion anisotropy, $A=0$. 
The phase diagram in the $K$-$H^z$ plane and a part of the results were shown in the previous study by the authors~\cite{Hayami_PhysRevB.95.224424}. 
We here discuss the changes of the spin and chirality structures in detail. 
Figure~\ref{fig:SIA_Hz_A=0} shows the magnetic field dependence of the spin and chirality components at $K=0$, $0.1$, and $0.3$. 
Note that the following results in this section are the same for the magnetic field along any direction due to the spin rotational symmetry. 

At $K=0$, the magnetic state at zero field is the single-$Q$ spiral state, whose spiral plane is arbitrary due to the spin rotational symmetry. 
When applying the magnetic field in the $z$ direction, the spiral plane is fixed in the $xy$ plane, 
and the spin pattern is characterized by $(m^{xy}_{\bm{Q}_1})^2$ in addition to the uniform component of the magnetization along the $z$ direction, $m^z_0$, as shown in the upper two panels of Fig.~\ref{fig:SIA_Hz_A=0}(a). 
This corresponds to the single-$Q$ conical spiral where the spiral plane is perpendicular to the field direction. 
Reflecting the noncoplanar spin structure, this single-$Q$ conical state exhibits a staggered arrangement of nonzero local scalar chirality between the upward and downward triangles, as signaled by nonzero $(\chi_0)^2$ shown in the top panel of Fig.~\ref{fig:SIA_Hz_A=0}(a). 
Note that the scalar chirality cancels out between the staggered components. 
While increasing $H^z$, the single-$Q$ conical state continuously changes into the fully-polarized state at $H^z=2$. 

At $K=0.1$, the double-$Q$ chiral stripe state is stabilized at zero field, as discussed in the previous section. 
In the presence of the magnetic field, this state survives up to $H^z \simeq 0.4$, as shown in Fig.~\ref{fig:SIA_Hz_A=0}(b). 
We note that $(\chi_0)^2$ takes a small nonzero value for $0 < H^z \lesssim 0.4$ because of the nonzero staggered chirality induced by the magnetic field, similar to the single-$Q$ state at $K=0$ above.  
While increasing $H^z$, this state is replaced with the $n_{\rm sk}=1$ skyrmion crystal at $H^z \simeq 0.5$ with a finite jump of $(\chi_0)^2$. 
The $n_{\rm sk}=1$ skyrmion crystal is characterized by the triple-$Q$ peak structures for both $xy$ and $z$ components in the spin structure, 
as shown in the two middle panels of Fig.~\ref{fig:SIA_Hz_A=0}(b). 
It also exhibits the triple-$Q$ peak structures in the chirality as shown in the lowest panel of Fig.~\ref{fig:SIA_Hz_A=0}(b), in addition to $(\chi_0)^2$.  
Thus, both spin and chirality configurations in real space have threefold rotational symmetry, as shown in Figs.~\ref{fig:Spin_nsk1SkX}(a) and \ref{fig:Spin_nsk1SkX}(b). 
When further increasing $H^z$, the system undergoes a first-order phase transition to a triple-$Q$ state at $H^z \simeq 1$, which has double-$Q$ peaks in the $xy$ component and a single-$Q$ peak in the $z$ component of the magnetic moments. 
This triple-$Q$ state accompanies the single-$Q$ chirality density wave with $\bm{Q}_3$. 
The triple-$Q$ state turns into the fully-polarized state at $H^{z}= 2$. 

Figure~\ref{fig:SIA_Hz_A=0}(c) displays the result at $K=0.3$. 
The $n_{\rm sk}=2$ skyrmion crystal at zero field is replaced with a triple-$Q$ state at $H^z \simeq 0.2$, which is similar to the high-field state at $K=0.1$. 
While further increasing $H^z$, it turns into the $n_{\rm sk}=1$ skyrmion crystal at $H^z \simeq 0.4$. 
After that, the phase sequence is similar to that at $K=0.1$. 

\begin{figure}[htb!]
\begin{center}
\includegraphics[width=1.0 \hsize]{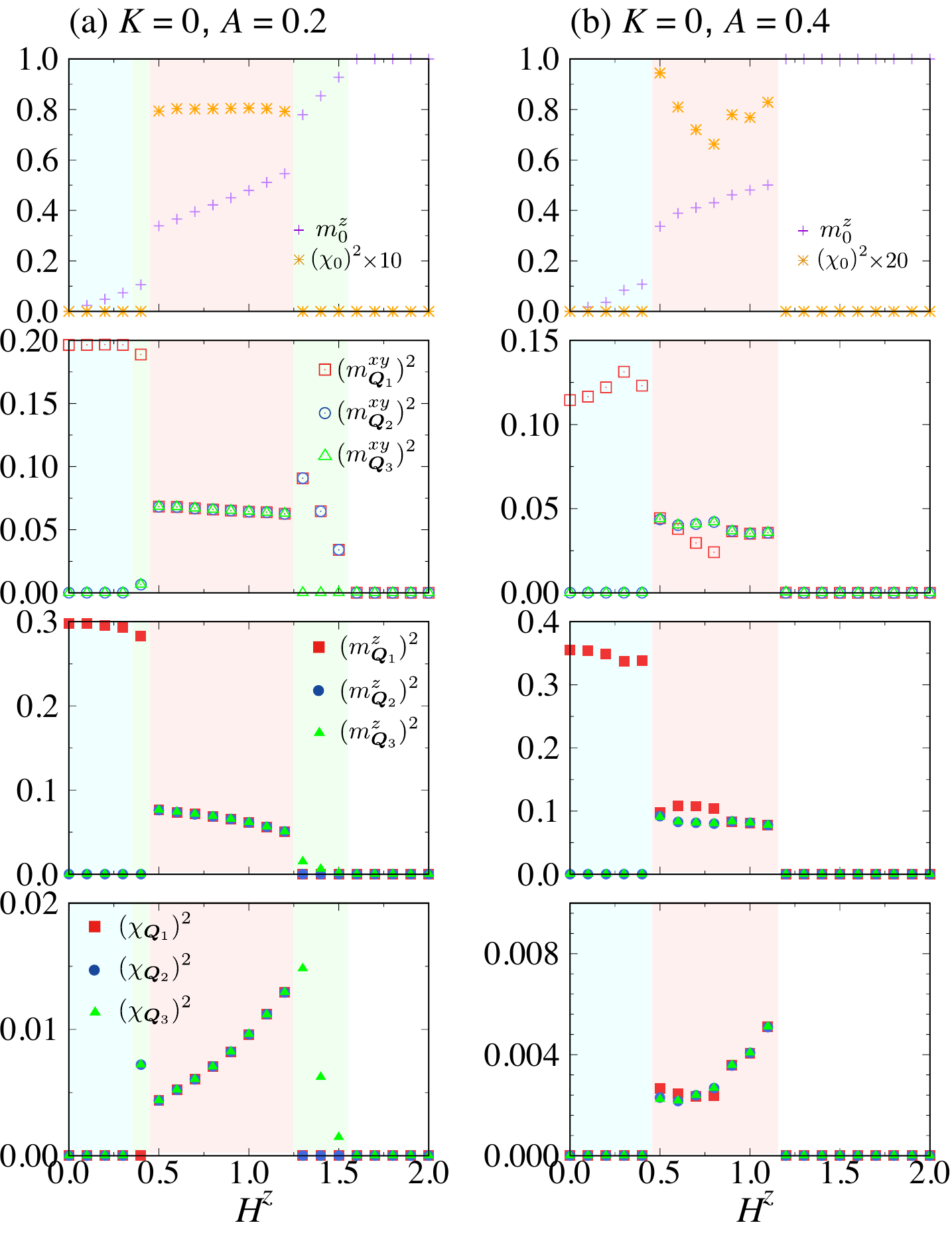} 
\caption{
\label{fig:SIA_Hz_A=posi_K=0}
$H^z$ dependence of (first row) $m^z_0$ and $(\chi_0)^2$, (second row) $(m^{xy}_{\bm{Q}_\nu})^2$, (third row) $(m^{z}_{\bm{Q}_\nu})^2$, and (fourth row) $ (\chi_{\bm{Q}_\nu})^2$ for $K=0$ at (a) $A=0.2$ and (b) $A=0.4$. 
}
\end{center}
\end{figure}

\begin{figure}[htb!]
\begin{center}
\includegraphics[width=1.0 \hsize]{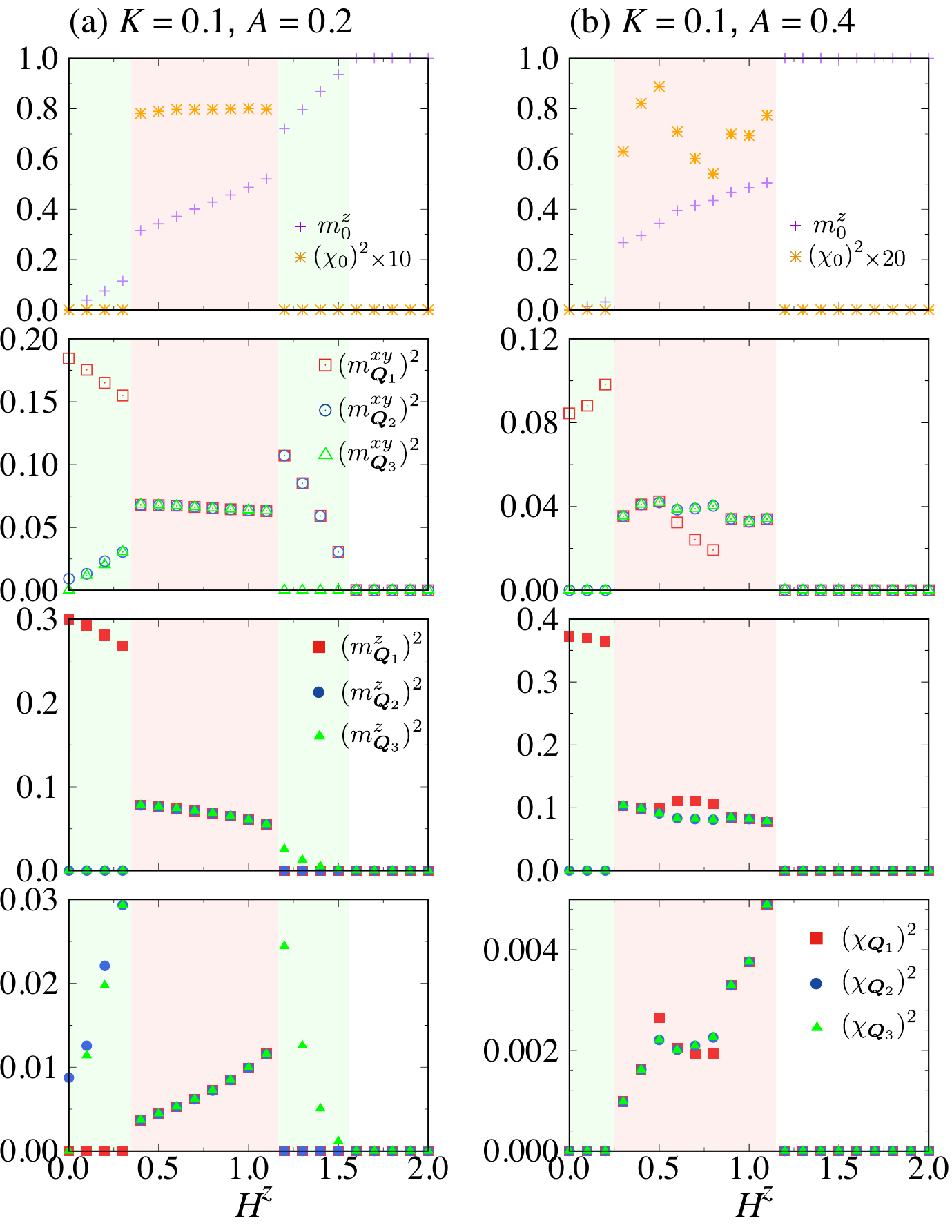} 
\caption{
\label{fig:SIA_Hz_A=posi_K=01}
The same plots as in Fig.~\ref{fig:SIA_Hz_A=posi_K=0} for $K=0.1$. 
}
\end{center}
\end{figure}

\begin{figure}[htb!]
\begin{center}
\includegraphics[width=1.0 \hsize]{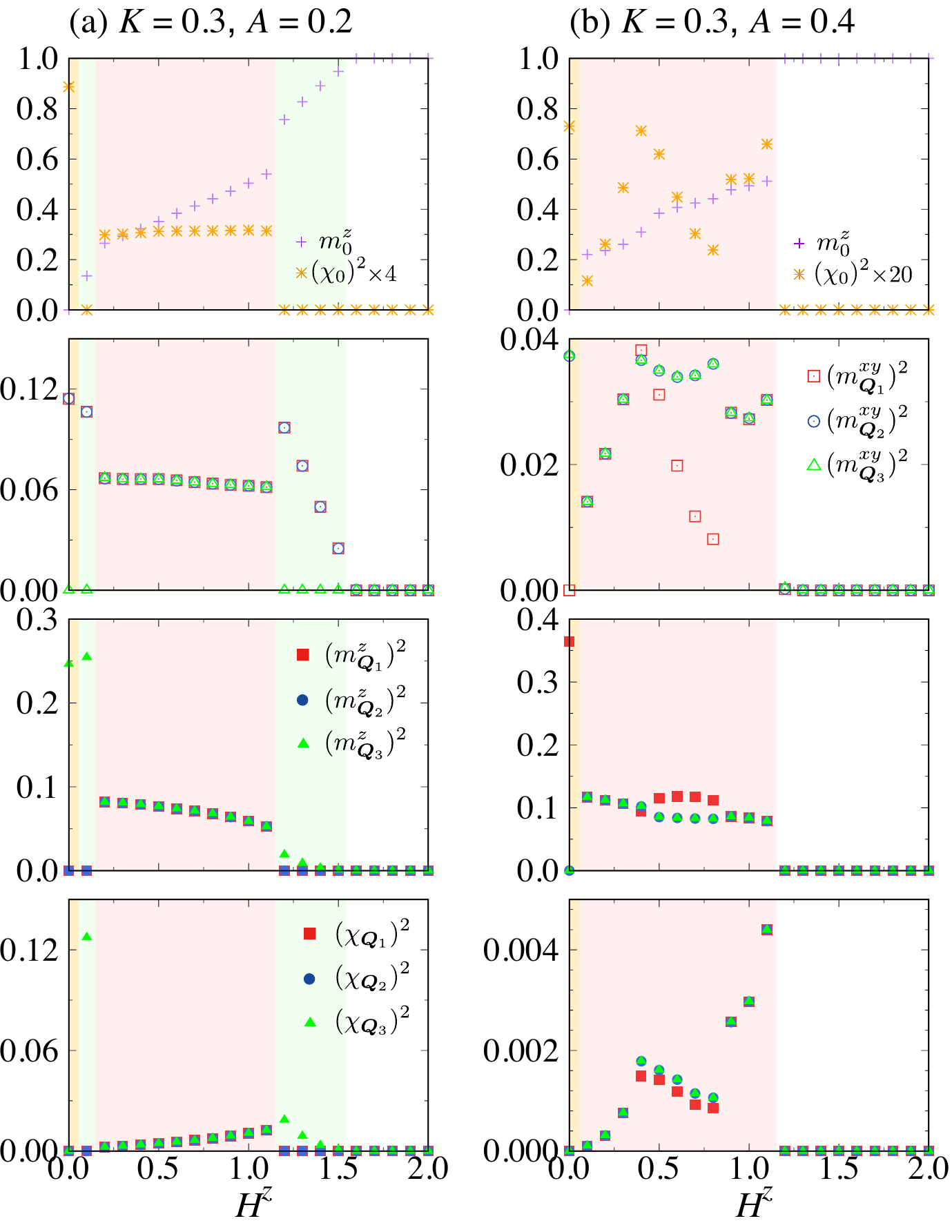} 
\caption{
\label{fig:SIA_Hz_A=posi_K=03}
The same plots as in Fig.~\ref{fig:SIA_Hz_A=posi_K=0} for $K=0.3$. 
}
\end{center}
\end{figure}

\subsubsection{With easy-axis anisotropy}
\label{With easy-axis anisotropy_Hz}

\begin{figure}[htb!]
\begin{center}
\includegraphics[width=1.0 \hsize]{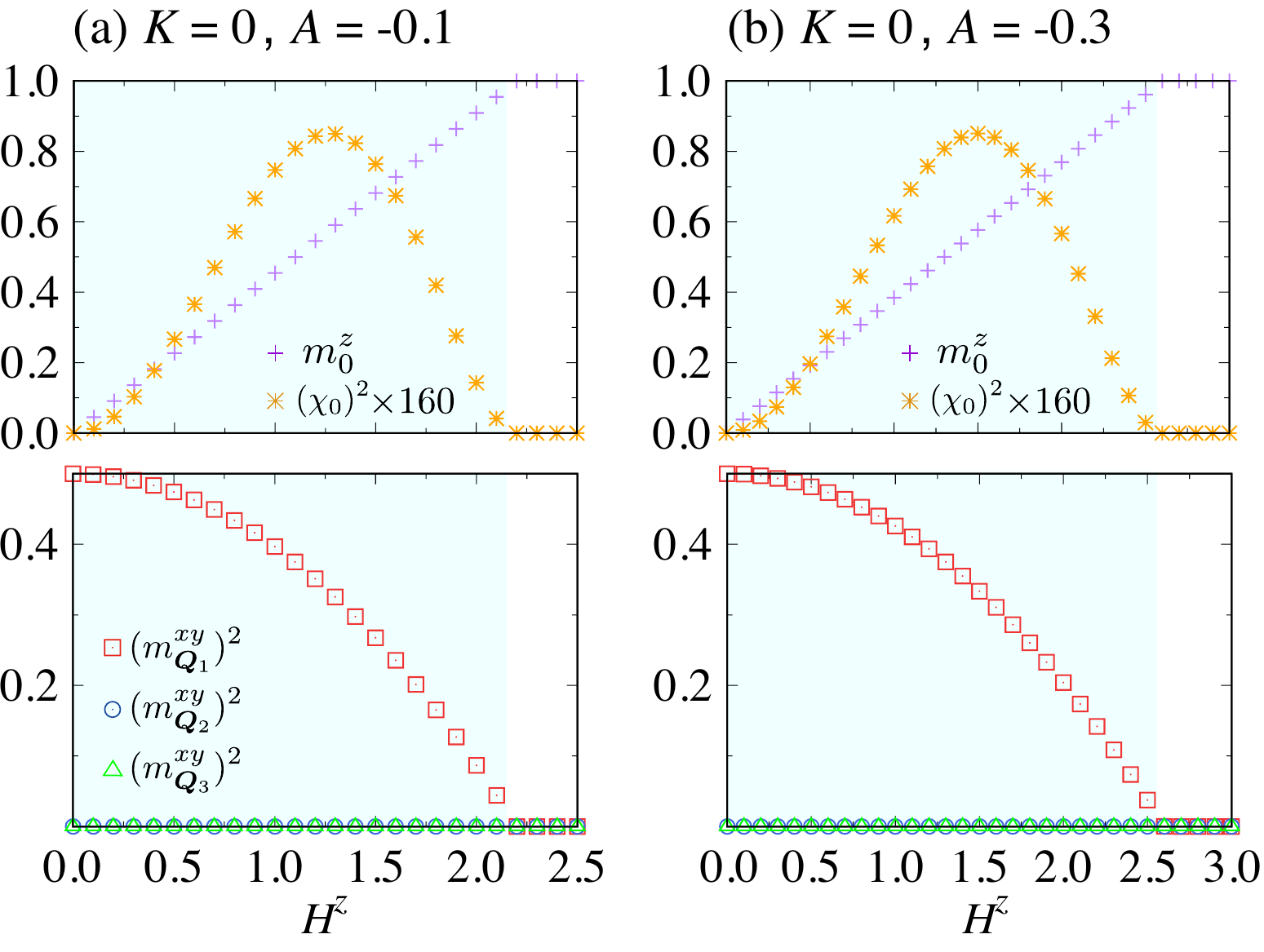} 
\caption{
\label{fig:SIA_Hz_A=nega_K=0}
$H^z$ dependence of (upper) $m^z_0$ and $(\chi_0)^2$ and (lower) $(m^{xy}_{\bm{Q}_\nu})^2$ for $K=0$ at (a) $A=-0.1$ and (b) $A=-0.3$. 
}
\end{center}
\end{figure}

\begin{figure}[htb!]
\begin{center}
\includegraphics[width=1.0 \hsize]{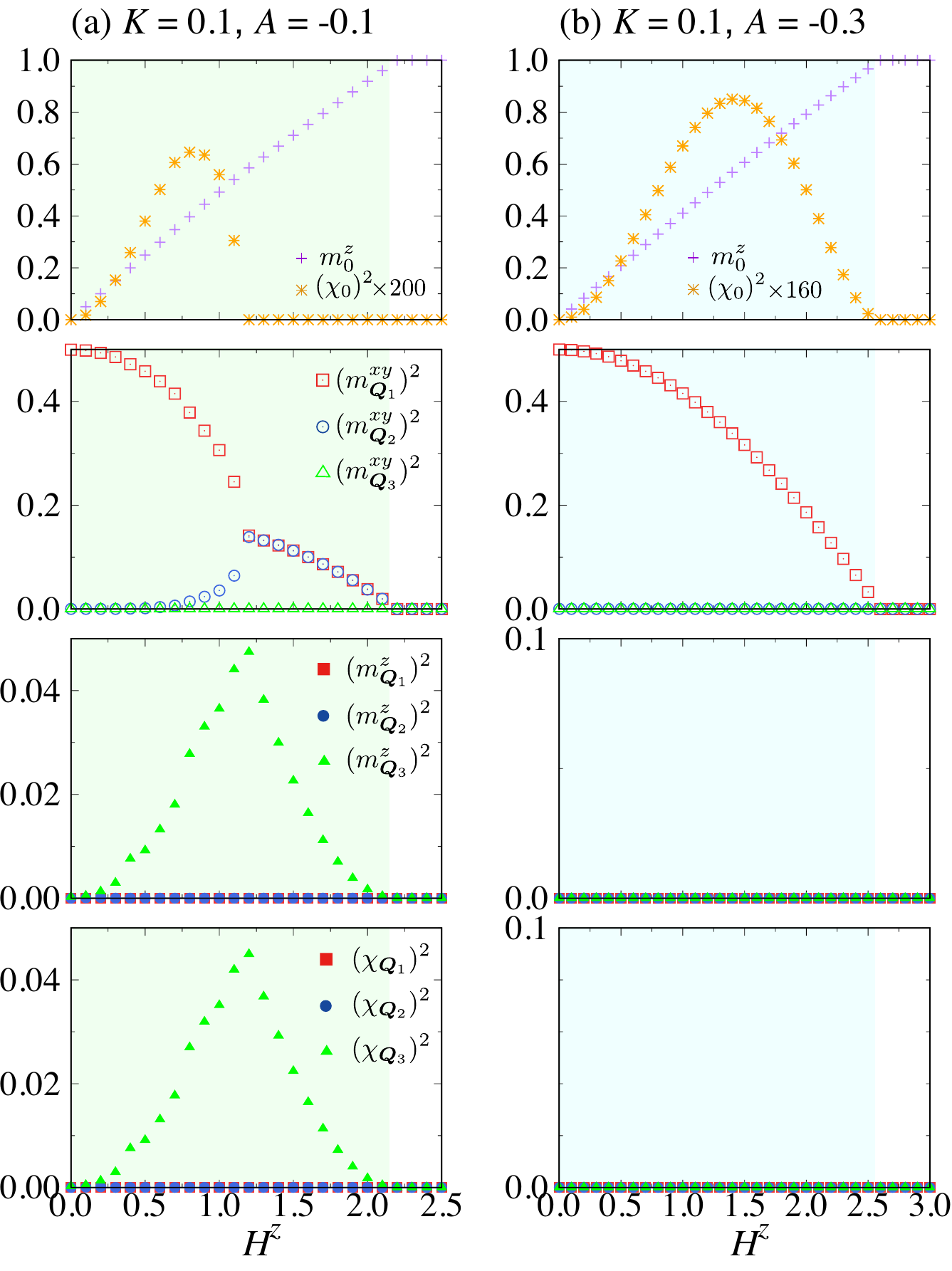} 
\caption{
\label{fig:SIA_Hz_A=nega_K=01}
$H^z$ dependence of (first row) $m^z_0$ and $(\chi_0)^2$, (second row) $(m^{xy}_{\bm{Q}_\nu})^2$, (third row) $(m^{z}_{\bm{Q}_\nu})^2$, and (fourth row) $ (\chi_{\bm{Q}_\nu})^2$ for $K=0.1$ at (a) $A=-0.1$ and (b) $A=-0.3$. 
}
\end{center}
\end{figure}

\begin{figure}[htb!]
\begin{center}
\includegraphics[width=1.0 \hsize]{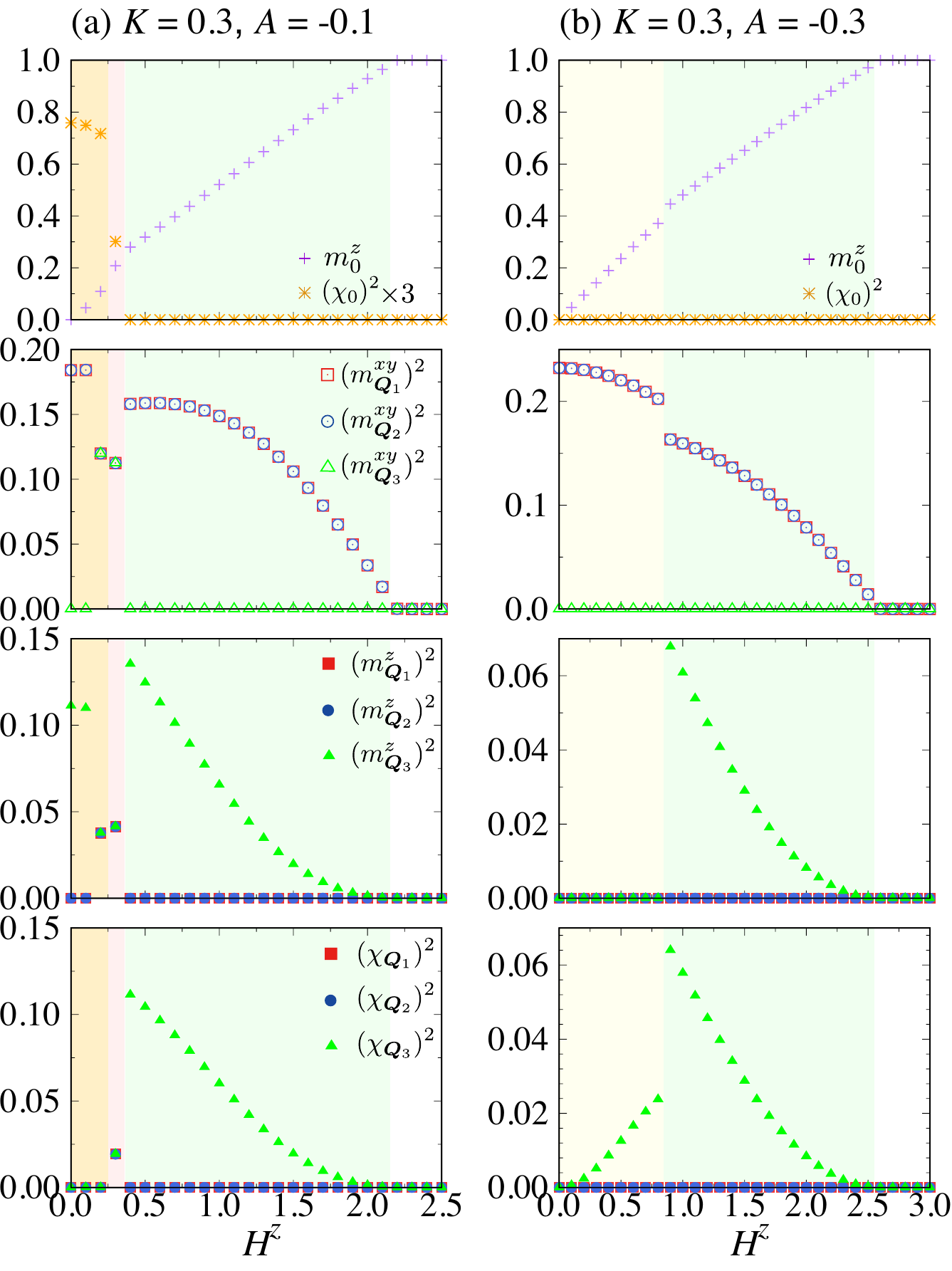} 
\caption{
\label{fig:SIA_Hz_A=nega_K=03}
The same plots as in Fig.~\ref{fig:SIA_Hz_A=nega_K=01} for $K=0.3$. 
}
\end{center}
\end{figure}

Next, we discuss the effect of the magnetic field along the $z$ direction, $H^z$, in the presence of the single-ion anisotropy for several $K$ and $A$. 
We show the results for the easy-axis anisotropy $A>0$ in Figs.~\ref{fig:SIA_Hz_A=posi_K=0}-\ref{fig:SIA_Hz_A=posi_K=03} in this section and for the easy-plane anisotropy $A<0$ in Figs.~\ref{fig:SIA_Hz_A=nega_K=0}-\ref{fig:SIA_Hz_A=nega_K=03} in the next section. 

Figures~\ref{fig:SIA_Hz_A=posi_K=0}(a) and \ref{fig:SIA_Hz_A=posi_K=0}(b) show the results at $A=0.2$ and $A=0.4$ with $K=0$, respectively. 
The main difference from the isotropic case with $A=0$ is found in the emergence of the $n_{\rm sk}=1$ skyrmion crystal in the intermediate-field region. 
This indicates that the easy-axis anisotropy can stabilize the $n_{\rm sk}=1$ skyrmion crystal even without the biquadratic interaction, consistent with the previous result in Ref.~\onlinecite{Wang_PhysRevLett.124.207201}. 
It is also found that $(\chi_0)^2$ becomes smaller for larger $A$, since the positive $A$ tends to align the spins along the $z$ direction, namely, it enhances $(m_{\bm{Q}_\nu}^z)^2$ and suppresses $(m_{\bm{Q}_\nu}^{xy})^2$. 
For $A=0.4$, there are two types of the $n_{\rm sk}=1$ skyrmion crystal, which are almost energetically degenerate: One shows weak anisotropy in both spin and chirality structures for $0.5 \lesssim H^z \lesssim 0.9$, and the other has the isotropic intensities for $0.9 \lesssim H^z \lesssim 1.1$.
Such quasi-degenerate skyrmion crystals have also been found in an itinerant 
electron model~\cite{Hayami_PhysRevB.99.094420} and a localized spin model~\cite{sotnikov2020quantum}, which indicates that optimized spin configurations in the skyrmion crystal is determined from a subtle balance among different interaction energies. 

Differences from the result at $A=0$ are also found in the low- and high-field regions. 
In the low-field region, the single-$Q$ spiral state is realized similar to the $A=0$ case, but $(m^{z}_{\bm{Q}_1})^2$ becomes nonzero for $A>0$, as shown in the middle row of Fig.~\ref{fig:SIA_Hz_A=posi_K=0}. 
In addition, at $A=0.2$, $(m^{xy}_{\bm{Q}_2})^2$ and $(m^{xy}_{\bm{Q}_3})^2$ become nonzero in the vicinity of the phase boundary at $H\simeq 0.4 $, as shown in Fig.~\ref{fig:SIA_Hz_A=posi_K=0}(a), suggesting a narrow intermediate phase between the single-$Q$ spiral state and the $n_{\rm sk}=1$ skyrmion crystal. 
Meanwhile, in the high-field region, the triple-$Q$ state, which is similar to that obtained at $A=0$ and $K>0$ in Figs.~\ref{fig:SIA_Hz_A=0}(b) and \ref{fig:SIA_Hz_A=0}(c), is stabilized for $1.3 \lesssim H^z \lesssim 1.6$ at $A=0.2$ without $K$, as shown in Fig.~\ref{fig:SIA_Hz_A=posi_K=0}(a). 
This state is shrunk and eventually vanishes while increasing $A$; 
the $n_{\rm sk}=1$ skyrmion crystal directly turns into the fully-polarized state at $A=0.4$, as shown in Fig.~\ref{fig:SIA_Hz_A=posi_K=0}(b). 
This indicates that the energy gain by $A$ in this triple-$Q$ state is smaller than that in the $n_{\rm sk}=1$ skyrmion crystal and the fully-polarized state. 

Figure~\ref{fig:SIA_Hz_A=posi_K=01} shows the results for $K=0.1$. 
At $A=0.2$, the sequence of the magnetic phases is similar to that for $A=0$ in Fig.~\ref{fig:SIA_Hz_A=0}(b). 
Comparing Figs.~\ref{fig:SIA_Hz_A=posi_K=0}(a) and \ref{fig:SIA_Hz_A=posi_K=01}(a), nonzero $K$ replaces the single-$Q$ state in the low-field region by the triple-$Q$ state with the dominant single-$Q$ peak in the $z$ spin component. 
Furthermore, $K$ extends the region of the $n_{\rm sk}=1$ skyrmion crystal, as clearly seen at $A=0.4$ in Fig.~\ref{fig:SIA_Hz_A=posi_K=01}(b), while $(\chi_0)^2$ is suppressed by increasing $A$. 

When further increasing $K$, the zero-field phase becomes the $n_{\rm sk}=2$ skyrmion crystal for $K\gtrsim 0.25$ as shown in Fig.~\ref{fig:souzu_A-K}. 
The field-induced phases are similar to those for $A=0$ in Fig.~\ref{fig:SIA_Hz_A=0}(c). 
The results are shown in Fig.~\ref{fig:SIA_Hz_A=posi_K=03} for $K=0.3$. 
Both $n_{\rm sk}=2$ and $n_{\rm sk}=1$ skyrmion crystals remain for $A=0.2$ and $A=0.4$; the region for the $n_{\rm sk}=2$ skyrmion crystal appears to be independent of $A$, whereas that for the $n_{\rm sk}=1$ skyrmion crystal is extended by increasing $A$. 
In particular, in the case of $A=0.4$ shown in Fig.~\ref{fig:SIA_Hz_A=posi_K=03}(b), the triple-$Q$ state between the $n_{\rm sk}=2$ and $n_{\rm sk}=1$ skyrmion crystals vanishes, and the $n_{\rm sk}=1$ skyrmion crystal is stabilized for $0.1 \lesssim H^z \lesssim 1.2$. 
At the same time, the triple-$Q$ state without $(\chi_0)^2$ appearing for $1 \lesssim H^z \lesssim 2$ in Fig.~\ref{fig:SIA_Hz_A=0}(c) is suppressed and vanishes while increasing $A$.

\begin{figure}[htb!]
\begin{center}
\includegraphics[width=1.0 \hsize]{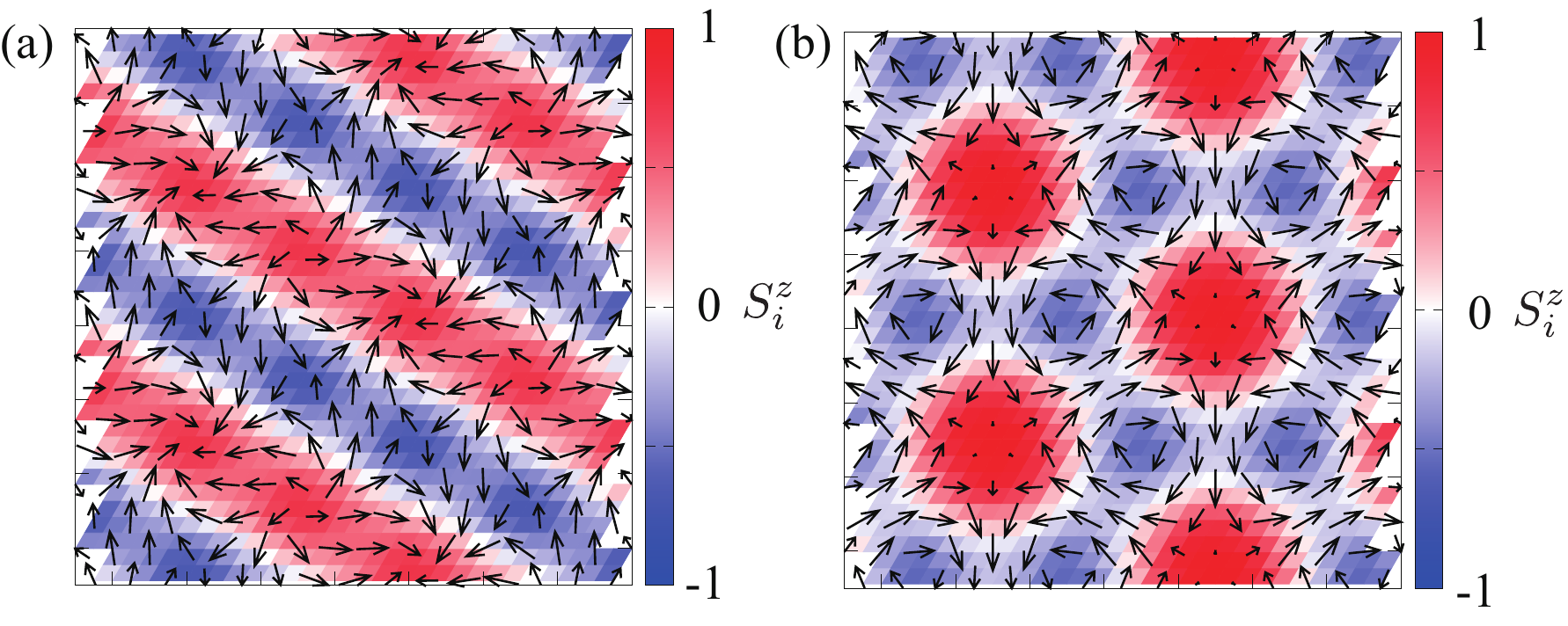} 
\caption{
\label{fig:Spin_n2SkX}
Real-space spin configuration of the $n_{\rm sk}=2$ skyrmion crystals at (a) $H^z=0.1$ and (b) $H^z=0.2$.
The contour shows the $z$ component of the spin moment, and the arrows represent the $xy$ components. 
}
\end{center}
\end{figure}

\subsubsection{With easy-plane anisotropy}
\label{sec:With easy-plane anisotropy_Hz}
We turn to the case with the easy-plane anisotropy, $A<0$. 
Figures~\ref{fig:SIA_Hz_A=nega_K=0}-\ref{fig:SIA_Hz_A=nega_K=03} show $H^z$ dependences of the spin and chirality related quantities for $K=(0,0.1,0.3)$ and $A=(-0.1, -0.3)$. 

At $K=0$, as shown in Fig.~\ref{fig:SIA_Hz_A=nega_K=0}, there is no qualitative change from the result at $A=0$ in Fig.~\ref{fig:SIA_Hz_A=0}(a) by introducing the easy-plane anisotropy. 
Meanwhile, when we turn on $K$, the system shows qualitatively different behavior with the instability toward the multiple-$Q$ states. 
As shown in Fig.~\ref{fig:SIA_Hz_A=nega_K=01}(a) for $K=0.1$ and $A=-0.1$, the single-$Q$ conical state at $K=0$ is replaced with two triple-$Q$ states by introducing the magnetic field. 
The lower-field one appearing for $0<H^z \lesssim 1.1$ shows a dominant contribution from $(m^{xy}_{\bm{Q}_1})^2$ accompanied by small $(m^{xy}_{\bm{Q}_2})^2$ and $(m^{z}_{\bm{Q}_3})^2$; while increasing $H^z$, the former decreases but the latter two increase. 
Accordingly, $(\chi_{\bm{Q}_3})^2$ becomes nonzero and shows similar $H^z$ dependence to $(m^{z}_{\bm{Q}_3})^2$, as shown in the lowest panel of Fig.~\ref{fig:SIA_Hz_A=nega_K=01}(a). 
Thus, the low-field phase is characterized by the anisotropic triple-$Q$ peaks with different intensities at ${\bm{Q}_1}$, ${\bm{Q}_2}$, and ${\bm{Q}_3}$ in the spin structure and the single peak at ${\bm{Q}_3}$ in the chirality. 
On the other hand, the higher-field state for $1.1 \lesssim H^z \lesssim 2.2$ shows $(m^{xy}_{\bm{Q}_1})^2 = (m^{xy}_{\bm{Q}_2})^2$, similar to the high-field triple-$Q$ phase at $K=0.1$ and $A=0$ in Fig.~\ref{fig:SIA_Hz_A=0}(b). 
The intensities of $(m_{\bm{Q}_1}^{xy})^2$, $(m_{\bm{Q}_2}^{xy})^2$, $(m^{z}_{\bm{Q}_3})^2$, and $(\chi_{\bm{Q}_3})^2$ become smaller as increasing $H^z$, and the system continuously changes into the fully-polarized state at $H^z \simeq 2.2$. 
For stronger easy-plane anisotropy, however, these triple-$Q$ states disappear as shown in Fig.~\ref{fig:SIA_Hz_A=nega_K=01}(b), and instead, the single-$Q$ state similar to that at $K=0$ in Fig.~\ref{fig:SIA_Hz_A=nega_K=0}(b) is recovered. 
Thus, the easy-plane anisotropy suppresses the multiple-$Q$ instability in the model in Eq.~(\ref{eq:Ham}), as seen in frustrated localized spin models~\cite{leonov2015multiply,Lin_PhysRevB.93.064430,Hayami_PhysRevB.93.184413}.

The results for a larger $K=0.3$ are shown in Fig.~\ref{fig:SIA_Hz_A=nega_K=03}.  
At $A=-0.1$, as shown in Fig.~\ref{fig:SIA_Hz_A=nega_K=03}(a), there is a phase transition within the low-field $n_{\rm sk}=2$ skyrmion crystal at $H^z \simeq 0.2$. 
The spin texture for $H^z \lesssim 0.2$ is characterized by the double-$Q$ peak structure with equal intensities at $\bm{Q}_1$ and $\bm{Q}_2$, while the $z$ component shows the single-$Q$ peak structure at $\bm{Q}_3$, as shown in Fig.~\ref{fig:Spin_zerofield_SIA_chiral}(b). 
This spin texture is similar to that in the case of easy-axis anisotropy in Figs.~\ref{fig:SIA_Hz_A=posi_K=03}(a) and \ref{fig:SIA_Hz_A=posi_K=03}(b). 
Meanwhile, the spin texture for $ H^{z} \simeq 0.2 $ has the triple-$Q$ peak structure for both $xy$ and $z$ components, as shown in the middle two panels of Fig.~\ref{fig:SIA_Hz_A=nega_K=03}(a). 
Comparison of the real-space spin configurations between the two types of the $n_{\rm sk}=2$ skyrmion crystal is shown in Fig.~\ref{fig:Spin_n2SkX}. 
The low-field one in Fig.~\ref{fig:Spin_n2SkX}(a) breaks the threefold rotational symmetry due to inequivalent $(\bm{m}_{\bm{Q}_{1,2}})^2$ and $(\bm{m}_{\bm{Q}_3})^2$, while the high-field one in Fig.~\ref{fig:Spin_n2SkX}(b) preserves the threefold rotational symmetry, as indicated in the data in Fig.~\ref{fig:SIA_Hz_A=nega_K=03}. 
It is noted that these two spin textures are connected by global spin rotation, and their energies are degenerate at $A=H^z=0$; the former spin texture is approximately given by
\begin{align}
\label{eq:nsk2_SkX1}
\bm{S}_i 
\propto \left(
    \begin{array}{c}
     \cos\bm{Q}_1 \cdot \bm{r}_i  \\
     \cos\bm{Q}_2 \cdot \bm{r}_i  \\
      \cos\bm{Q}_3 \cdot \bm{r}_i  
          \end{array}
  \right)^{\rm T}, 
\end{align}
and the latter is 
by 
\begin{align}
\label{eq:nsk2_SkX2}
\bm{S}_i 
\propto
\left(
    \begin{array}{c}
\frac{\sqrt{3}}{2} (\cos \bm{Q}_2 \cdot \bm{r}_i -\cos \bm{Q}_3 \cdot \bm{r}_i) \\
\cos \bm{Q}_1 \cdot \bm{r}_i
- \frac{1}{2}(
\cos \bm{Q}_2 \cdot \bm{r}_i
+\cos \bm{Q}_3 \cdot \bm{r}_i
) \\
\frac{1}{\sqrt{2}}(\cos \bm{Q}_1 \cdot \bm{r}_i+\cos \bm{Q}_2 \cdot \bm{r}_i+\cos \bm{Q}_3 \cdot \bm{r}_i)
          \end{array}
  \right)^{\rm T}. 
\end{align}
The result indicates that the spin texture in Eq.~(\ref{eq:nsk2_SkX1}) is chosen in the presence of $A$ and small $H^z$, while that in Eq.~(\ref{eq:nsk2_SkX2}) is chosen for moderate $H^z$ presumably due to subtle balance among different interaction energies. 
While further increasing $H^z$, the $n_{\rm sk}=2$ skyrmion crystal changes into the $n_{\rm sk}=1$ skyrmion crystal at $H^{z} \simeq 0.3$. 
Thus, in this case, there are three different skyrmion crystals in the low-field region. 
At $H^z \sim 0.4$, the system undergoes a transition to the anisotropic triple-$Q$ state which is the same as that found for $K=0.1$ in Fig.~\ref{fig:SIA_Hz_A=nega_K=01}(a). 

For stronger easy-plane anisotropy, the low-field skyrmion crystals are all replaced with the double-$Q$ state with equal intensities of $(m^{xy}_{\bm{Q}_{1}})^2$ and $(m^{xy}_{\bm{Q}_{2}})^2$, as shown in Fig.~\ref{fig:SIA_Hz_A=nega_K=03}(b) for $A=-0.3$. 
This turns into the triple-$Q$ state for $H^z\gtrsim 0.9$ with a small additional contribution from $(m^{z}_{\bm{Q}_{3}})^2$. 
Thus, the $z$-spin component of the low-field double-$Q$ state is uniform, while that of the high-field triple-$Q$ state exhibits the sinusoidal modulation along the $\bm{Q}_3$ direction. 
Both states show the single-$Q$ chirality density wave at $\bm{Q}_3$, as shown in the lowest panel of Fig.~\ref{fig:SIA_Hz_A=nega_K=03}(b).

\subsubsection{Discussion}
\label{sec:Summary of this section_1}

\begin{figure}[htb!]
\begin{center}
\includegraphics[width=0.8 \hsize]{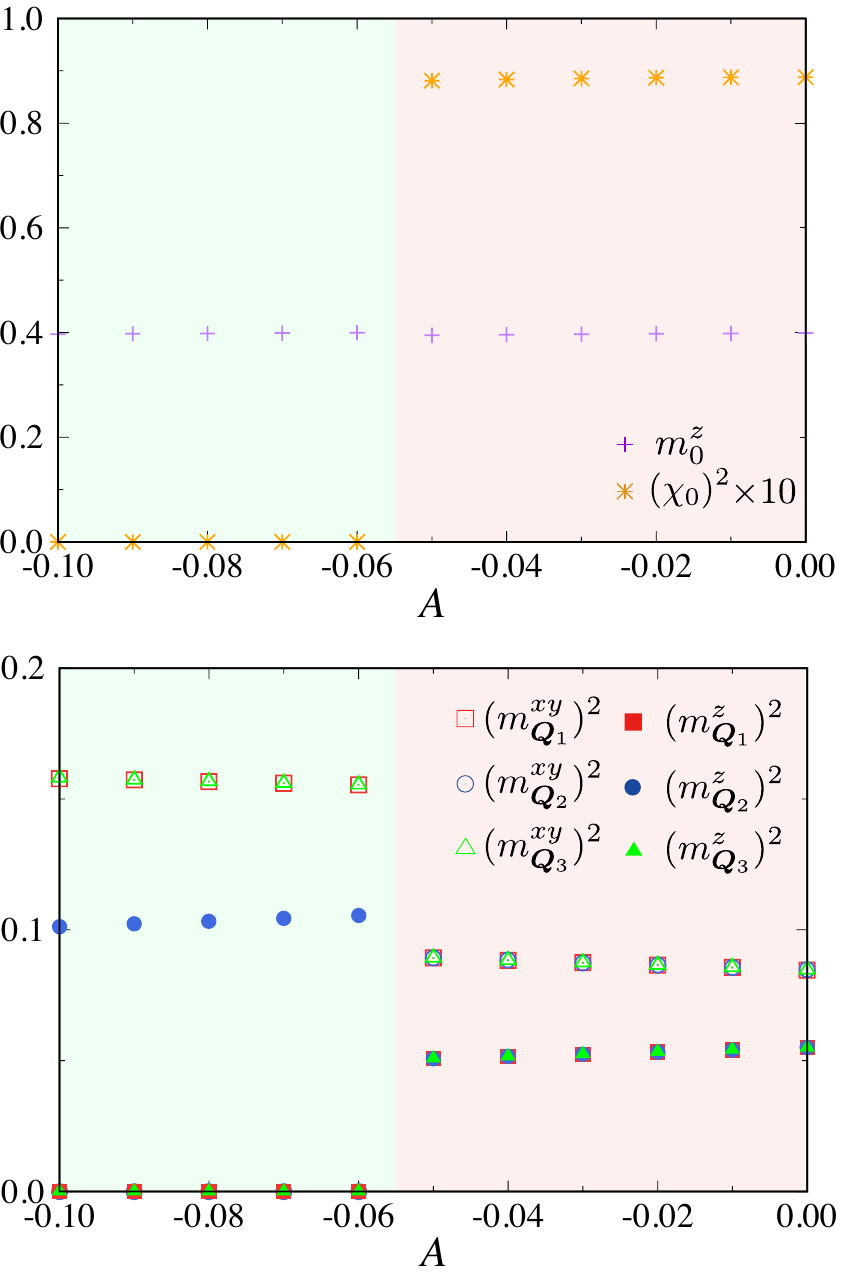} 
\caption{
\label{fig:Easy_plane_dep.pdf}
$A$ dependence of (upper) $m^z_0$ and $(\chi_0)^2$ and (lower) $(m^{xy}_{\bm{Q}_\nu})^2$ and $(m^{z}_{\bm{Q}_\nu})^2$ at $K=0.3$ and $H^z=0.7$. 
}
\end{center}
\end{figure}

The results obtained in this section are summarized in Fig.~\ref{fig:Summary}(a). 
We found a variety of multiple-$Q$ instabilities in the presence of the single-ion anisotropy $A$ under the [001] magnetic field. 
The triple-$Q$ states including the skyrmion crystals with $n_{\rm sk}=1$ and $2$ are stabilized by the biquadratic interaction $K$ even at $A=0$, but they show contrastive responses to the easy-axis ($A>0$) or easy-plane ($A<0$) anisotropy. 
In the following, we discuss the differences focusing on the skyrmion crystals. 

In the case of the $n_{\rm sk}=2$ skyrmion crystal, 
although the stable region in the presence of the easy-axis anisotropy at zero field is wider than that 
for the easy-plane anisotropy, e.g., $-0.25 \lesssim A \lesssim 0.45$ at $K=0.3$ in Fig.~\ref{fig:souzu_A-K}, the robustness against the magnetic field tends to be opposite: 
The critical field to destabilize the $n_{\rm sk}=2$ skyrmion crystal is larger for $A<0$ compared to that for $A>0$ [see Figs.~\ref{fig:SIA_Hz_A=posi_K=03}(a) and \ref{fig:SIA_Hz_A=nega_K=03}(a)]. 
For $A<0$, we found two different type of the $n_{\rm sk}=2$ skyrmion crystal depending on $H^z$, as shown in Fig.~\ref{fig:Spin_n2SkX}.

Meanwhile, the stable field range of the $n_{\rm sk}=1$ skyrmion crystal changes more sensitively depending on the sign of $A$. 
The range is extended by increasing positive $A$ for small $A$ [see Figs.~\ref{fig:SIA_Hz_A=0}(b), \ref{fig:SIA_Hz_A=0}(c), \ref{fig:SIA_Hz_A=posi_K=01}(a) and \ref{fig:SIA_Hz_A=posi_K=03}(a)], but it is rapidly shrunk by decreasing negative $A$ and does not appear in Figs.~\ref{fig:SIA_Hz_A=nega_K=0}-\ref{fig:SIA_Hz_A=nega_K=03}.  
We show the stability in the small negative $A$ region at $K=0.3$ and $H^z=0.7$ in Fig.~\ref{fig:Easy_plane_dep.pdf}. 
The result clearly indicates that the $n_{\rm sk}=1$ skyrmion crystal is very weak against the easy-plane anisotropy; 
it is destabilized at $A \simeq -0.06$, while it remains stable for much stronger easy-axis anisotropy, as exemplified in Fig.~\ref{fig:SIA_Hz_A=posi_K=03}(b) for $A=0.4$. 
The results are qualitatively consistent with those obtained for the Kondo lattice model~\cite{Hayami_PhysRevB.99.094420}. 
Despite the narrow stable region, it is worth noting that the $n_{\rm sk}=1$ skyrmion crystal for $A<0$ is one of the good indicators for the importance of the spin-charge coupling, since it is hardly stabilized in the localized spin models with the easy-plane anisotropy~\cite{leonov2015multiply,Lin_PhysRevB.93.064430,Hayami_PhysRevB.93.184413}. 

Let us comment on the model parameters in relation to experiments. 
The $n_{\rm sk}=2$ skyrmion crystal is realized only for nonzero $K$, while the $n_{\rm sk}=1$ one is stabilized even without $K$~\cite{Wang_PhysRevLett.124.207201} or $A$~\cite{Hayami_PhysRevB.95.224424}. 
This indicates that the phase diagram against $H^z$ in experiments provides information whether $K$ and/or $A$ are important. 
For example, in the skyrmion-hosting centrosymmetric materials such as Gd$_2$PdSi$_3$~\cite{kurumaji2019skyrmion,Hirschberger_PhysRevLett.125.076602,Hirschberger_PhysRevB.101.220401} 
and Gd$_3$Ru$_4$Al$_{12}$~\cite{hirschberger2019skyrmion}, the effect of magnetic anisotropy might be significant rather than $K$, since the zero-field phase does not correspond to the $n_{\rm sk}=2$ skyrmion crystal. 
Nevertheless, the chemical substitution or carrier doping would result in the $n_{\rm sk}=2$ skyrmion crystal, since $K$ is sensitive to the electronic band structure~\cite{Hayami_PhysRevB.95.224424}.

\subsection{Field along the $x$ direction}
\label{sec:Field along the $x$ direction_SIA}

\begin{figure}[htb!]
\begin{center}
\includegraphics[width=1.0 \hsize]{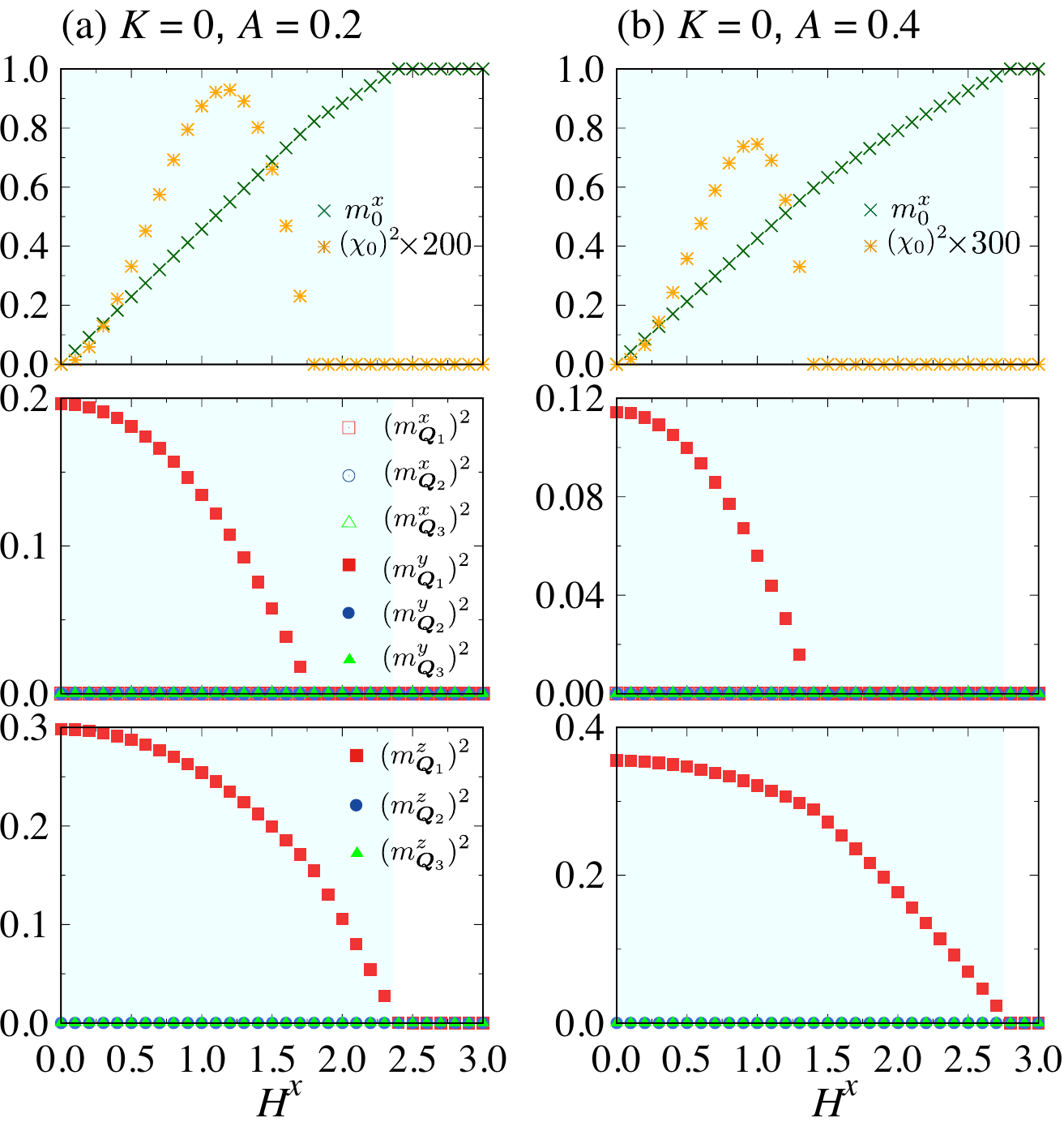} 
\caption{
\label{fig:SIA_Hx_A=posi_K=0}
$H^x$ dependence of (first row) $m^x_0$ and $(\chi_0)^2$, (second row) $(m^{x}_{\bm{Q}_\nu})^2$ and $(m^{y}_{\bm{Q}_\nu})^2$, and (third row) $(m^{z}_{\bm{Q}_\nu})^2$ for $K=0$ at (a) $A=0.2$ and (b) $A=0.4$. 
}
\end{center}
\end{figure}

\begin{figure}[htb!]
\begin{center}
\includegraphics[width=1.0 \hsize]{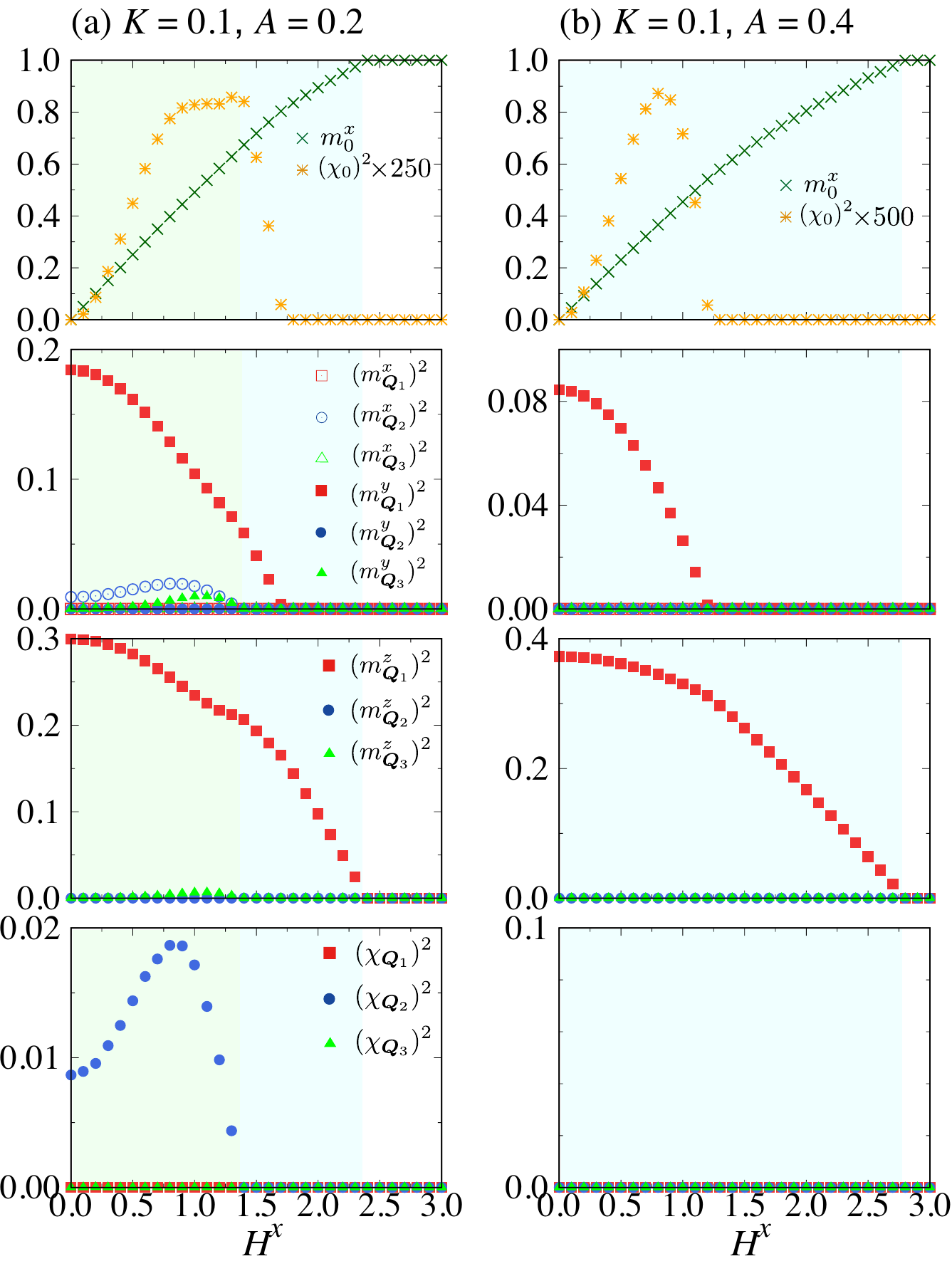} 
\caption{
\label{fig:SIA_Hx_A=posi_K=01}
$H^x$ dependence of (first row) $m^x_0$ and $(\chi_0)^2$, (second row) $(m^{x}_{\bm{Q}_\nu})^2$ and $(m^{y}_{\bm{Q}_\nu})^2$, (third row) $(m^{z}_{\bm{Q}_\nu})^2$, and (fourth row) $ (\chi_{\bm{Q}_\nu})^2$ for $K=0.1$ at (a) $A=0.2$ and (b) $A=0.4$. 
}
\end{center}
\end{figure}

\begin{figure}[htb!]
\begin{center}
\includegraphics[width=1.0 \hsize]{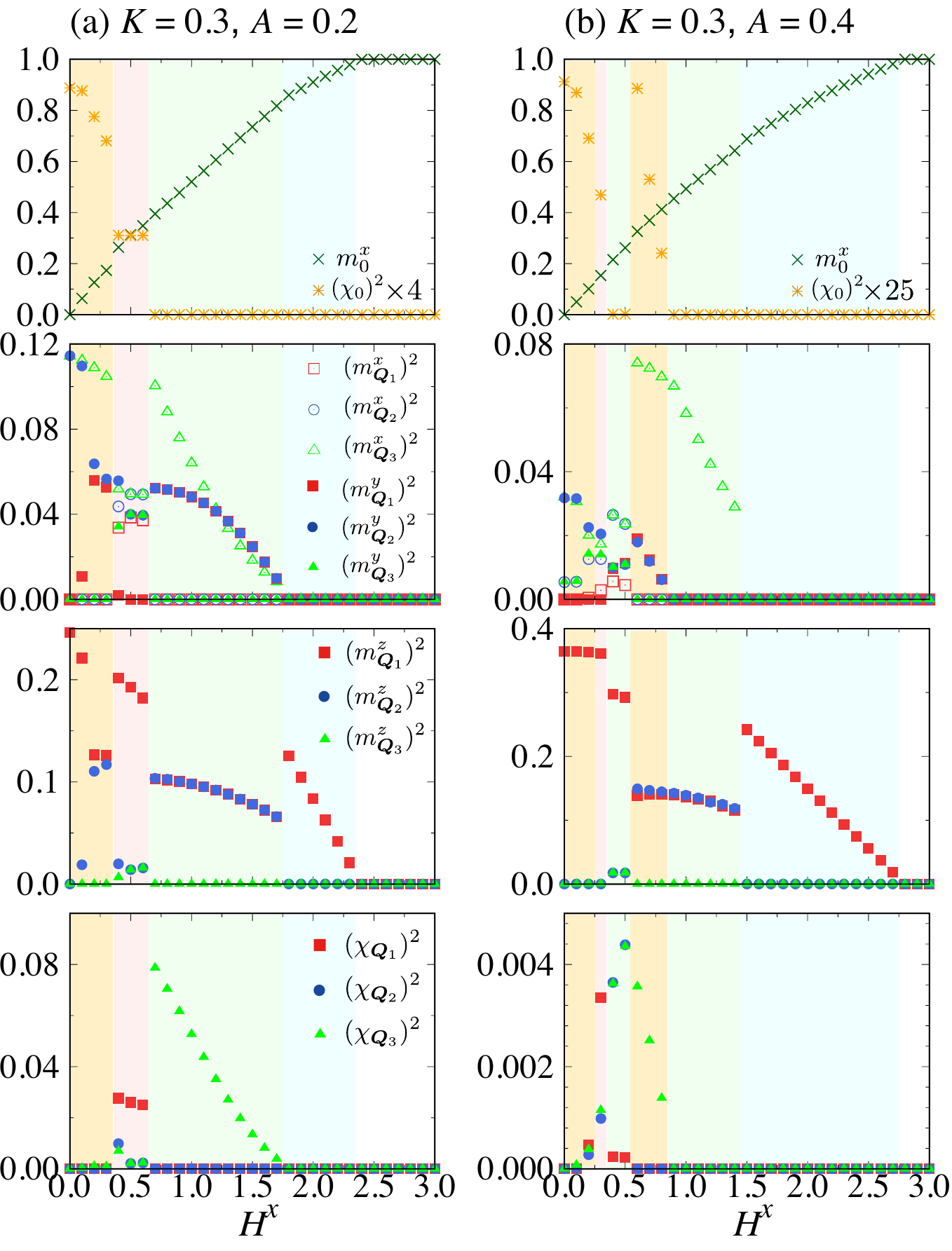} 
\caption{
\label{fig:SIA_Hx_A=posi_K=03}
The same plots as in Fig.~\ref{fig:SIA_Hx_A=posi_K=01} for $K=0.3$. 
}
\end{center}
\end{figure}

\begin{figure}[htb!]
\begin{center}
\includegraphics[width=1.0 \hsize]{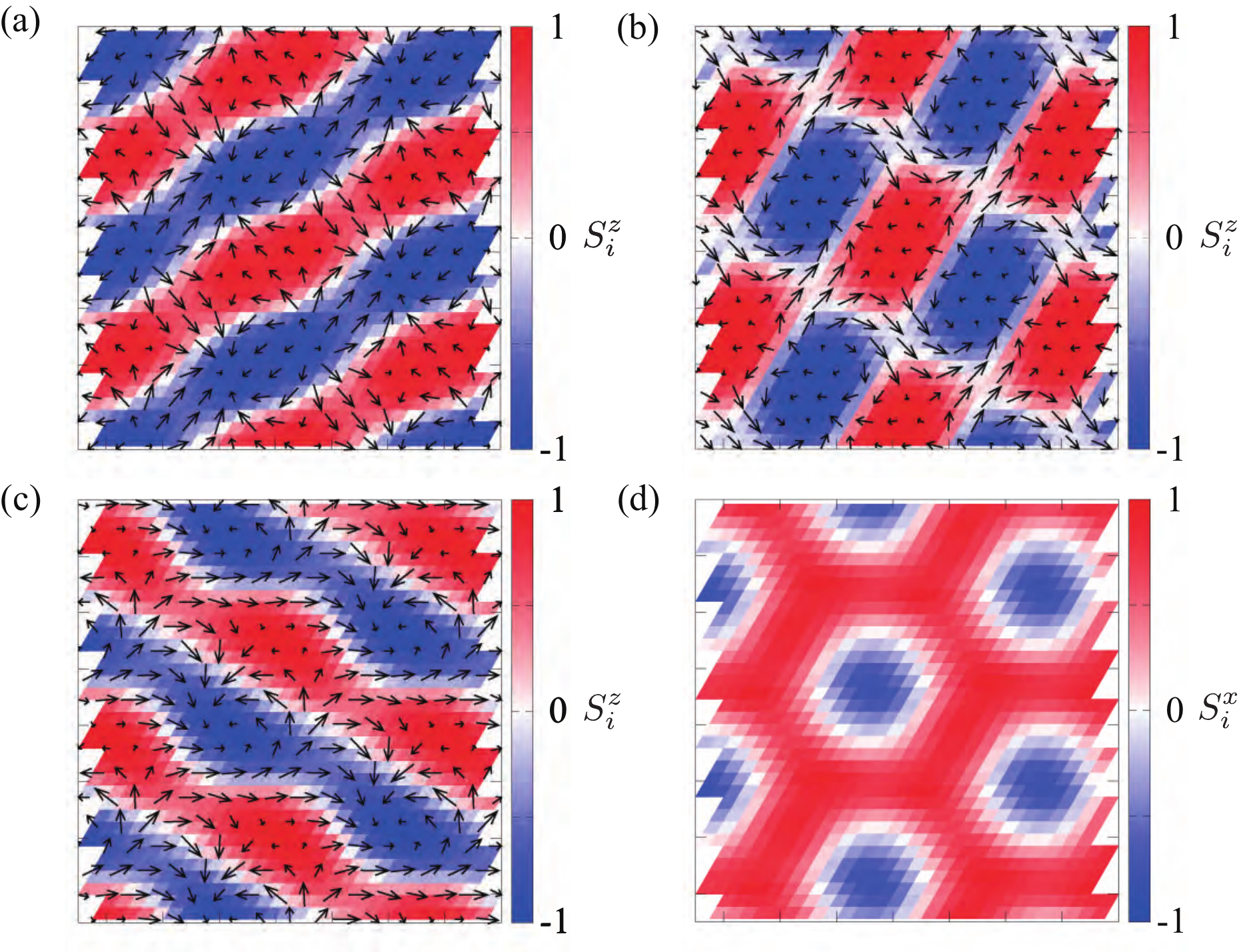} 
\caption{
\label{fig:Spin_SkX_Hx}
Real-space spin configuration of the $n_{\rm sk}=2$ skyrmion crystals at (a) $H^x=0.1$ and (b) $H^x=0.3$, and (c), (d) the $n_{\rm sk}=2$ skyrmion crystal at $H^x=0.5$ for $K=0.3$ and $A=0.2$.
In (a)-(c), the contour shows the $z$ component of the spin moment, and the arrows represent the $xy$ components. 
In (d), the contour shows the $x$ component of the spin moment. 
}
\end{center}
\end{figure}

\begin{figure}[htb!]
\begin{center}
\includegraphics[width=1.0 \hsize]{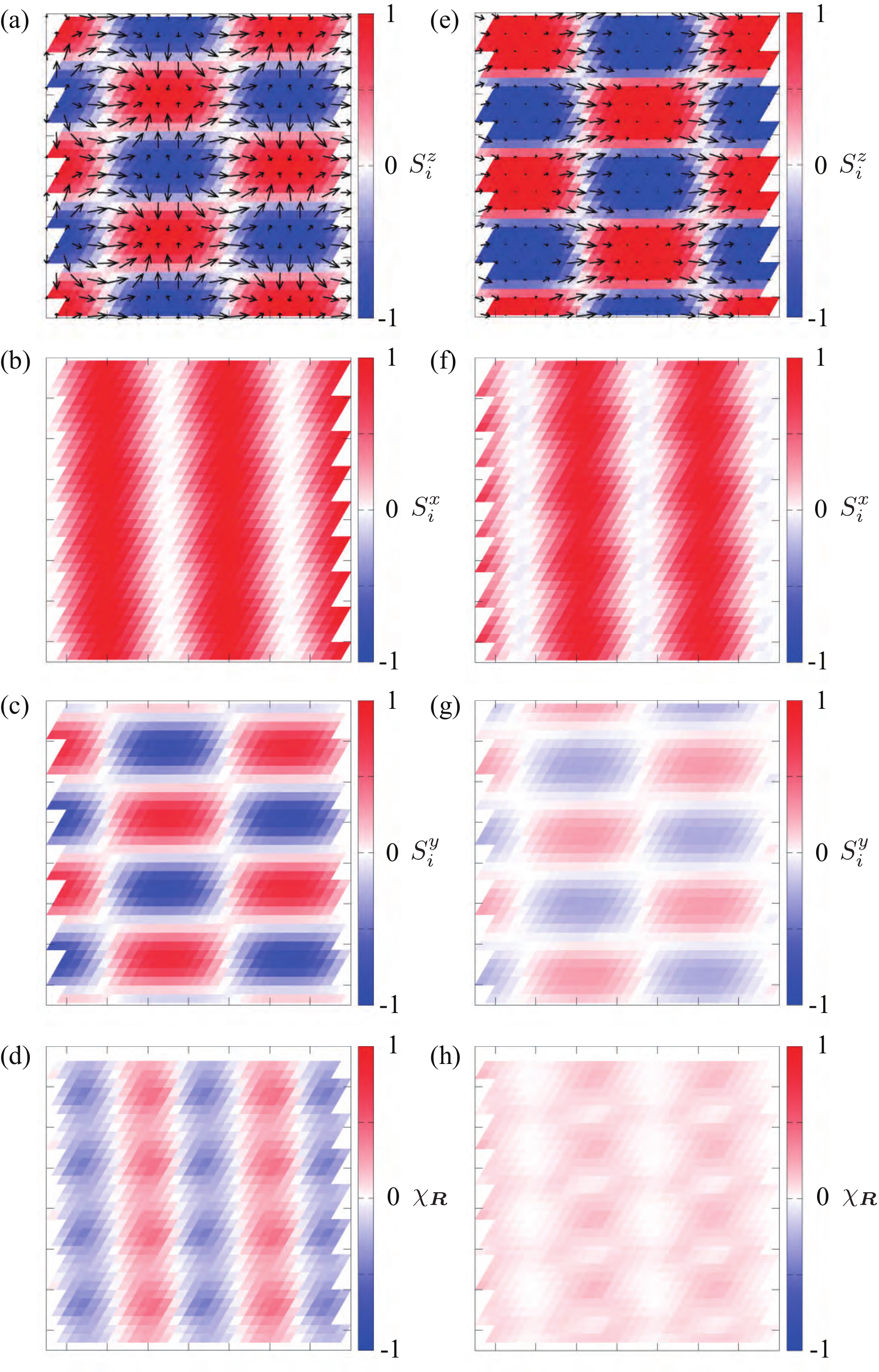} 
\caption{
\label{fig:Spin_compare_Hx}
Real-space spin and chirality configurations of (a)-(d) the anisotropic triple-$Q$ state at $H^x=0.9$ and $A=0.2$ and (e)-(h) the $n_{\rm sk}=2$ skyrmion crystal at $H^x=0.8$ and $A=0.4$. 
The contour shows the (a) and (e) $z$, (b) and (f) $x$, and (c) and (g) $y$ components of the spin moment, and the arrows in (a) and (e) represent the $xy$ components. 
In (d) and (h), the contour shows the scalar chirality. 
}
\end{center}
\end{figure}

Next, we discuss the result in the presence of the magnetic field along the $x$ direction by taking $\bm{H}=(H^x, 0, 0)$ in the Zeeman Hamiltonian $H^{\rm Z}$ in Eq.~(\ref{eq:Ham_Z}). 
As the result for the isotropic case with $A=0$ is the same (by replacing $H^x$ with $H^z$) as that in Sec.~\ref{sec:Isotropic case_Hz} due to spin rotational symmetry, we show the results for the anisotropic cases with $A>0$ in Sec.~\ref{sec:With easy-axis anisotropy_Hx} and $A<0$ in Sec.~\ref{sec:With easy-plane anisotropy_Hx}. 
We discuss the results in this section in Sec.~\ref{sec:Summary of this section_2}.

\subsubsection{With easy-axis anisotropy}
\label{sec:With easy-axis anisotropy_Hx}

Figures~\ref{fig:SIA_Hx_A=posi_K=0}-\ref{fig:SIA_Hx_A=posi_K=03} show $H^x$ dependences of the spin and chirality related quantities in the case of the easy-axis anisotropy, $A=0.2$ and $0.4$  for $K=0$, $0.1$, and $0.3$.  
At $K=0$ and $A=0.2$, the spiral plane of the single-$Q$ helical state is locked in the $yz$ plane for nonzero $H^x$, as shown in the lower two panels of Fig.~\ref{fig:SIA_Hx_A=posi_K=0}(a). 
By increasing $H^x$, $(m^{y}_{\bm{Q}_{1}})^2$ and $(m^{z}_{\bm{Q}_{1}})^2$ become smaller and vanish at $H^x \simeq 1.8$ and $H^x \simeq 2.4$, respectively. 
The difference of the critical field is due to the presence of the easy-axis anisotropy favoring the $z$-spin component. 
Thus, the magnetic state for $0<H^x \lesssim 1.8$ is the single-$Q$ conical state with the anisotropic spiral in the $yz$ plane and the magnetic state for $1.8 \lesssim H^x \lesssim 2.4$ is the single-$Q$ fan state consisting of the collinear $z$ spin and the uniform magnetization $m^x_0$. 
The single-$Q$ conical state in the low-field region has nonzero $(\chi_0)^2$ as plotted in the top panel of Fig.~\ref{fig:SIA_Hx_A=posi_K=0}(a), but this is not a uniform but staggered component. 
For larger $A=0.4$ in Fig.~\ref{fig:SIA_Hx_A=posi_K=0}(b), the phase sequence is similar to that for $A=0.2$. 
The critical field between the single-$Q$ conical and fan states becomes smaller, while that between the single-$Q$ fan and fully-polarized states becomes larger; namely, the fan state is extended by increasing $A$. 
This tendency is naturally understood from the fact that the easy-axis anisotropy prefers to align the spins parallel to the $z$ direction.   

By introducing $K$, several multiple-$Q$ instabilities appear in the presence of $H^x$ as in the case of $H^z$ in Sec.~\ref{sec:Field along the $z$ direction_SIA}. 
Figure~\ref{fig:SIA_Hx_A=posi_K=01}(a) shows the result for $K=0.1$ and $A=0.2$. 
While the single-$Q$ fan state is stabilized for $1.8 \lesssim H^x \lesssim 2.4$ similar to the case for $K=0$ in Fig.~\ref{fig:SIA_Hx_A=posi_K=0}(a), the major part of the lower-field single-$Q$ conical state is replaced with a multiple-$Q$ state. 
At $H^x=0$, the spin configuration is modulated from the single-$Q$ cycloidal spin structure at $K=0$ so that the $x$-spin component has $\bm{Q}_2$ modulation, which corresponds to the double-$Q$ chiral stripe state with nonzero $(\chi_{\bm{Q}_{2}})^2$, as shown in the lowest panel of Fig.~\ref{fig:SIA_Hx_A=posi_K=01}(a).
When $H^x$ is applied, $(\chi_0)^2$ by a staggered chirality configuration, $(m^{y}_{\bm{Q}_{3}})^2$, and $(m^{z}_{\bm{Q}_{3}})^2$ are induced with similar $H^x$ dependence and shows a broad peak structure around $H^x \sim 1$.  
At $H^x \simeq 1.4$, this state turns into the single-$Q$ conical state which was found in Fig.~\ref{fig:SIA_Hx_A=posi_K=0}(a). 
For larger $A$, however, as shown in Fig.~\ref{fig:SIA_Hx_A=posi_K=01}(b), 
the double-$Q$ state is suppressed and the phase sequence becomes similar to that for $K=0$ in Fig.~\ref{fig:SIA_Hx_A=posi_K=0}(b). 

In the case of larger $K=0.3$ and $A=0.2$ in Fig.~\ref{fig:SIA_Hx_A=posi_K=03}(a), the spin structure in the $n_{\rm sk}=2$ skyrmion crystal stabilized at $H^x=0$ is modulated from the triple-$Q$ sinusoidal structure with $(m^{x}_{\bm{Q}_{3}})^2$, $(m^{y}_{\bm{Q}_{2}})^2$, and $(m^{z}_{\bm{Q}_{1}})^2$ so as to possess nonzero $(m^{y}_{\bm{Q}_{1}})^2$ and $(m^{z}_{\bm{Q}_{2}})^2$ [or $(m^{z}_{\bm{Q}_{1}})^2$ and $(m^{y}_{\bm{Q}_{2}})^2$]. 
The real-space spin configuration at $H^x=0.1$ is shown in Fig.~\ref{fig:Spin_SkX_Hx}(a). 
For $0.2\lesssim H^x  \lesssim 0.4$, the additional component $(m^{y}_{\bm{Q}_{1}})^2$ has a similar value to $(m^{y}_{\bm{Q}_{2}})^2$. 
In other words, the real-space spin structure in this field region is characterized by the single-$Q$ sinusoidal modulation along the field direction and the double-$Q$ checker-board-type modulation perpendicular to the field direction. The real-space spin configuration is presented in Fig.~\ref{fig:Spin_SkX_Hx}(b). 

While further increasing $H^x$, $(\chi_0)^2$ jumps at $H^x\simeq 0.4$, and the $n_{\rm sk}=1$ skyrmion crystal is realized for $0.4\lesssim H^z \lesssim 0.7$, similar to that for $A>0$ and $H^z>0$ in Sec.~\ref{With easy-axis anisotropy_Hz}. 
In this case, however, the skyrmion core has $S^x \simeq -1$ and the spin structure breaks threefold rotational symmetry due to the in-plane field, as shown in Figs.~\ref{fig:Spin_SkX_Hx}(c) and \ref{fig:Spin_SkX_Hx}(d). 
For $0.5\lesssim H^x \lesssim 0.7$, $(\bm{m}_{\bm{Q}_{2}})^2$ and $(\bm{m}_{\bm{Q}_{3}})^2$ take the same value, which are smaller than $(\bm{m}_{\bm{Q}_{1}})^2$, while $(\bm{m}_{\bm{Q}_{1}})^2$, $(\bm{m}_{\bm{Q}_{2}})^2$, and $(\bm{m}_{\bm{Q}_{3}})^2$ are all different for smaller $H^x$. 
This suggests that there are two regions in the $n_{\rm sk}=1$ skyrmion crystal with slightly different multiple-$Q$ structures. 

At $H^x \simeq 0.7$, the $n_{\rm sk}=1$ skyrmion crystal turns into the anisotropic triple-$Q$ state, which is characterized by the equal intensities in $(m^{\alpha}_{\bm{Q}_{1}})^2$ and $(m^{\alpha}_{\bm{Q}_{2}})^2$ for $\alpha=y$ and $z$, in addition to $(m^x_{\bm{Q}_{3}})^2$, as shown in the middle two panels of Fig.~\ref{fig:SIA_Hx_A=posi_K=03}(a). 
This triple-$Q$ state has a single-$Q$ chirality modulation at $\bm{Q}_3$, as shown in the lowest panel of Fig.~\ref{fig:SIA_Hx_A=posi_K=03}(a). 
The spin and chirality configurations are shown in Figs.~\ref{fig:Spin_compare_Hx}(a)-\ref{fig:Spin_compare_Hx}(d). 
These are similar to those in the high-field region in Figs.~\ref{fig:SIA_Hz_A=nega_K=03}(a) and \ref{fig:SIA_Hz_A=nega_K=03}(b) by a 
replacement of the spin components $(x, y, z) \to (z, y, x)$. 
This indicates that the effect of $H^x$ for $A>0$ is similar to that of $H^z$ for $A<0$. 
With a further increase of $H^x$, this triple-$Q$ state turns into the single-$Q$ fan state at $H^x \simeq 1.8$, and finally becomes the fully-polarized state at $H^x \simeq 2.4$. 

The result for larger $A=0.4$ at $K=0.3$ is shown in Fig.~\ref{fig:SIA_Hx_A=posi_K=03}(b). 
The $n_{\rm sk}=2$ and $n_{\rm sk}=1$ skyrmion crystals appear in a similar manner to the case with $A=0.2$ in Fig.~\ref{fig:SIA_Hx_A=posi_K=03}(a), but in a narrower field range for $0\leq H^x \lesssim 0.4$. 
The state stabilized for $0.4 \lesssim H^x \lesssim 0.6$, where $(\chi_0)^2$ vanishes, is dominantly characterized by a sinusoidal spin structure with $(m^z_{\bm{Q}_{1}})^2$ with small additional intensities at $(\bm{m}_{\bm{Q}_{2}})^2$ and $(\bm{m}_{\bm{Q}_{3}})^2$, as shown in the middle two panels of Fig.~\ref{fig:SIA_Hx_A=posi_K=03}(b). 
This is a different triple-$Q$ state from the anisotropic one found for $A=0.2$. 
While increasing $H^x$, $(\chi_0)^2$ as well as $\chi_{\bm{Q}_3}$ becomes nonzero again for $0.6 \lesssim H^x \lesssim 0.9$. 
In this region, the spin structure has double-$Q$ modulations for the $y$ and $z$ components and the single-$Q$ modulation for the $x$ component. 
This is regarded as a square-type vortex crystal with nonzero uniform scalar chirality, whose real-space spin and chirality structures are plotted in Figs.~\ref{fig:Spin_compare_Hx}(e)-\ref{fig:Spin_compare_Hx}(h). 

It is interesting to note that this triple-$Q$ vortex crystal is hardly distinguished from that found for $0.7 \lesssim H^x \lesssim 1.8$ at $A=0.2$ solely from the spin structure. 
As shown in Figs.~\ref{fig:Spin_compare_Hx}(a)-\ref{fig:Spin_compare_Hx}(c) and \ref{fig:Spin_compare_Hx}(e)-\ref{fig:Spin_compare_Hx}(g), their spin patterns appear to be similar: Both are represented by the checker-board-type modulation in the $y$ and $z$ components and the sinusoidal modulation in the $x$ component. 
The difference, however, lies in the relative phases among the constituent waves. 
For $H^x=0.9$ and $A=0.2$ [Figs.~\ref{fig:Spin_compare_Hx}(a)-\ref{fig:Spin_compare_Hx}(c)], $S^x$ shows the maximum value where $S^y$ becomes zero, while $S^x$ and $|S^y|$ take their maximum at the same positions for $H^x=0.8$ and $A=0.4$ [Figs.~\ref{fig:Spin_compare_Hx}(e)-\ref{fig:Spin_compare_Hx}(g)]. 
Thus, these two double-$Q$ states are distinguished by the phase shift among the constituent triple-$Q$ waves~\cite{hayami2020phase}. 
Reflecting the phase shift, the chirality behaves differently between the two states: 
The positive and negative contributions of the scalar chirality are canceled out for the former state, while there is no cancelation for the latter state, as shown in Figs.~\ref{fig:Spin_compare_Hx}(d) and \ref{fig:Spin_compare_Hx}(h), respectively. 
By calculating the skyrmion number for the latter state, we find that it exhibits the skyrmion number of two in the magnetic unit cell. 
This indicates that the obtained square-type vortex crystal can also be regarded as the $n_{\rm sk}=2$ skyrmion crystal, although the skyrmion cores are arranged in a one-dimensional way rather than a threefold-symmetric way. 

While further increasing $H^x$ in Fig.~\ref{fig:SIA_Hx_A=posi_K=03}(b), $(m^y_{\bm{Q}_{1}})^2$ and $(m^y_{\bm{Q}_{2}})^2$ decrease, and the system undergoes a phase transition to another triple-$Q$ state without $(\chi_0)^2$ and $\chi_{\bm{Q}_3}$ at $H^x \simeq 0.9$. 
At $H^x \simeq 1.5$, the triple-$Q$ state turns into the single-$Q$ fan state, and finally into the fully-polarized state at $H^x \simeq 2.8$.

\begin{figure}[htb!]
\begin{center}
\includegraphics[width=1.0 \hsize]{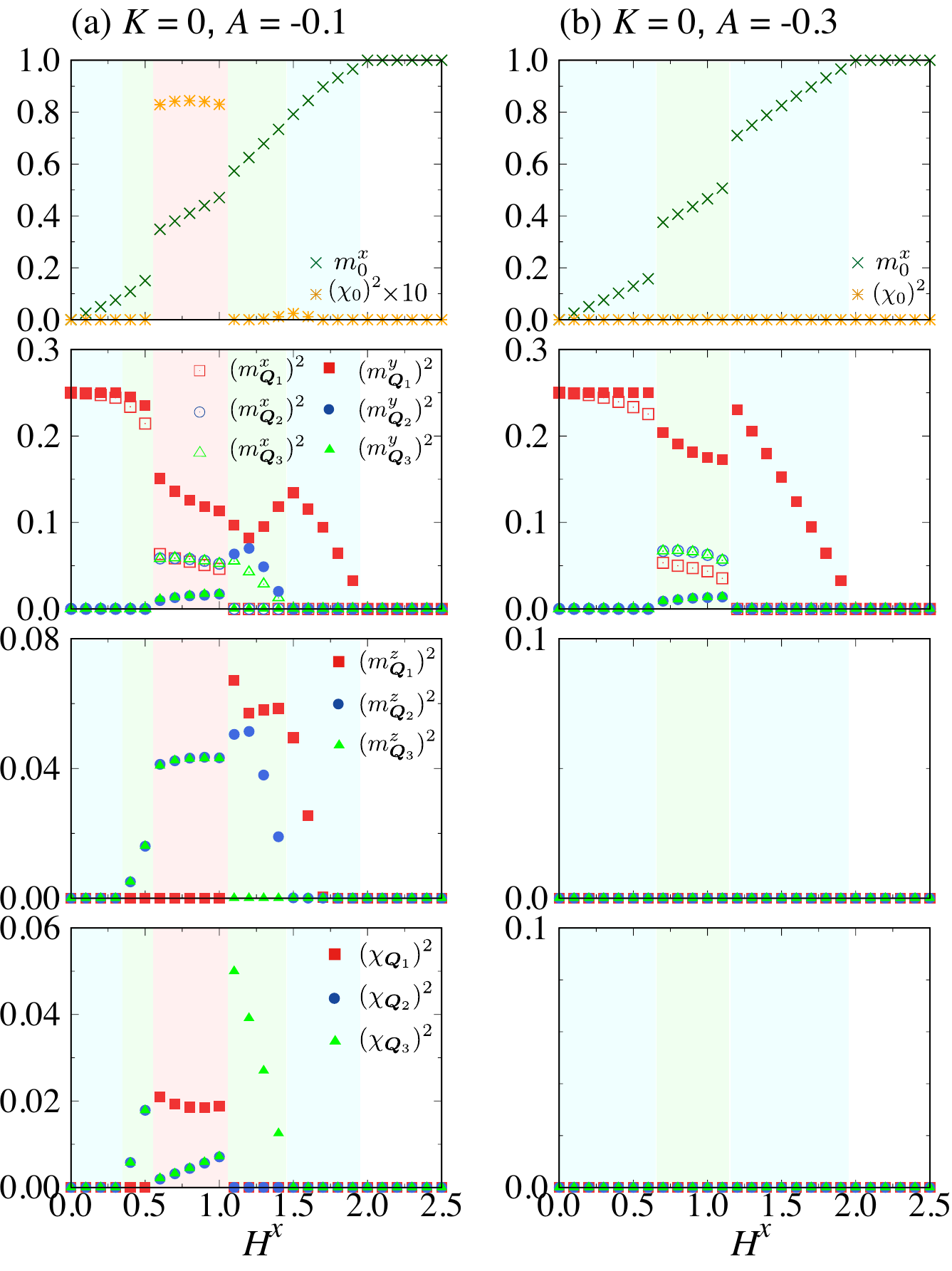} 
\caption{
\label{fig:SIA_Hx_A=nega_K=0}
$H^x$ dependence of (first row) $m^x_0$ and $(\chi_0)^2$, (second row) $(m^{x}_{\bm{Q}_\nu})^2$ and $(m^{y}_{\bm{Q}_\nu})^2$, (third row) $(m^{z}_{\bm{Q}_\nu})^2$, and (fourth row) $ (\chi_{\bm{Q}_\nu})^2$ for $K=0$ at (a) $A=-0.1$ and (b) $A=-0.3$. 
}
\end{center}
\end{figure}

\begin{figure}[htb!]
\begin{center}
\includegraphics[width=1.0 \hsize]{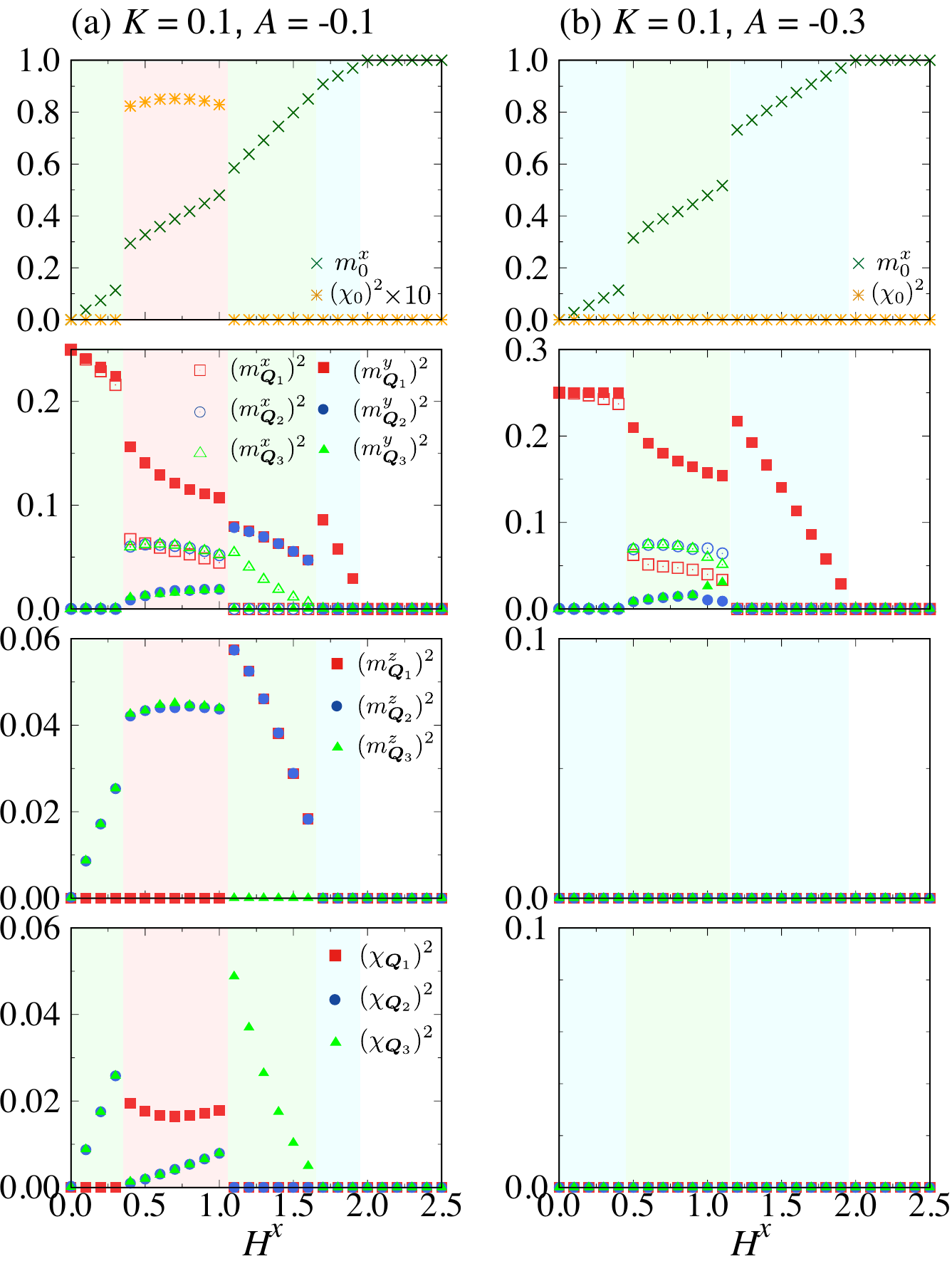} 
\caption{
\label{fig:SIA_Hx_A=nega_K=01}
The same plots as in Fig.~\ref{fig:SIA_Hx_A=nega_K=0} for $K=0.1$.
}
\end{center}
\end{figure}

\begin{figure}[htb!]
\begin{center}
\includegraphics[width=1.0 \hsize]{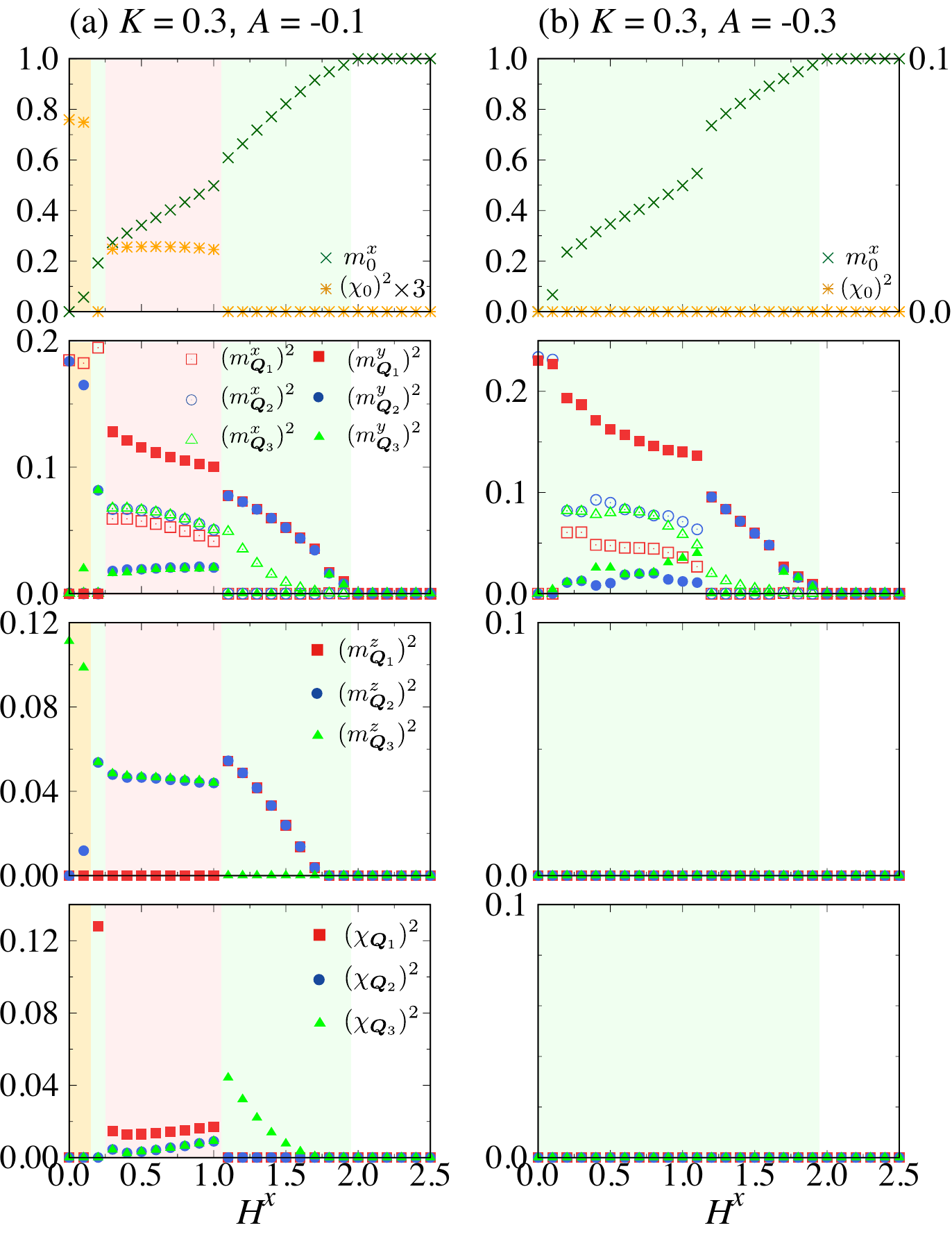} 
\caption{
\label{fig:SIA_Hx_A=nega_K=03}
The same plots as in Fig.~\ref{fig:SIA_Hx_A=nega_K=0} for $K=0.3$.
}
\end{center}
\end{figure}

\begin{figure}[htb!]
\begin{center}
\includegraphics[width=1.0 \hsize]{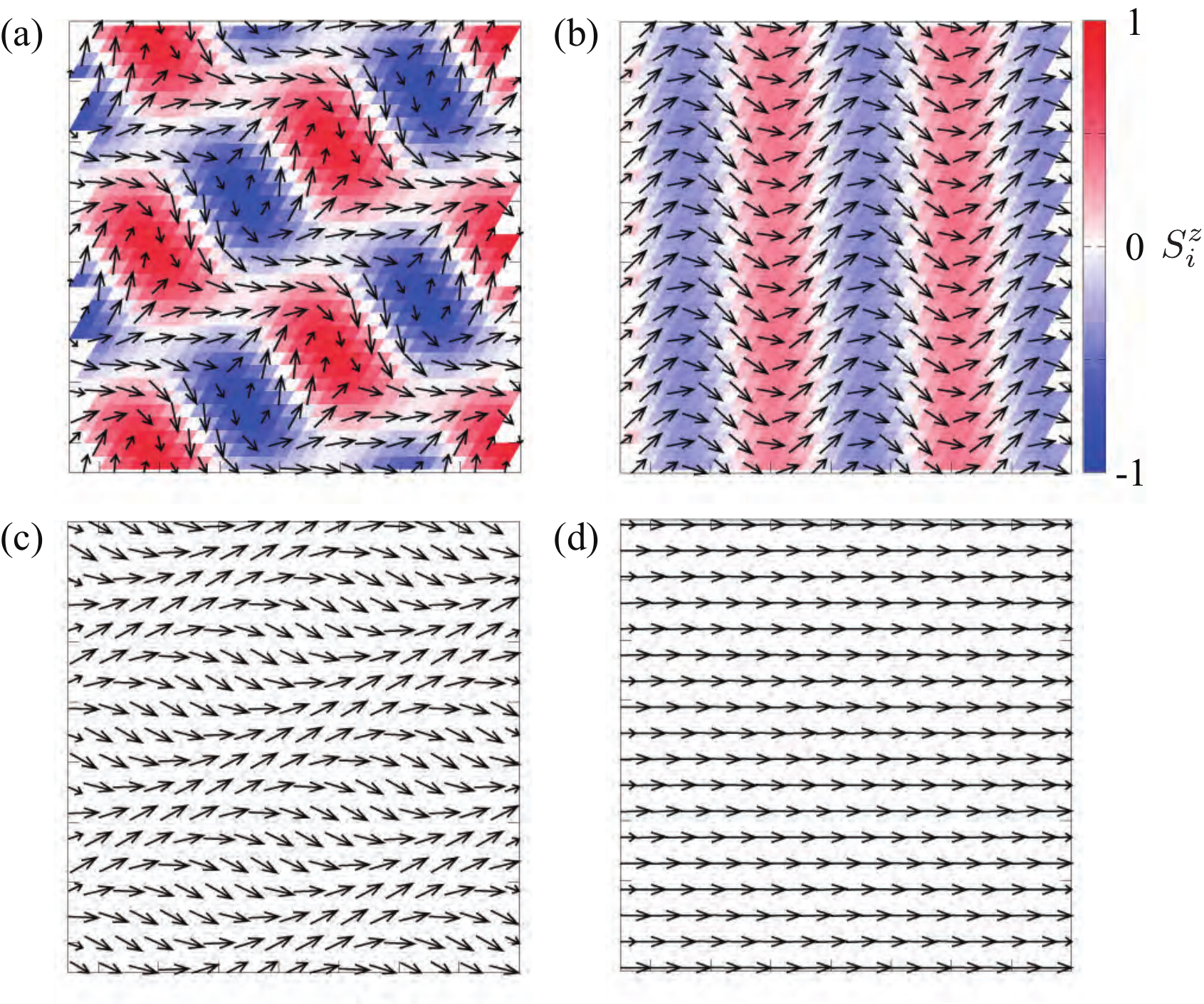} 
\caption{
\label{fig:Spin_Hx_EPA}
Real-space spin configurations of (a) the anisotropic triple-$Q$ state at $H^x=1.1$, (b) the single-$Q$ conical state at $H^x=1.6$, (c) the single-$Q$ fan state at $H^x=1.8$, and (d) the fully-polarized state at $H^x = 2$ for $K=0$ and $A=-0.1$. 
The contour shows the $z$ component of the spin moment, and the arrows represent the $xy$ components. 
}
\end{center}
\end{figure}

\begin{figure}[htb!]
\begin{center}
\includegraphics[width=1.0 \hsize]{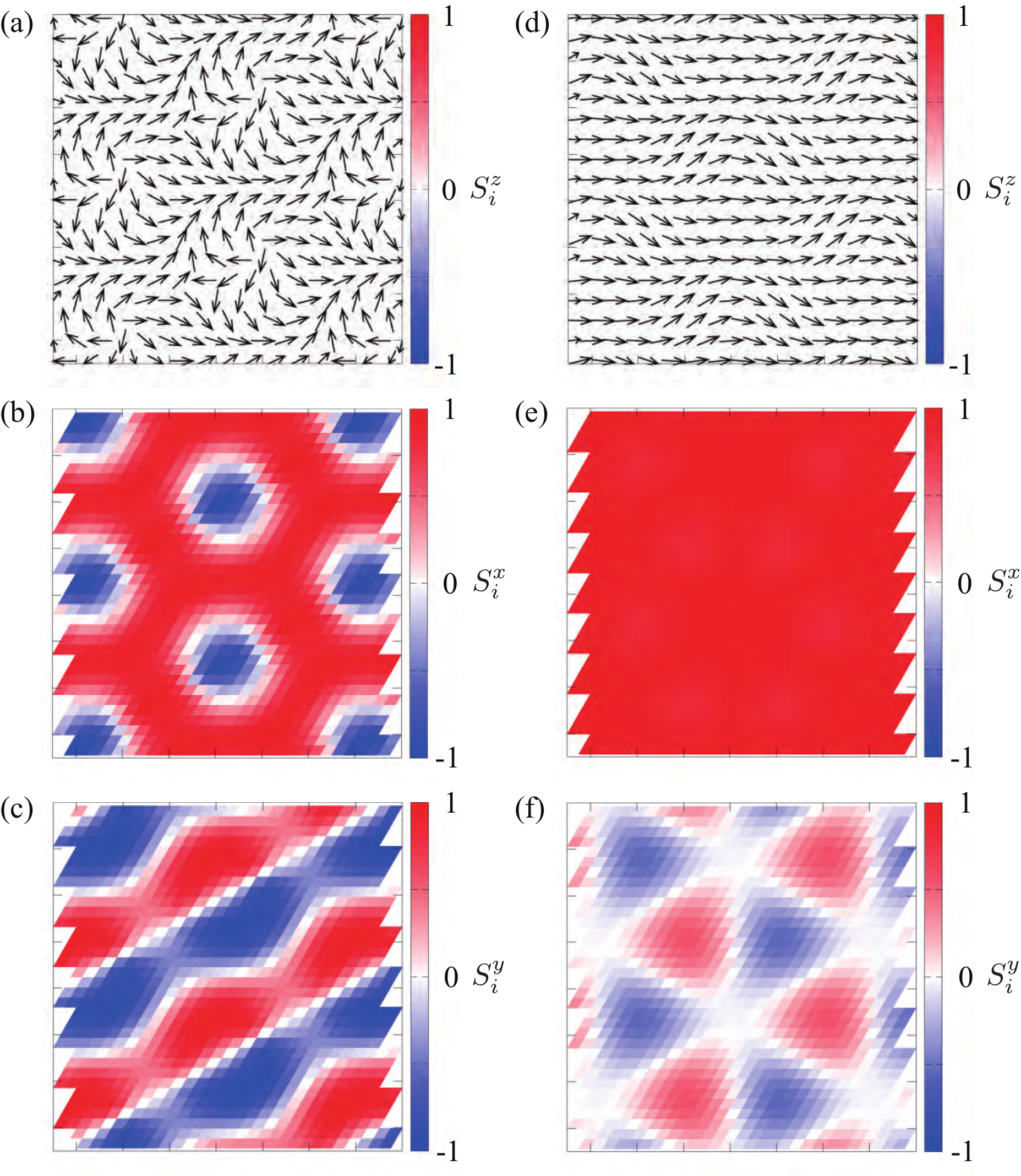} 
\caption{
\label{fig:Spin_Hx_K=0_A=-03}
Real-space spin configurations of (a)-(c) the triple-$Q$ coplanar bubble crystal at $K=0$, $A=-0.3$, and $H^x=1$ and (d)-(e) the triple-$Q$ coplanar fan state at $K=0.3$, $A=-0.1$, and $H^x=1.8$. 
The contour shows the (a) and (d) $z$, (b) and (e) $x$, and (c) and (f) $y$ components of the spin moment, and the arrows in (a) and (d) represent the $xy$ components. 
}
\end{center}
\end{figure}

\subsubsection{With easy-plane anisotropy}
\label{sec:With easy-plane anisotropy_Hx}

Next, we investigate the case of the easy-plane anisotropy under $H^x$. 
Figures~\ref{fig:SIA_Hx_A=nega_K=0}-\ref{fig:SIA_Hx_A=nega_K=03} show the results for $A=-0.1$ and $-0.3$. 
First, we discuss the result at $A=-0.1$ and $K=0$ in Fig.~\ref{fig:SIA_Hx_A=nega_K=0}(a). 
By introducing $H^x$, $(m^x_{\bm{Q}_{1}})^2$ and $(m^y_{\bm{Q}_{1}})^2$ become inequivalent in the single-$Q$ helical spiral state in the low-field region. 
While further increasing $H^x$, the single-$Q$ state turns into the triple-$Q$ state for $0.4 \lesssim H^x \lesssim 0.6$ where $(m^z_{\bm{Q}_{2}})^2$ and $(m^z_{\bm{Q}_{3}})^2$ are slightly induced in addition to $(\chi_{\bm{Q}_{2}})^2$ and $(\chi_{\bm{Q}_{3}})^2$. 
At $H^x \simeq 0.6$, the system undergoes a phase transition to the $n_{\rm sk}=1$ skyrmion crystal. 
The skyrmion core has $S^x \simeq -1$ similar to that stabilized by the easy-axis anisotropy in Figs.~\ref{fig:Spin_SkX_Hx}(c) and \ref{fig:Spin_SkX_Hx}(d). 
This result indicates that the skyrmion crystal can be stabilized by an in-plane magnetic field in itinerant magnets with the easy-plane anisotropy even without $K$. 
For larger $H^x$, the $n_{\rm sk}=1$ skyrmion crystal is replaced with other states: the anisotropic triple-$Q$ state for $1.1 \lesssim H^x \lesssim 1.5$, the single-$Q$ conical state for $1.5 \lesssim H^x \lesssim 1.7$, the single-$Q$ fan state for $1.7 \lesssim H^x \lesssim 2$, and the fully-polarized state for $H^x \gtrsim 2$. 
The spin configurations in these states are shown in Fig.~\ref{fig:Spin_Hx_EPA}. 
It is noted that the anisotropic triple-$Q$ and single-$Q$ conical states have small but nonzero $(\chi_0)^2$ as shown in the top panel of Fig.~\ref{fig:SIA_Hx_A=nega_K=0}(a), due to the staggered arrangement of the scalar chirality.  

By increasing the easy-plane anisotropy, $(m^z_{\bm{Q}_{\nu}})^2$ are suppressed as shown in Fig.~\ref{fig:SIA_Hx_A=nega_K=0}(b) in the case of $A=-0.3$. 
The low-field state for $0<H^x \lesssim 0.7$ remains unchanged from the single-$Q$ helical spiral state at $A=-0.1$ for $0 < H^x \lesssim 0.3$. 
On the other hand, the intermediate phase for $0.7 \lesssim H^x \lesssim 1.2$ is a different triple-$Q$ state from those for $A=-0.1$ because of the absence of $(m^z_{\bm{Q}_{\nu}})^2$, in spite of a similar magnetization curve to that for $A=-0.1$ as shown in the top row of Fig.~\ref{fig:SIA_Hx_A=nega_K=0}. 
This triple-$Q$ state has zero $(\chi_0)^2$ and a similar in-plane vortex structure of $(m^x_{\bm{Q}_{\nu}})^2$ and $(m^y_{\bm{Q}_{\nu}})^2$ to that in the $n_{\rm sk}=1$ skyrmion crystal at $A=-0.1$. 
The real-space spin configuration is shown in Figs.~\ref{fig:Spin_Hx_K=0_A=-03}(a)-\ref{fig:Spin_Hx_K=0_A=-03}(c). 
Interestingly, the bubble structure appears in the $x$-spin component in Fig.~\ref{fig:Spin_Hx_K=0_A=-03}(b) where the cores with $S^x \simeq -1$ form an anisotropic triangular lattice. 
We note that similar bubble structures were obtained for an out-of-plane magnetic field in frustrated magnets~\cite{Hayami_PhysRevB.93.184413} and itinerant magnets~\cite{Su_PhysRevResearch.2.013160} with strong easy-axis anisotropy. 
However, the present bubble state exhibits a coplanar spin structure with additional modulation in the $y$ component, in contrast to the collinear bubble structures for the easy-axis anisotropy. 
The in-plane component orthogonal to the magnetic field gains the energy under the easy-plane anisotropy, and contributes to the stabilization of the coplanar bubble state. 

While further increasing $H^x$, the system undergoes a phase transition to the single-$Q$ fan state at $H^x \simeq 1.2$, as shown in Fig.~\ref{fig:SIA_Hx_A=nega_K=0}(b). 
Finally, the system turns into the fully-polarized state for $H^x \gtrsim 2$. 

When we introduce $K$, the multiple-$Q$ states found for $K=0$ tend to be more stabilized, as shown in Fig.~\ref{fig:SIA_Hx_A=nega_K=01} for $K=0.1$. 
For $A=-0.1$ in Fig.~\ref{fig:SIA_Hx_A=nega_K=01}(a), the low-field single-$Q$ state is suppressed and the $n_{\rm sk}=1$ skyrmion crystal is stabilized from a smaller $H^x$ compared to the $K=0$ case in Fig.~\ref{fig:SIA_Hx_A=nega_K=0}(a). 
In the higher-field region, the anisotropic triple-$Q$ state is also extended up to a larger $H^x$, while the spin structure is modulated from the $K=0$ case so that the dominant peaks at $\bm{Q}_1$ and $\bm{Q}_2$ have the same intensities. 
Meanwhile, for $A=-0.3$ shown in Fig.~\ref{fig:SIA_Hx_A=nega_K=01}(b), there are no additional phases compared to the $K=0$ case in Fig.~\ref{fig:SIA_Hx_A=nega_K=0}(b), while the region of the triple-$Q$ state is extended. 

For larger $K$, the multiple-$Q$ states are more stabilized and take over the single-$Q$ states, as shown in Fig.~\ref{fig:SIA_Hx_A=nega_K=03} for $K=0.3$. 
In addition, for $A=-0.1$, the $n_{\rm sk}=2$ skyrmion crystal appears for $0\leq H^x \lesssim 0.2$, as shown in Fig.~\ref{fig:SIA_Hx_A=nega_K=03}(a). 
The spin texture is modulated in an anisotropic manner with larger intensities for the $xy$ components than the $z$ component, which is opposite to the case with the easy-axis anisotropy in Fig.~\ref{fig:SIA_Hx_A=posi_K=03}(a). 
The narrow triple-$Q$ state for $H^x \simeq 0.2$ has different spin and chirality textures from those for the lower-field state; 
it is characterized by nonzero $(m_{\bm{Q}_2}^y)^2 = (m_{\bm{Q}_3}^y)^2$ and $(m_{\bm{Q}_2}^z)^2 = (m_{\bm{Q}_3}^z)^2$ in addition to $(m^x_{\bm{Q}_{1}})^2$. 
The triple-$Q$ state turns into the $n_{\rm sk}=1$ skyrmion crystal at $H^x \simeq 0.3$ with a finite jump of $(\chi_0)^2$, as shown in the top panel of Fig.~\ref{fig:SIA_Hx_A=nega_K=03}(a). 
The $n_{\rm sk}=1$ skyrmion crystal is stabilized for $0.3\lesssim H^x  \lesssim 1$, whose region is larger compared to that at $K=0.1$ in Fig.~\ref{fig:SIA_Hx_A=nega_K=01}(a). 
For $1.1 \lesssim H^x \lesssim 1.8$, we find a triple-$Q$ state with similar spin and chirality structures to the state for $H^x \simeq 0.2$. 
In the higher-field region for $1.8 \lesssim H^x \lesssim 2$, 
a different type of the triple-$Q$ state appears, which is characterized by the triple-$Q$ fan structure by superposing $(m^y_{\bm{Q}_{\nu}})^2$ for $\nu=1$-$3$ with equal intensities, as shown in Fig.~\ref{fig:SIA_Hx_A=nega_K=03}(a). 
Figures~\ref{fig:Spin_Hx_K=0_A=-03}(d)-\ref{fig:Spin_Hx_K=0_A=-03}(f) show the real-space spin textures of the triple-$Q$ fan state. 
The result indicates that there are no modulations for $x$- and $z$-spin components, while the $y$ component forms a staggered hexagonal lattice satisfying threefold rotational symmetry. 

In the case of $A=-0.3$ in Fig.~\ref{fig:SIA_Hx_A=nega_K=03}(b), the behavior of the $xy$ components is qualitatively similar to that for 
$A=-0.1$ in Fig.~\ref{fig:SIA_Hx_A=nega_K=03}(a), except for the anisotropic triple-$Q$ state at $H^x \simeq 0.2$ and $0.3 \lesssim H^x \lesssim 1.1$. 
For $0<H^x \lesssim 0.2$, the in-plane anisotropic triple-$Q$ state with nonzero $(m^y_{\bm{Q}_{1}})^2$, $(m^x_{\bm{Q}_{2}})^2$, and $(m^y_{\bm{Q}_{3}})^2$ is stabilized.
For $0.2 \lesssim H^x \lesssim 1.2$, the $xy$-spin components are similar to those in the state for $0.3 \lesssim H^x \lesssim 1.1$ at $A=-0.1$ in Fig.~\ref{fig:SIA_Hx_A=nega_K=03}(a), i.e., the $xy$-spin structures are characterized by the dominant $(m^y_{\bm{Q}_{1}})^2$ and the subdominant $(m^x_{\bm{Q}_{1}})^2$, $(m^x_{\bm{Q}_{2}})^2$, $(m^y_{\bm{Q}_{2}})^2$, $(m^x_{\bm{Q}_{3}})^2$, and $(m^y_{\bm{Q}_{3}})^2$. 
We note that there are two types of the triple-$Q$ state for $0.2 \lesssim H^x \lesssim 1.2$, which are almost energetically degenerate: One has equal intensities with $(m^x_{\bm{Q}_{2}})^2$ and $(m^x_{\bm{Q}_{3}})^2$ [and $(m^y_{\bm{Q}_{2}})^2$ and $(m^y_{\bm{Q}_{3}})^2$], and the other does not. 
These two states are interchanged with each other depending on the value of $H^x$. 
Meanwhile, the states for $1.2 \lesssim H^x \lesssim 1.7$ and $1.7 \lesssim H^x \lesssim 2$ have similar spin structures to the triple-$Q$ state for $1.1 \lesssim H^x \lesssim 1.8$ and the triple-$Q$ fan state for $1.8 \lesssim H^x \lesssim 2$ at $A=-0.1$, respectively.

\subsubsection{Discussion}
\label{sec:Summary of this section_2}

The results obtained in this section are summarized in Fig.~\ref{fig:Summary}(b). 
Similar to the case with the magnetic field along the $z$ direction in Sec.~\ref{sec:Field along the $z$ direction_SIA}, we found a variety of multiple-$Q$ instabilities in the presence of the single-ion anisotropy $A$ by applying the magnetic field along the $x$ direction.  
Among them, we obtained both $n_{\rm sk}=2$ and $n_{\rm sk}=1$ skyrmion crystals, although their spin and chirality textures are different from those in Sec.~\ref{sec:Field along the $z$ direction_SIA}.
In the following, we discuss the characteristics of the skyrmion crystals comparing the effects of easy-axis and easy-plane anisotropy. 

The $n_{\rm sk}=2$ skyrmion crystal is widely stabilized for large $K$ under easy-axis anisotropy $A>0$, as shown in Fig.~\ref{fig:SIA_Hx_A=posi_K=03}. 
The in-plane magnetic field modulates its spin patterns from the triple-$Q$ sinusoidal waves to the single-$Q$ sinusoidal and double-$Q$ checker-board-type waves, as shown in Fig.~\ref{fig:Spin_SkX_Hx}. 
Meanwhile, for easy-plane anisotropy $A<0$, the $n_{\rm sk}=2$ skyrmion crystal is limited to large $K$ and small $|A|$, as shown in Fig.~\ref{fig:SIA_Hx_A=nega_K=03}(a). 
The critical field to destabilize the $n_{\rm sk}=2$ skyrmion crystal is larger for $A>0$ [Fig.~\ref{fig:SIA_Hx_A=posi_K=03}(a)] than $A<0$ [Fig.~\ref{fig:SIA_Hx_A=nega_K=03}(a)]. 

We also obtained the $n_{\rm sk}=2$ skyrmion crystal in the intermediate-field region for $K=0.3$ and $A=0.4$ [Fig.~\ref{fig:SIA_Hx_A=posi_K=03}(b)], where the spin texture is characterized by a superposition of the single-$Q$ sinusoidal and double-$Q$ checker-board-type waves in Figs.~\ref{fig:Spin_compare_Hx}(e)-\ref{fig:Spin_compare_Hx}(h). 
This indicates that the materials with easy-axis anisotropy may show the $n_{\rm sk}=2$ skyrmion crystal in the in-plane magnetic field.

On the other hand, the $n_{\rm sk}=1$ skyrmion crystal is found only for $K=0.3$ and $A=0.2$ under easy-axis anisotropy, as shown in Fig.~\ref{fig:SIA_Hx_A=posi_K=03}(a). 
This suggests that large $K$ and moderate $A$ are necessary to stabilize the $n_{\rm sk}=1$ skyrmion crystal under the in-plane field, which is in contrast to the case under the out-of-plane field in Sec.~\ref{With easy-axis anisotropy_Hz}. 
On the other hand, the $n_{\rm sk}=1$ skyrmion crystal appears for small $|A|$ irrespective of $K$ for easy-plane anisotropy with $A<0$, as shown in Figs.~\ref{fig:SIA_Hx_A=nega_K=0}(a), \ref{fig:SIA_Hx_A=nega_K=01}(a), and \ref{fig:SIA_Hx_A=nega_K=03}(a). 
Furthermore, the $n_{\rm sk}=1$ skyrmion crystal is stabilized even without $K$ for $A<0$, similar to the situation for $A>0$ and $H^z>0$ in Sec.~\ref{With easy-axis anisotropy_Hz}. 
These results indicate that the materials showing a single-$Q$ spiral state in the $xy$ plane at zero field under easy-axis anisotropy are potential candidates for the field-induced $n_{\rm sk}=1$ skyrmion crystal in the in-plane magnetic field. 

Besides the skyrmion crystals, we found several intriguing multiple-$Q$ spin textures under the in-plane magnetic field. 
In particular, for $A<0$, we found two types of interesting magnetic structures without scalar chirality: 
the triple-$Q$ coplanar bubble crystal with additional in-plane modulations for large $|A|$ irrespective of $K$ shown in Figs.~\ref{fig:Spin_Hx_K=0_A=-03}(a)-\ref{fig:Spin_Hx_K=0_A=-03}(c), and the triple-$Q$ coplanar fan state for large $K$ shown in Figs.~\ref{fig:Spin_Hx_K=0_A=-03}(d)-\ref{fig:Spin_Hx_K=0_A=-03}(f). 
While the former triple-$Q$ coplanar bubble crystal shows a similar magnetization curve to the $n_{\rm sk}=1$ skyrmion crystal, it is useful to measure the topological Hall effect to distinguish the triple-$Q$ state with and without uniform scalar chirality. 
Meanwhile, the latter triple-$Q$ coplanar fan state appears only for large $K$, its observation provides an evidence of the importance of the itinerant nature of electrons.

\subsection{Field rotation in the $xz$ plane}
\label{sec:Field rotation in the $zx$ plane_SIA}

\begin{figure}[htb!]
\begin{center}
\includegraphics[width=1.0 \hsize]{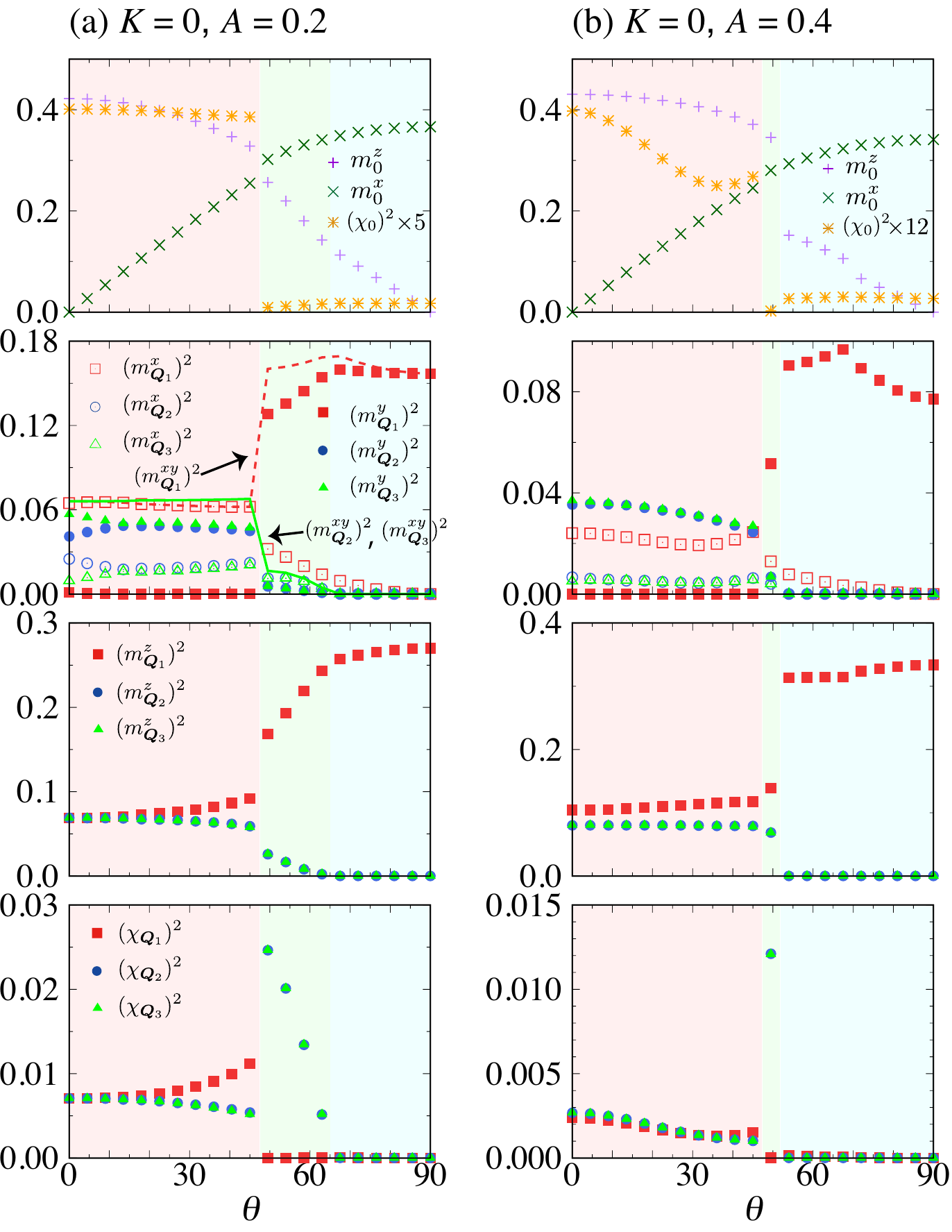} 
\caption{
\label{fig:SIA_Hrot_A=posi_K=0}
$\theta$ dependence of (first row) $m^z_0$, $m^x_0$, and $(\chi_0)^2$, (second row) $(m^{x}_{\bm{Q}_\nu})^2$ and $(m^{y}_{\bm{Q}_\nu})^2$, (third row) $(m^{z}_{\bm{Q}_\nu})^2$, and (fourth row) $ (\chi_{\bm{Q}_\nu})^2$ for $K=0$ at (a) $A=0.2$ and (b) $A=0.4$. 
In the second row in (a), the dashed line represents $(m^{xy}_{\bm{Q}_1})^2$, while the solid lines represent $(m^{xy}_{\bm{Q}_2})^2$ and $(m^{xy}_{\bm{Q}_3})^2$. 
The magnitude of the magnetic field is fixed at $H=0.8$. 
}
\end{center}
\end{figure}

\begin{figure}[htb!]
\begin{center}
\includegraphics[width=1.0 \hsize]{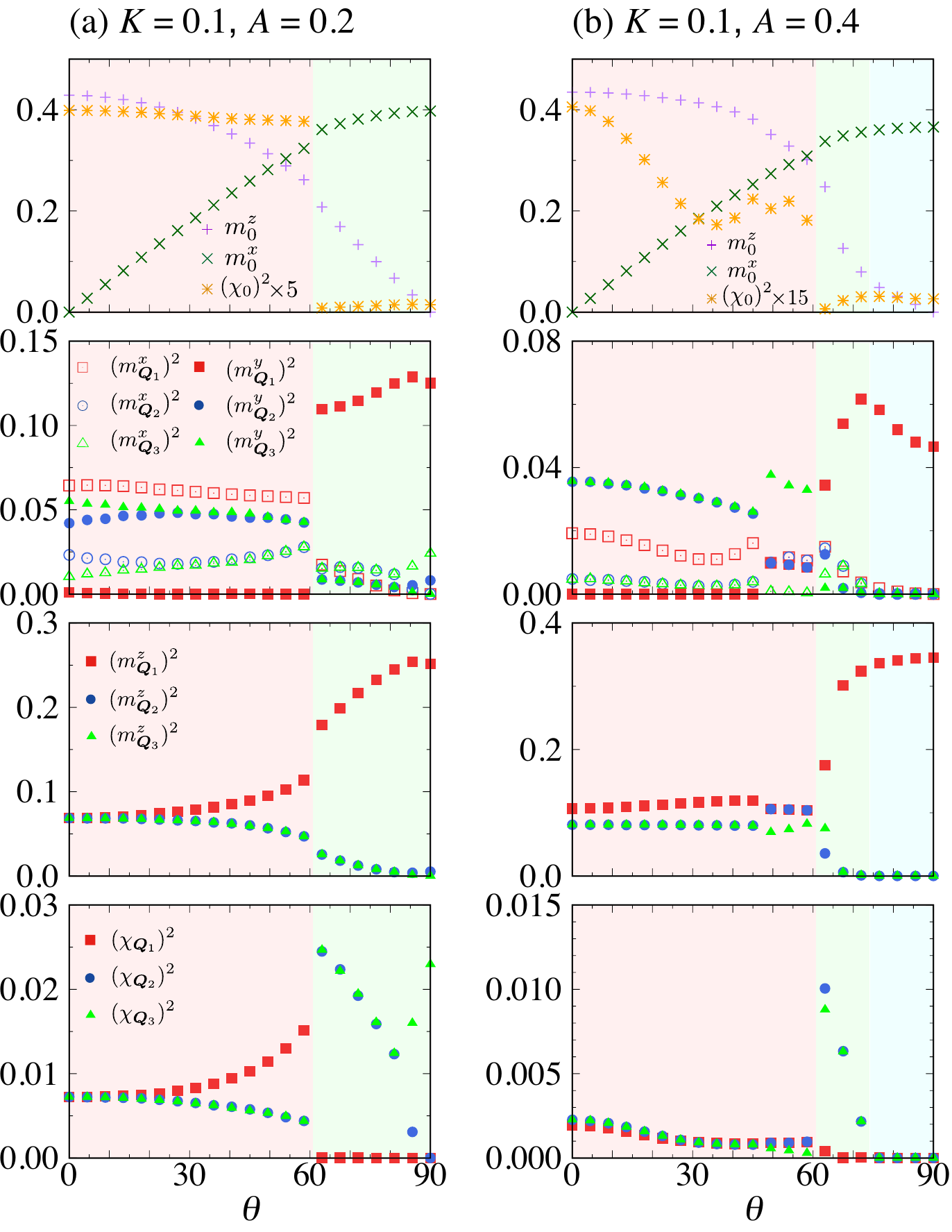} 
\caption{
\label{fig:SIA_Hrot_A=posi_K=01}
The same plots as in Fig.~\ref{fig:SIA_Hrot_A=posi_K=0} for $K=0.1$.
}
\end{center}
\end{figure}

\begin{figure}[htb!]
\begin{center}
\includegraphics[width=1.0 \hsize]{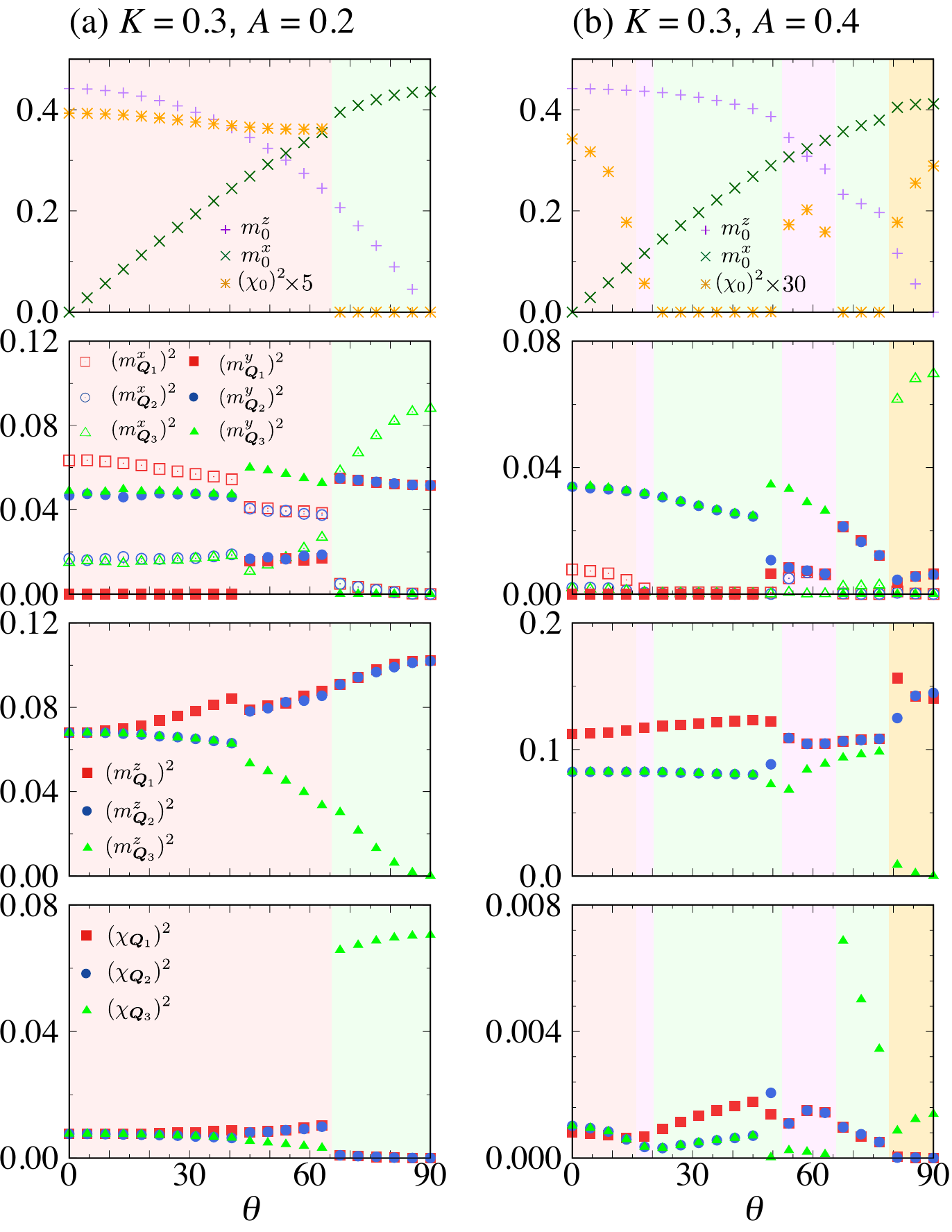} 
\caption{
\label{fig:SIA_Hrot_A=posi_K=03}
The same plots as in Fig.~\ref{fig:SIA_Hrot_A=posi_K=0} for $K=0.3$.
}
\end{center}
\end{figure}

\begin{figure}[htb!]
\begin{center}
\includegraphics[width=1.0 \hsize]{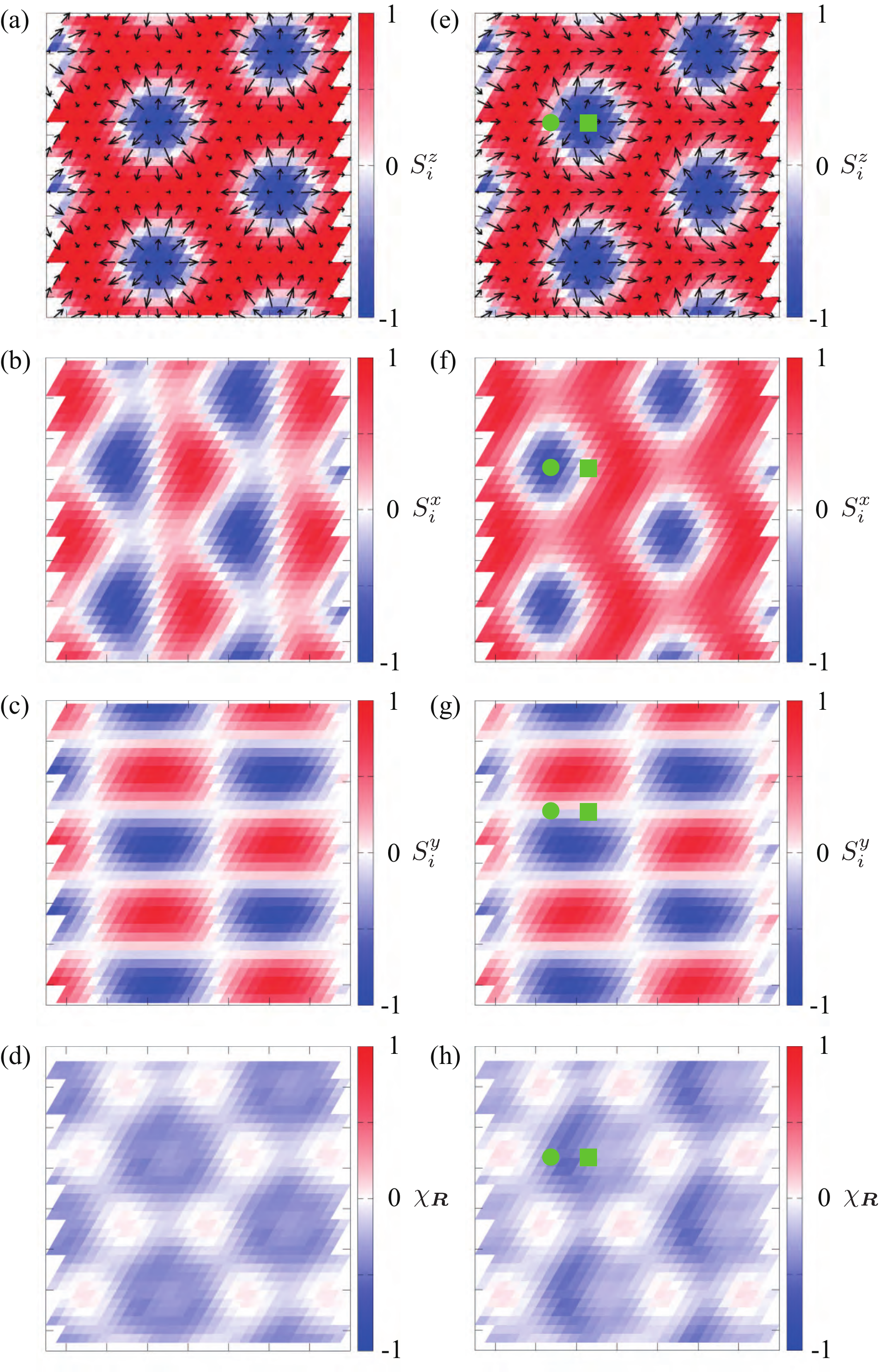} 
\caption{
\label{fig:Spin_Hrot_K=0_A=02}
Real-space spin and chirality configurations of the $n_{\rm sk}=1$ skyrmion crystals at (a)-(d) $\theta=5^{\circ}$ and (e)-(h) $\theta=45^{\circ}$ for $K=0$ and $A=0.2$.
The contour shows the (a) and (e) $z$, (b) and (f) $x$, and (c) and (g) $y$ components of the spin moment, and the arrows in (a) and (e) represent the $xy$ components. 
In (d) and (h), the contour shows the scalar chirality. 
In (e)-(h), the green squares and circles represent the positions of the minima of $S^z$ and $S^x$, respectively. 
}
\end{center}
\end{figure}

\begin{figure}[htb!]
\begin{center}
\includegraphics[width=1.0 \hsize]{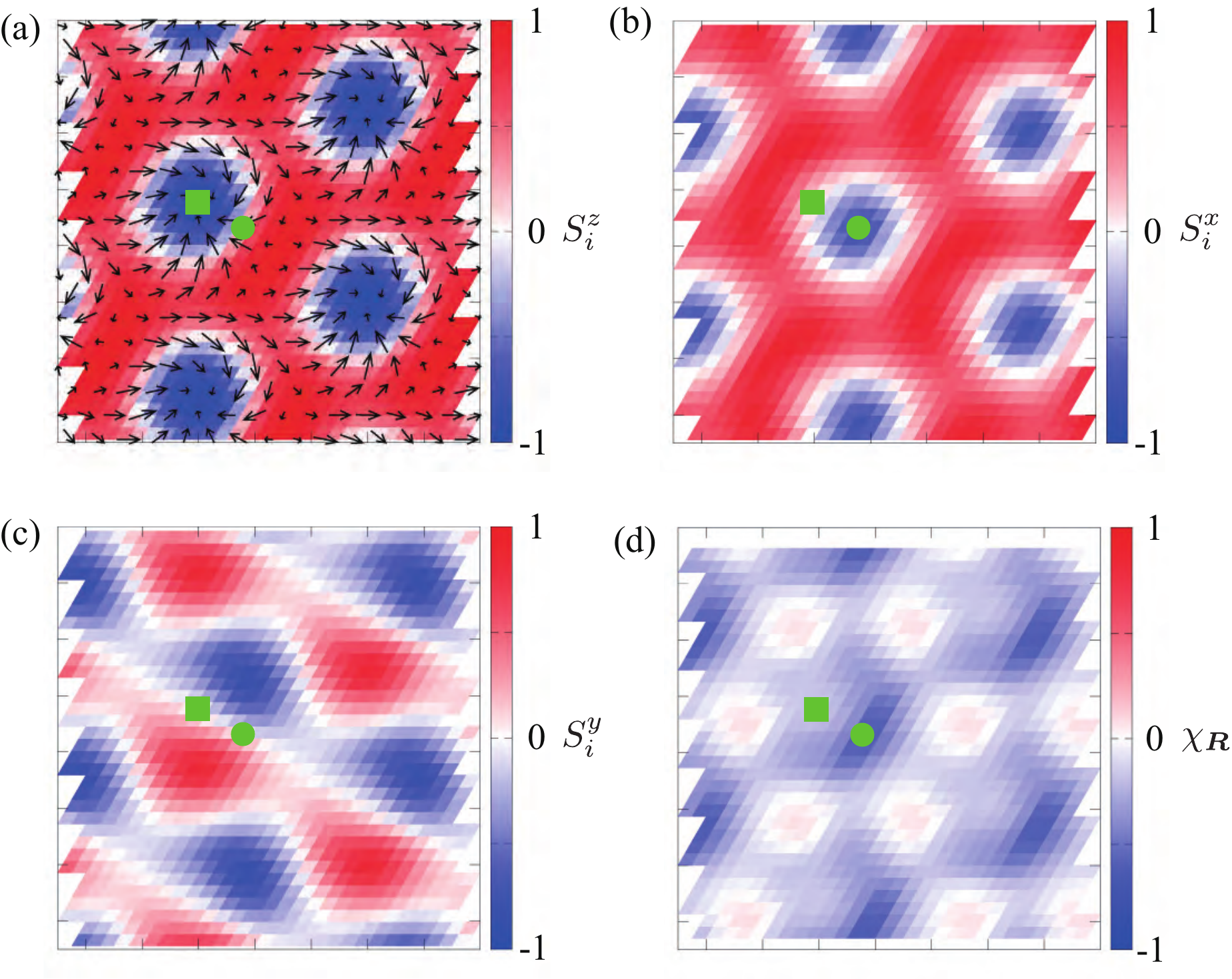} 
\caption{
\label{fig:Spin_Hrot_K=03_A=02}
Real-space spin and chirality configurations of the $n_{\rm sk}=1$ skyrmion crystal at $\theta=63^{\circ}$ for $K=0.3$ and $A=0.2$.
The contour shows the (a) $z$, (b) $x$, and (c) $y$ components of the spin moment, and the arrows in (a) represent the $xy$ components. 
In (d), the contour shows the scalar chirality. 
The green squares and circles represent the positions of the minima of $S^z$ and $S^x$, respectively.  
}
\end{center}
\end{figure}

\begin{figure}[htb!]
\begin{center}
\includegraphics[width=1.0 \hsize]{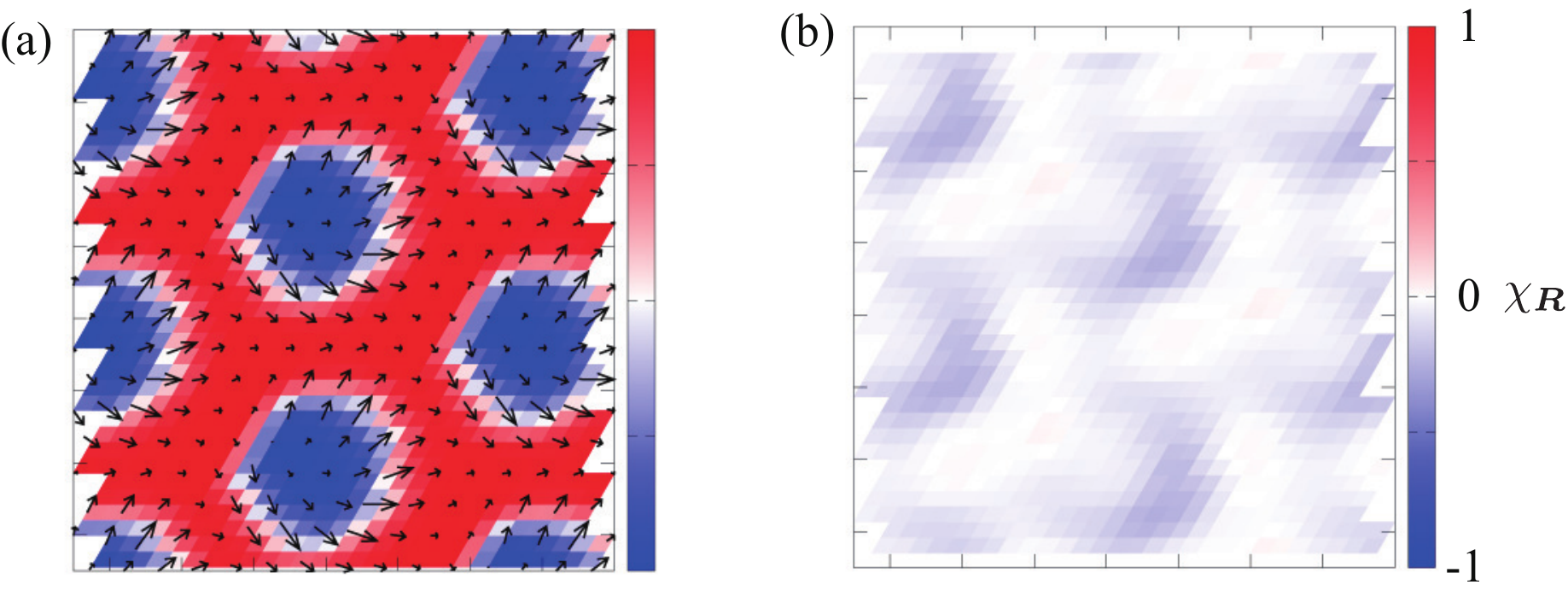} 
\caption{
\label{fig:Spin_Hrot_K=01_A=04}
Real-space spin and chirality configurations of the triple-$Q$ state with nonzero $(\chi_0)^2$ at $\theta=58.5^{\circ}$ for $K=0.3$ and $A=0.4$.
The contour shows (a) the $z$ component of the spin moment, and the arrows represent the $xy$ components. 
In (b), the contour shows the scalar chirality. 
}
\end{center}
\end{figure}

\begin{figure}[htb!]
\begin{center}
\includegraphics[width=1.0 \hsize]{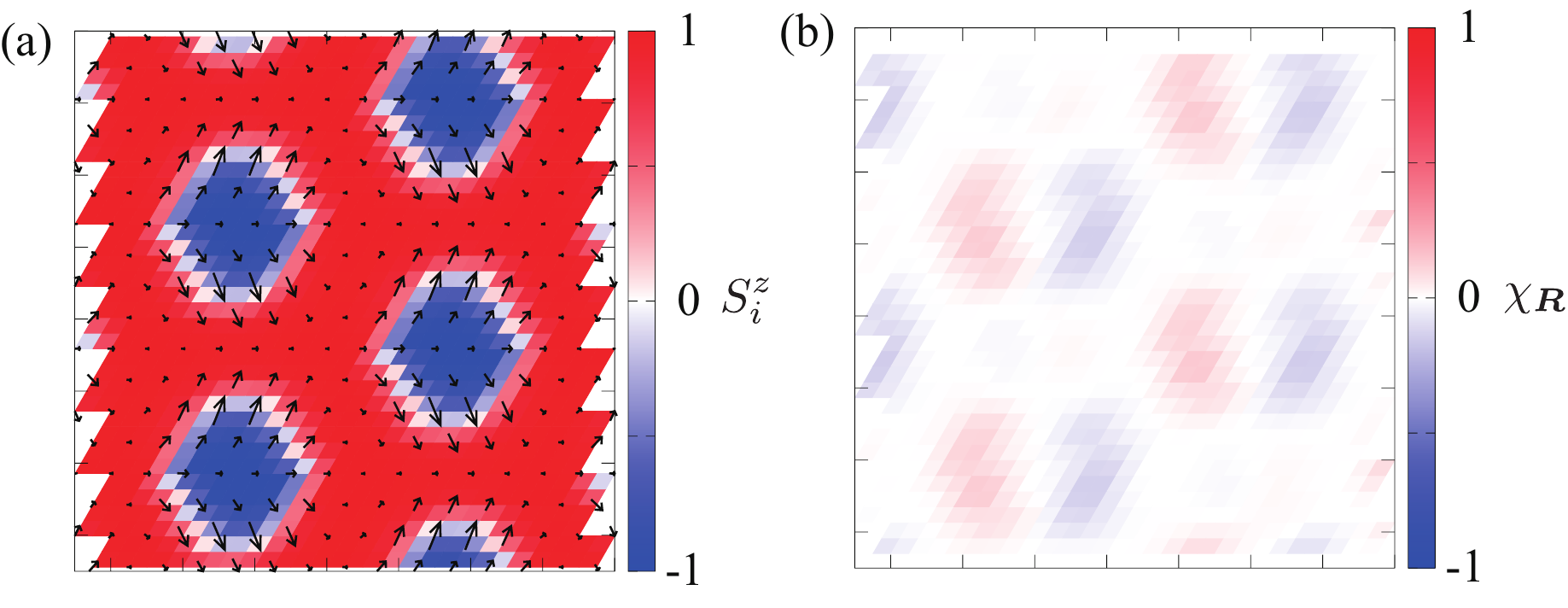} 
\caption{
\label{fig:Spin_Hrot_K=03_A=04}
Real-space spin and chirality configurations of the triple-$Q$ state with the bubble crystal like structure at $\theta=31.5^{\circ}$ for $K=0.3$ and $A=0.4$.
The contour shows (a) the $z$ component of the spin moment, and the arrows represent the $xy$ components. 
In (b), the contour shows the scalar chirality. 
}
\end{center}
\end{figure}

In this section, we examine the multiple-$Q$ instability by rotating the magnetic field in the $xz$ plane as $\bm{H}=H (\sin\theta,0,\cos \theta)$ for $0^{\circ} \leq \theta \leq 90^{\circ}$.  
We fix the magnitude of the field at $H=0.8$ for which the skyrmion crystals are stabilized at $\theta=0^{\circ}$, i.e., for the [001] field, and $\theta=90^{\circ}$, i.e., for the [100] field. 
The results are the same for the magnetic field rotated in the $yz$ plane due to spin rotational symmetry in the $xy$ plane in the absence of the bond-dependent anisotropy, i.e., $\mathcal{H}^{\rm BA}=0$. 
We show the results under the easy-axis anisotropy in Sec.~\ref{sec:With easy-axis anisotropy_Hrot} and the easy-plane anisotropy in Sec.~\ref{With easy-plane anisotropy_Hrot}. 
We discuss the results in this section in Sec.~\ref{sec:Summary of this section_3}. 

\subsubsection{With easy-axis anisotropy}
\label{sec:With easy-axis anisotropy_Hrot}

First, we discuss the results for easy-axis anisotropy with $A=0.2$ shown 
in Figs.~\ref{fig:SIA_Hrot_A=posi_K=0}(a), \ref{fig:SIA_Hrot_A=posi_K=01}(a), and \ref{fig:SIA_Hrot_A=posi_K=03}(a).  
Figure~\ref{fig:SIA_Hrot_A=posi_K=0}(a) shows the result for $K=0$. 
The $n_{\rm sk}=1$ skyrmion crystal with $(\bm{m}_{\bm{Q}_{1}})^2=(\bm{m}_{\bm{Q}_{2}})^2=(\bm{m}_{\bm{Q}_{3}})^2$ is stabilized at $\theta=0^{\circ}$, whose spin texture is presented in Fig.~\ref{fig:Spin_nsk1SkX}(a).
With an increase of $\theta$, the intensities of the triple-$Q$ peaks in the spin structure factor become different and split into two and one; 
see the dashed and solid lines in the second panel of Fig.~\ref{fig:SIA_Hrot_A=posi_K=0}(a). 
This means that the threefold rotational symmetry is broken by the in-plane component of the applied field. 
The symmetry breaking can be clearly seen in the real-space spin and chirality configurations, exemplified for $\theta=5^{\circ}$ and $
45^{\circ}$ in Fig.~\ref{fig:Spin_Hrot_K=0_A=02}. 
While increasing $\theta$, the almost circular skyrmions in Fig.~\ref{fig:Spin_Hrot_K=0_A=02}(a) are slightly deformed in an elliptical form along the $\bm{Q}_1$ direction, as shown in Fig.~\ref{fig:Spin_Hrot_K=0_A=02}(e). 
At the same time, the $x$-spin component shows an elongated hexagonal crystal of the bubbles as shown in Fig.~\ref{fig:Spin_Hrot_K=0_A=02}(f), whose centers defined by the minima 
of $S^x$ are different from those of the skyrmions with the minima of $S^z$, as indicated by the green squares and circles in Figs.~\ref{fig:Spin_Hrot_K=0_A=02}(e) and \ref{fig:Spin_Hrot_K=0_A=02}(f).  
The $y$-spin component shows checker-board type modulation for both $\theta=5^{\circ}$ and $\theta=45^{\circ}$, as shown in Figs.~\ref{fig:Spin_Hrot_K=0_A=02}(c) and \ref{fig:Spin_Hrot_K=0_A=02}(g), respectively. 
These spin configurations imply that the spin axis at the skyrmion cores is tilted from the $z$ to $x$ direction by increasing $\theta$. 
Accordingly, the texture of the scalar chirality is modulated in an asymmetric form in the $x$ direction while increasing $\theta$, as shown in Figs.~\ref{fig:Spin_Hrot_K=0_A=02}(d) and \ref{fig:Spin_Hrot_K=0_A=02}(h). 
This is due to the fact that the intensities of $(\chi_{\bm{Q}_{\nu}})^2$ are also split into two and one, as shown in the lowest panel of Fig.~\ref{fig:SIA_Hrot_A=posi_K=0}(a). 

The $n_{\rm sk}=1$ skyrmion crystal remains stable against the rotation of the magnetic field up to $\theta \simeq 49^{\circ}$, and then changes into a different triple-$Q$ 
state, as shown in Fig.~\ref{fig:SIA_Hrot_A=posi_K=0}(a).  
The triple-$Q$ state for $ 49^{\circ} \lesssim \theta \lesssim 63^{\circ}$ is characterized by the dominant peak at $\bm{Q}_1$ and two subdominant peaks at $\bm{Q}_2$ and $\bm{Q}_3$ in the spin and by the two peaks at $\bm{Q}_2$ and $\bm{Q}_3$ in the chirality. 
While further increasing $\theta$, this state smoothly changes into the single-$Q$ helical state discussed in Sec.~\ref{sec:With easy-axis anisotropy_Hx}. 
It should be noted that a small $(\chi_0)^2$ is induced for $\theta  \gtrsim 49^{\circ} $ due to the staggered arrangement of the scalar chirality. 

When we switch on $K$, the $n_{\rm sk}=1$ skyrmion crystal becomes more robust against $\theta$; it extends up to $\theta \simeq 63^{\circ}$ for $K=0.1$ as shown in Fig.~\ref{fig:SIA_Hrot_A=posi_K=01}(a). 
Meanwhile, the single-$Q$ helical state for large $\theta$ is unstable and taken over by the triple-$Q$ state found in the region for $49^{\circ} \lesssim \theta \lesssim 63^{\circ}$ in Fig.~\ref{fig:SIA_Hrot_A=posi_K=0}(a). 
In the large $\theta$ region, however, the triple-$Q$ state changes its symmetry for $\theta \gtrsim 85^\circ$ with the different intensities at $\bm{Q}_2$ and $\bm{Q}_3$ in both spin and chirality. 
The tendency that $K$ favors the multiple-$Q$ states is consistent with the results in Secs.~\ref{sec:Field along the $z$ direction_SIA} and \ref{sec:Field along the $x$ direction_SIA}. 

For larger $K$, the region where the $n_{\rm sk}=1$ skyrmion crystal is stabilized is further extended to larger $\theta \simeq 67^{\circ}$ for $K=0.3$, as shown in Fig.~\ref{fig:SIA_Hrot_A=posi_K=03}(a). 
Within the region, however, $\bm{m}_{\bm{Q}_\nu}$ and $\chi_{\bm{Q}_\nu}$ show discontinuity at $\theta \simeq 45^{\circ}$, while $(\chi_0)^2$ appears to be continuous. 
The discontinuity is ascribed to further deformation of the skyrmions. 
We show the real-space spin and chirality configurations at $\theta=63^{\circ}$ in Fig.~\ref{fig:Spin_Hrot_K=03_A=02}. 
Due to the anisotropic triple-$Q$ structure for the $y$-spin component in Fig.~\ref{fig:Spin_Hrot_K=03_A=02}(c) in contrast to the double-$Q$ structure in Fig.~\ref{fig:Spin_Hrot_K=0_A=02}(c), the positions of the minima of $S^x$ and $S^z$ are different in not only the $x$ but also $y$ direction, as shown by the green squares and circles in Figs.~\ref{fig:Spin_Hrot_K=03_A=02}(a) and \ref{fig:Spin_Hrot_K=03_A=02}(b). 
Accordingly, the scalar chirality is distributed in an asymmetric form in both $x$ and $y$ directions, as shown in Fig.~\ref{fig:Spin_Hrot_K=03_A=02}(d). 
These spin configurations imply that the spin axis at the skyrmion cores is tilted from the $z$ to both $x$ and $y$ directions. 
We thus deduce that the phase transition at $\theta \simeq 45^{\circ}$ is caused by a phase shift among the constituent waves, similar to that found for in Sec.~\ref{sec:With easy-axis anisotropy_Hx} (see Fig.~\ref{fig:Spin_compare_Hx}).  
For larger $\theta$, the $n_{\rm sk}=1$ skyrmion crystal changes into the anisotropic triple-$Q$ state at $\theta \simeq 67^{\circ}$, which smoothly turns into the state obtained at $\theta=90^{\circ}$ in Sec.~\ref{sec:With easy-axis anisotropy_Hx}. 

The results for larger single-ion anisotropy $A=0.4$ are shown in Figs.~\ref{fig:SIA_Hrot_A=posi_K=0}(b), \ref{fig:SIA_Hrot_A=posi_K=01}(b), and \ref{fig:SIA_Hrot_A=posi_K=03}(b). 
The critical angles where the $n_{\rm sk}=1$ skyrmion crystal is destabilized are almost the same as those at $A=0.2$ in the cases of $K=0$ and $K=0.1$, although $(\chi_0)^2$ is suppressed due to the reduction of $(m^{x}_{\bm{Q}_\nu})^2$ and $(m^{y}_{\bm{Q}_\nu})^2$, as shown in Figs.~\ref{fig:SIA_Hrot_A=posi_K=0}(b) and \ref{fig:SIA_Hrot_A=posi_K=01}(b).

Meanwhile, the situation for $K=0.3$ looks more complicated than for $K=0$ and $0.1$, as shown in Fig.~\ref{fig:SIA_Hrot_A=posi_K=03}(b). 
In this case, we find two skyrmion crystals: the $n_{\rm sk}=1$ skyrmion crystal for $0^{\circ} \lesssim  \theta \lesssim 18^{\circ}$ and the $n_{\rm sk}=2$ skyrmion crystal for $81^{\circ} \lesssim  \theta \lesssim 90^{\circ}$. 
The former is similar to that found at $K=0.1$ in Fig.~\ref{fig:SIA_Hrot_A=posi_K=01}(b).
We also obtain the other chiral magnetic states which are topologically trivial (the skyrmion number is zero) next to the skyrmion crystal ($18^\circ \lesssim \theta \lesssim 22^\circ$) and in the intermediate field region ($54^\circ \lesssim \theta \lesssim 67^\circ$).
The real-space spin and chirality configurations of the intermediate-field state are shown in Figs.~\ref{fig:Spin_Hrot_K=01_A=04}(a) and \ref{fig:Spin_Hrot_K=01_A=04}(b), respectively.
Although the spin texture looks similar to the $n_{\rm sk}=1$ skyrmion crystal in Fig.~\ref{fig:Spin_Hrot_K=03_A=02}(a), this state has zero skyrmion number. 
The results imply that the topological nature can be switched by keen competition among the different spin textures in the rotated magnetic field under the strong influence of itinerant nature of electrons.
Between these chiral states, we obtain two triple-$Q$ states with $(\chi_0)^2=0$ for $22^{\circ}\lesssim \theta \lesssim 54^\circ$ and $67^\circ \lesssim \theta \lesssim 81^\circ$. 
While the latter is similar to the one found in the case with $A=0.2$ in Fig.~\ref{fig:SIA_Hrot_A=posi_K=03}(a), the former appears only for larger $A$ and has a bubble crystal like structure. 
The spin and chirality configurations are shown in Figs.~\ref{fig:Spin_Hrot_K=03_A=04}(a) and \ref{fig:Spin_Hrot_K=03_A=04}(b), respectively. 
The $xy$-spin components do not rotate around the cores denoted by the blue regions in Fig.~\ref{fig:Spin_Hrot_K=03_A=04}(a); they rotate in an opposite way between the left and right sides of the cores. 
Thus, the local scalar chirality with the opposite sign is induced around the core, but they are canceled out with each other, as shown in Fig.~\ref{fig:Spin_Hrot_K=03_A=04}(b).

\begin{figure}[htb!]
\begin{center}
\includegraphics[width=1.0 \hsize]{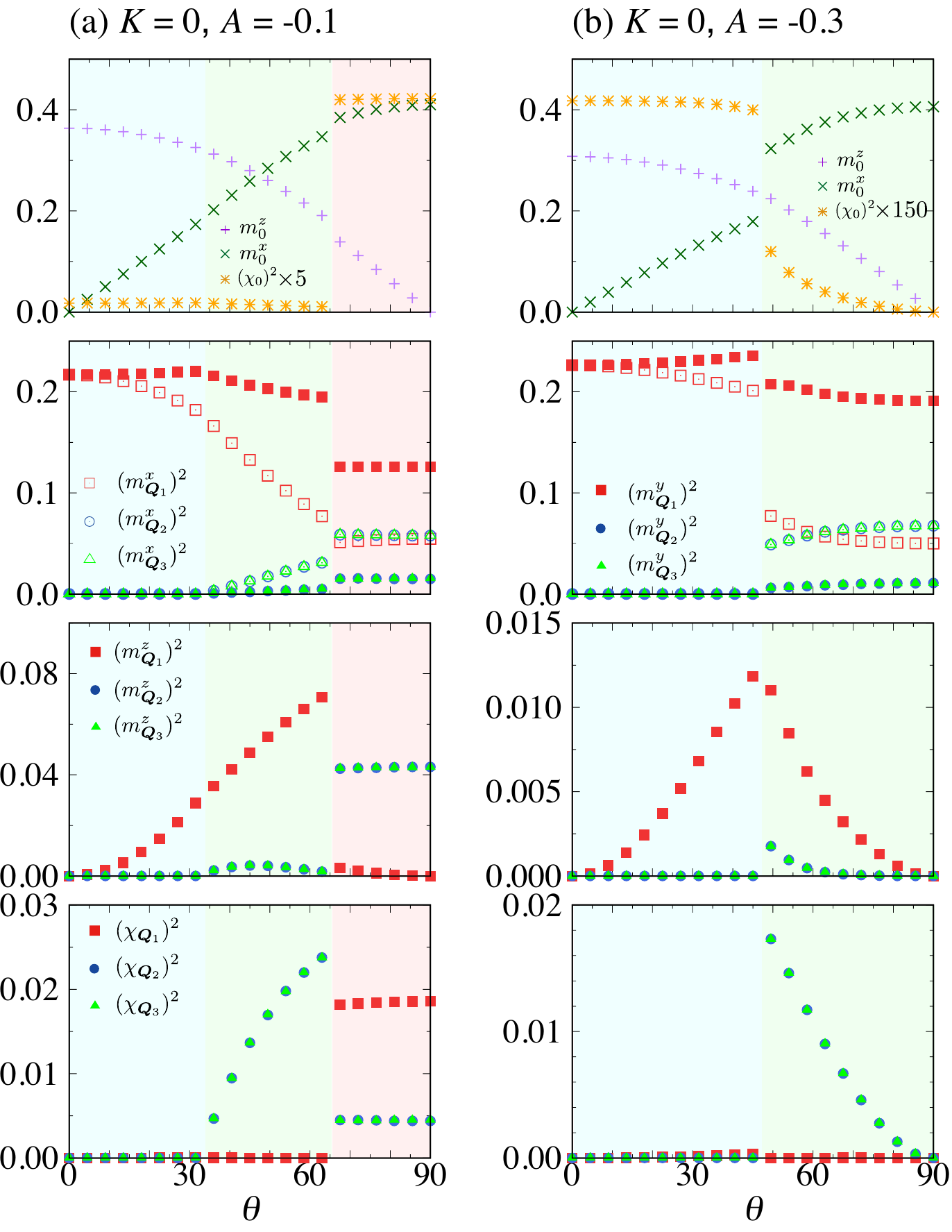} 
\caption{
\label{fig:SIA_Hrot_A=nega_K=0}
$\theta$ dependence of (first row) $m^z_0$, $m^x_0$, and $(\chi_0)^2$, (second row) $(m^{x}_{\bm{Q}_\nu})^2$ and $(m^{y}_{\bm{Q}_\nu})^2$, (third row) $(m^{z}_{\bm{Q}_\nu})^2$, and (fourth row) $ (\chi_{\bm{Q}_\nu})^2$ for $K=0$ at (a) $A=-0.1$ and (b) $A=-0.3$. 
The magnitude of the magnetic field is fixed at $H=0.8$.
}
\end{center}
\end{figure}

\begin{figure}[htb!]
\begin{center}
\includegraphics[width=1.0 \hsize]{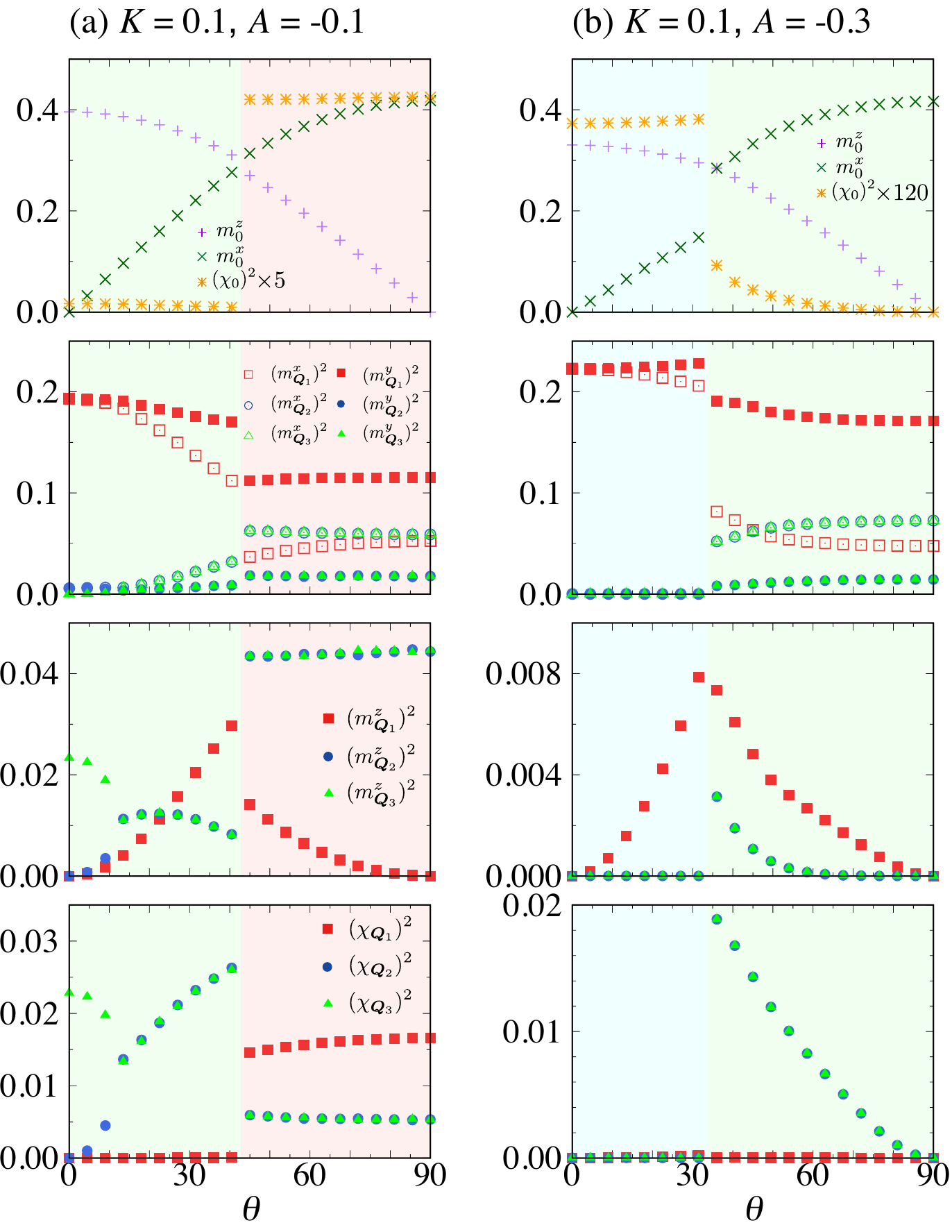} 
\caption{
\label{fig:SIA_Hrot_A=nega_K=01}
The same plots as in Fig.~\ref{fig:SIA_Hrot_A=nega_K=0} for $K=0.1$.
}
\end{center}
\end{figure}

\begin{figure}[htb!]
\begin{center}
\includegraphics[width=1.0 \hsize]{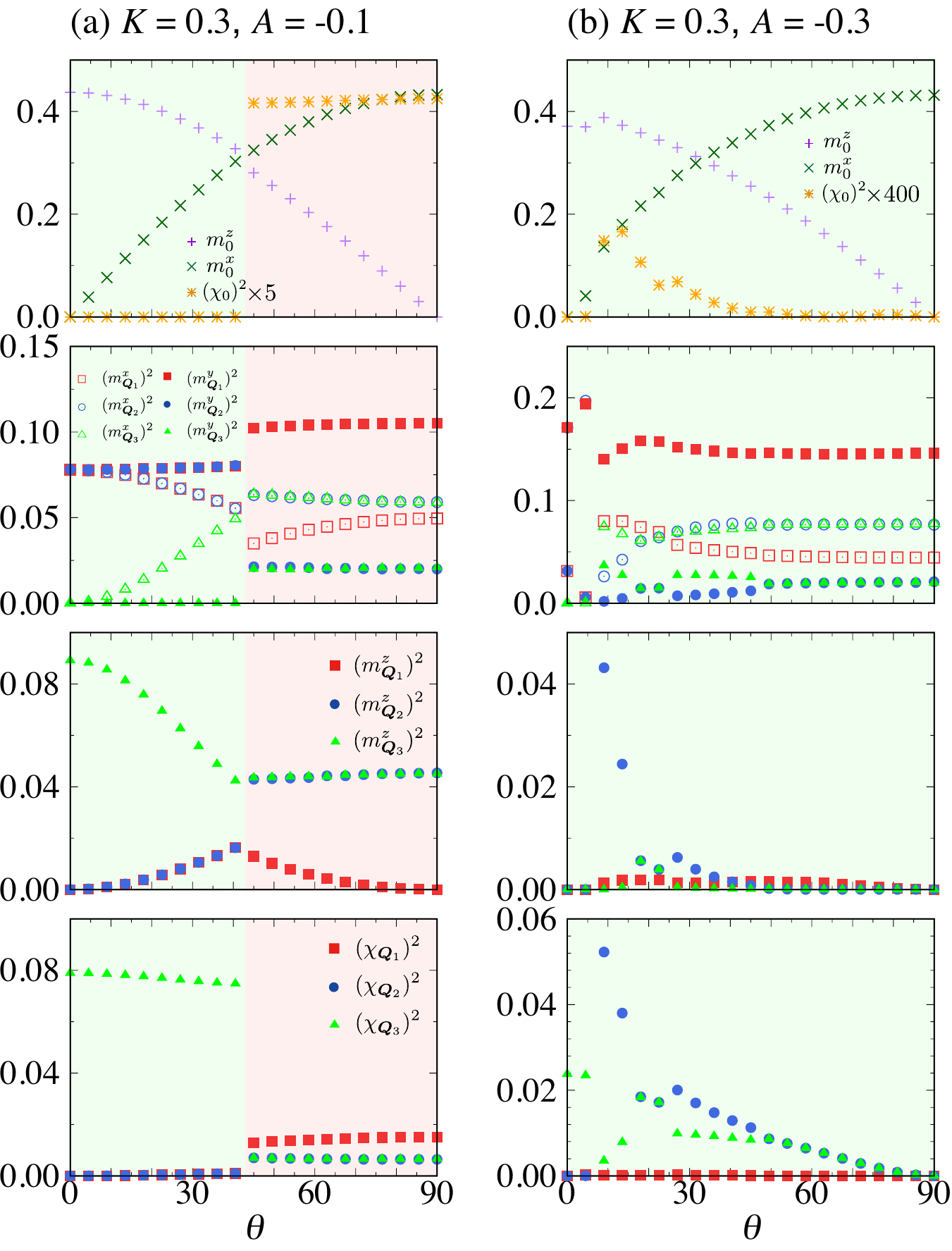} 
\caption{
\label{fig:SIA_Hrot_A=nega_K=03}
The same plots as in Fig.~\ref{fig:SIA_Hrot_A=nega_K=0} for $K=0.3$.
}
\end{center}
\end{figure}

\subsubsection{With easy-plane anisotropy}
\label{With easy-plane anisotropy_Hrot}

Figures~\ref{fig:SIA_Hrot_A=nega_K=0}-\ref{fig:SIA_Hrot_A=nega_K=03} show the results in the presence of easy-plane anisotropy when the field is rotated in the $xz$ plane. 
For $A=-0.1$, the $n_{\rm sk}=1$ skyrmion crystal is stabilized at $\theta=90^{\circ}$ irrespective of $K$, as shown in Figs.~\ref{fig:SIA_Hrot_A=nega_K=0}(a), \ref{fig:SIA_Hrot_A=nega_K=01}(a), and \ref{fig:SIA_Hrot_A=nega_K=03}(a). 
The critical angles where the $n_{\rm sk}=1$ skyrmion crystal is destabilized are $\theta \simeq 65^{\circ}$ for $K=0$, $\theta \simeq 42^{\circ}$ for $K=0.1$, and $\theta \simeq 42^{\circ}$ for $K=0.3$. 
This indicates that nonzero $K$ enhances the stability of the $n_{\rm sk}=1$ skyrmion crystal, while the critical angle appears to saturate for large values of $K$. 
By further tilting the magnetic field to the $z$ direction, the $n_{\rm sk}=1$ skyrmion crystal at $K=0$ is replaced with the triple-$Q$ state accompanied with the double-$Q$ chirality density wave and the staggered component of $(\chi_0)^2$ for $\theta \lesssim 65^{\circ}$, as shown in Fig.~\ref{fig:SIA_Hrot_A=nega_K=0}(a). 
While further decreasing $\theta$, the intensities of $(\bm{m}_{\bm{Q}_2})^2$ and $(\bm{m}_{\bm{Q}_3})^2$ are suppressed and become zero at $\theta \simeq 33^{\circ}$. 
In other words, the anisotropic triple-$Q$ state changes into the single-$Q$ state, in which $(m_{\bm{Q}_1}^\alpha)^2$ change gradually so that the spiral plane keeps being perpendicular to the field direction. 
For $K=0.1$, the behavior against $\theta$ is similar to that for $K=0$ except that another triple-$Q$ state with different intensities at $\bm{Q}_1$, $\bm{Q}_2$, and $\bm{Q}_3$ appears for $\theta \lesssim 11^{\circ}$, which continuously turns into the state at $\theta = 0^{\circ}$ obtained in Sec.~\ref{sec:With easy-plane anisotropy_Hz}, as shown in Fig.~\ref{fig:SIA_Hrot_A=nega_K=01}(a). 
In the case of $K=0.3$, as shown in Fig.~\ref{fig:SIA_Hrot_A=nega_K=03}(a), yet another triple-$Q$ state with the single-$Q$ chirality density wave is realized for $\theta \lesssim 42^{\circ}$, which also continuously turns into the state at $\theta = 0^{\circ}$ obtained in Sec.~\ref{sec:With easy-plane anisotropy_Hz}. 

When the easy-plane anisotropy becomes stronger, the $n_{\rm sk}=1$ skyrmion crystal is destabilized for all $K$, as shown in Figs.~\ref{fig:SIA_Hrot_A=nega_K=0}(b), \ref{fig:SIA_Hrot_A=nega_K=01}(b), and \ref{fig:SIA_Hrot_A=nega_K=03}(b) for $A=-0.3$. 
For $K=0$ and $0.1$, there remains a phase transition between the single-$Q$ conical state realized at $\theta=0^{\circ}$ and the anisotropic triple-$Q$ state realized at $\theta=90^{\circ}$, at $\theta \simeq 47^{\circ}$ for $K=0$ and $\theta \simeq 34^{\circ}$ for $K=0.1$, as shown in Figs.~\ref{fig:SIA_Hrot_A=nega_K=0}(b) and \ref{fig:SIA_Hrot_A=nega_K=01}(b), respectively. 
For $K=0.3$, however, the single-$Q$ conical state disappears and there are multiple phase transitions between three different types of triple-$Q$ states, as shown in Fig.~\ref{fig:SIA_Hrot_A=nega_K=03}(b). 
The state for $0^\circ < \theta \lesssim 7^\circ$ is continuously modulated from the anisotropic double-$Q$ state found for the $\theta=0$ case in Fig.~\ref{fig:SIA_Hz_A=nega_K=03}(b), by acquiring a small nonzero $(m^y_{\bm{Q}_3})^2$ for nonzero $\theta$. 
On the other hand, the state for $47^{\circ}\lesssim \theta < 90^{\circ}$ is also continuously modulated from the one for $\theta=90^\circ$ in Fig.~\ref{fig:SIA_Hx_A=nega_K=03}(b). 
For $7^\circ \lesssim \theta \lesssim 47^{\circ}$, the state for $47^{\circ}\lesssim \theta < 90^{\circ}$ is almost energetically degenerate with a different triple-$Q$ state, and the competition causes phase transitions at $\theta \simeq 16^\circ$ and $25^\circ$, as found in Fig.~\ref{fig:SIA_Hx_A=nega_K=03}(b) while changing $H^x$.

\subsubsection{Discussion}
\label{sec:Summary of this section_3}

The results obtained in this section are summarized in Fig.~\ref{fig:Summary}(c). 
We found similar tendency with respect to the stability of the $n_{\rm sk}=1$ skyrmion crystal for both $A>0$ and $A<0$; the range of the field angle $\theta$ for the $n_{\rm sk}=1$ skyrmion crystal becomes wider for larger $K$. 
We found, however, that the spin axis at the skyrmion cores is tilted from the $z$ direction to the $xy$ plane in the case of $A>0$, as shown in Figs.~\ref{fig:Spin_Hrot_K=0_A=02}(e), \ref{fig:Spin_Hrot_K=0_A=02}(f), \ref{fig:Spin_Hrot_K=03_A=02}(a), and \ref{fig:Spin_Hrot_K=03_A=02}(b).
In the case of $A<0$, a similar tilting occurs from the $x$ direction to the $yz$ plane (not shown).

In addition to the skyrmion crystal, we found a triple-$Q$ state with nonzero $(\chi_0)^2$ in the rotated field, as shown in Fig.~\ref{fig:Spin_Hrot_K=01_A=04}. 
We also found a bubble crystal with $(\chi_0)^2=0$ between the topological states, where the opposite sign of the scalar chirality is distributed around the single core, as shown in Fig.~\ref{fig:Spin_Hrot_K=03_A=04}. 
As these peculiar states are obtained only for large $K$ and $A$, it is desired to target the materials with large spin-charge coupling and easy-axis anisotropy for exploring them.

\section{Bond-dependent Anisotropy}
\label{sec:Bond-exchange Anisotropy}
In this section, we examine the effect of the bond-dependent exchange interaction $I^{\rm A}$ by considering the Hamiltonian $\mathcal{H}= \mathcal{H}^{\rm BBQ}+\mathcal{H}^{\rm BA}+\mathcal{H}^{\rm Z}$ (i.e., $\mathcal{H}^{\rm SIA}=0$). 
The magnetic phase diagram at zero magnetic field is shown in Sec.~\ref{sec:At zero field_bond}. 
In Sec.~\ref{sec:Field along the $z$ direction}, we present the results in the magnetic field applied to the $z$ direction. 
In contrast to the case with single-ion anisotropy, we could not find any instability toward the skyrmion crystals in the in-plane magnetic field, and hence, we do not show the results for the in-plane magnetic field as well as the rotated field. 
We discuss the results in this section in Sec.~\ref{sec:Summary of this section_4}.

\subsection{At zero field}
\label{sec:At zero field_bond}

\begin{figure}[htb!]
\begin{center}
\includegraphics[width=1.0 \hsize]{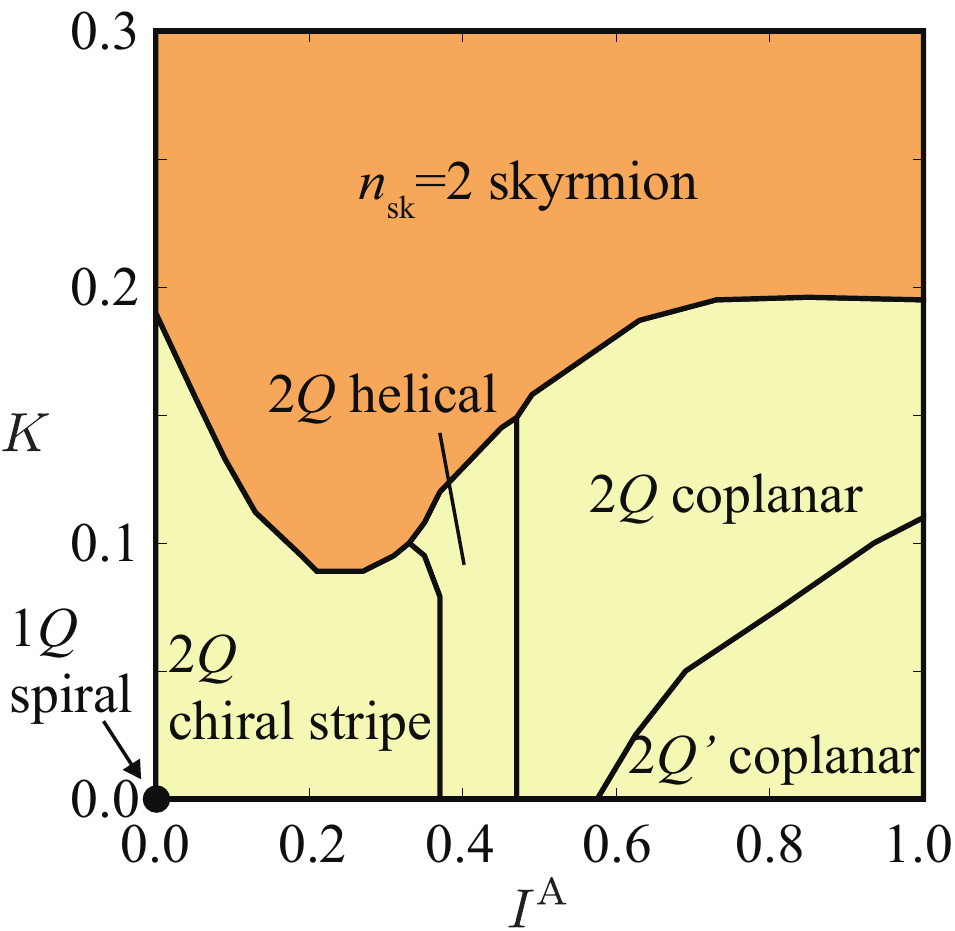} 
\caption{
\label{fig:souzu_bond-K}
Magnetic phase diagram of the model in Eq.~(\ref{eq:Ham}) with $\mathcal{H}^{\rm SIA} = \mathcal{H}^{\rm Z} = 0$ obtained by the simulated annealing at $T=0.01$.  
}
\end{center}
\end{figure}

\begin{figure}[htb!]
\begin{center}
\includegraphics[width=1.0 \hsize]{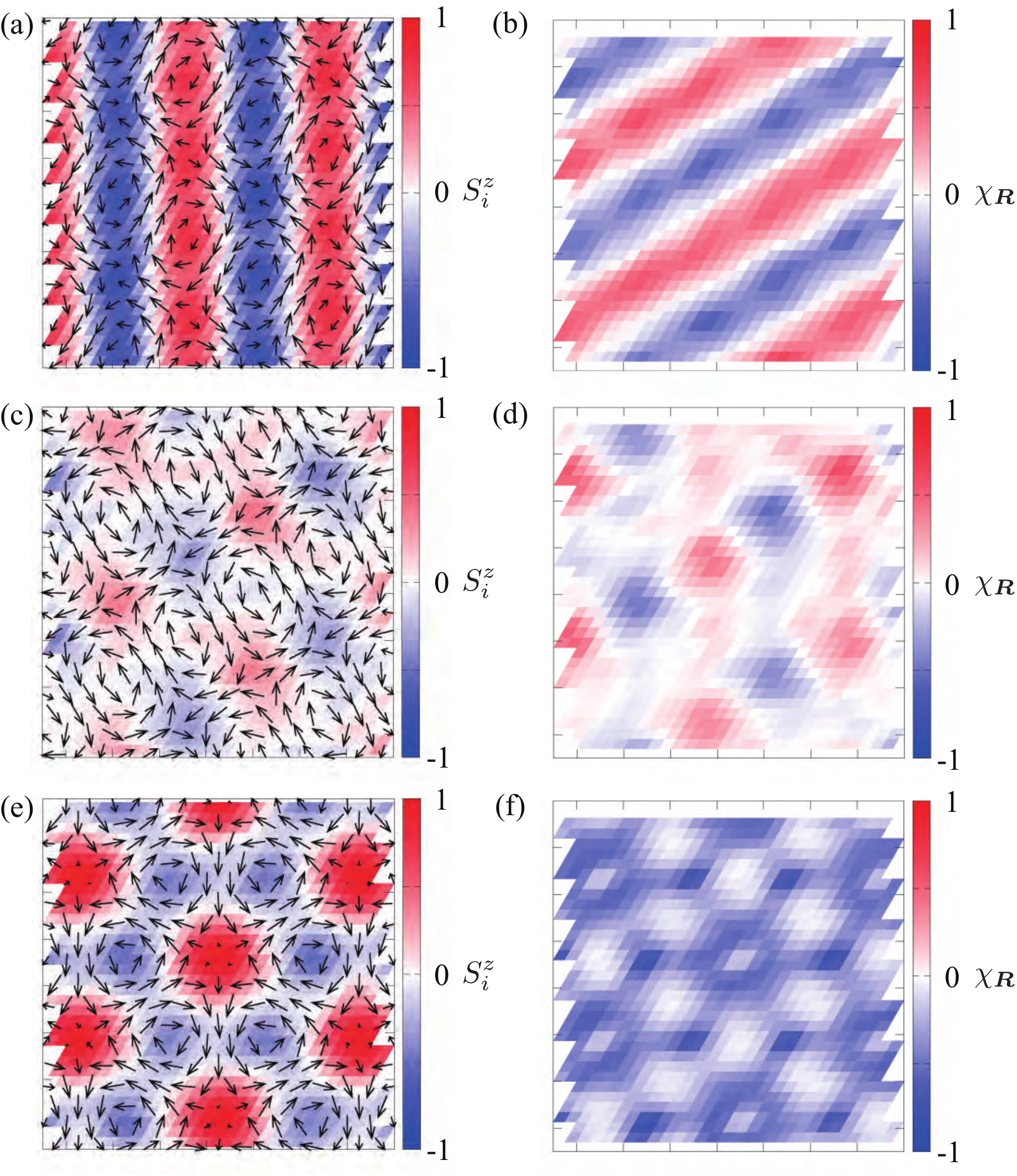} 
\caption{
\label{fig:Spin_bond_K=03_IA=05}
Real-space spin configurations of (a) the double-$Q$ (2$Q$) chiral stripe state at $K=0.1$ and $I^{\rm A}=0.1$, (c) the 2$Q$ helical state at $K=0.1$ and $I^{\rm A}=0.4$, and (e) the $n_{\rm sk}=2$ skyrmion crystal at $K=0.3$ and $I^{\rm A}=0.5$. 
The contour shows the $z$ component of the spin moment, and the arrows represent the $xy$ components. 
(b), (d), and (f) display the real-space chirality configurations corresponding to (a), (c), and (e), respectively. 
}
\end{center}
\end{figure}

\begin{figure}[htb!]
\begin{center}
\includegraphics[width=1.0 \hsize]{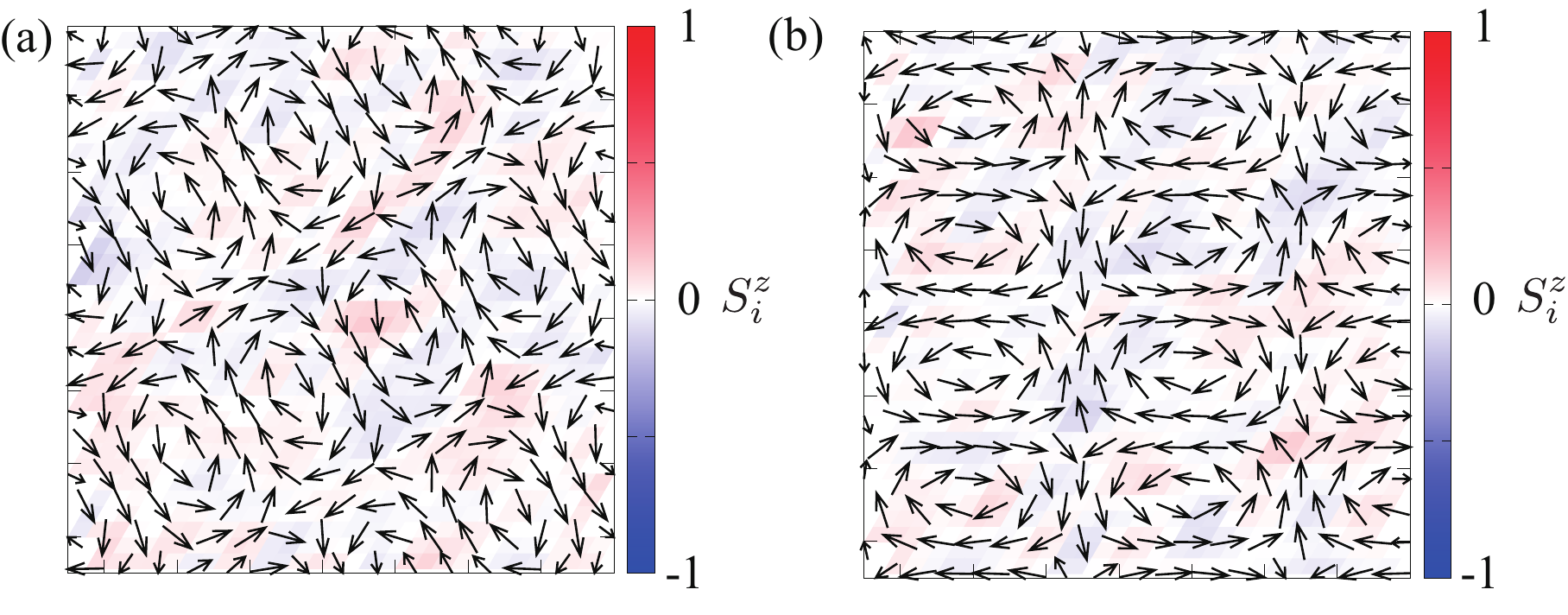} 
\caption{
\label{fig:Spin_bond_cop}
Real-space spin configurations of (a) the double-$Q$ (2$Q$) coplanar state at $K=0.1$ and $I^{\rm A}=0.6$ and (b) the anisotropic 2$Q$ (2$Q'$) coplanar state at $K=0.1$ and $I^{\rm A}=1$.
The contour shows the $z$ component of the spin moment, and the arrows represent the $xy$ components. 
}
\end{center}
\end{figure}

\begin{figure*}[htb!]
\begin{center}
\includegraphics[width=1.0 \hsize]{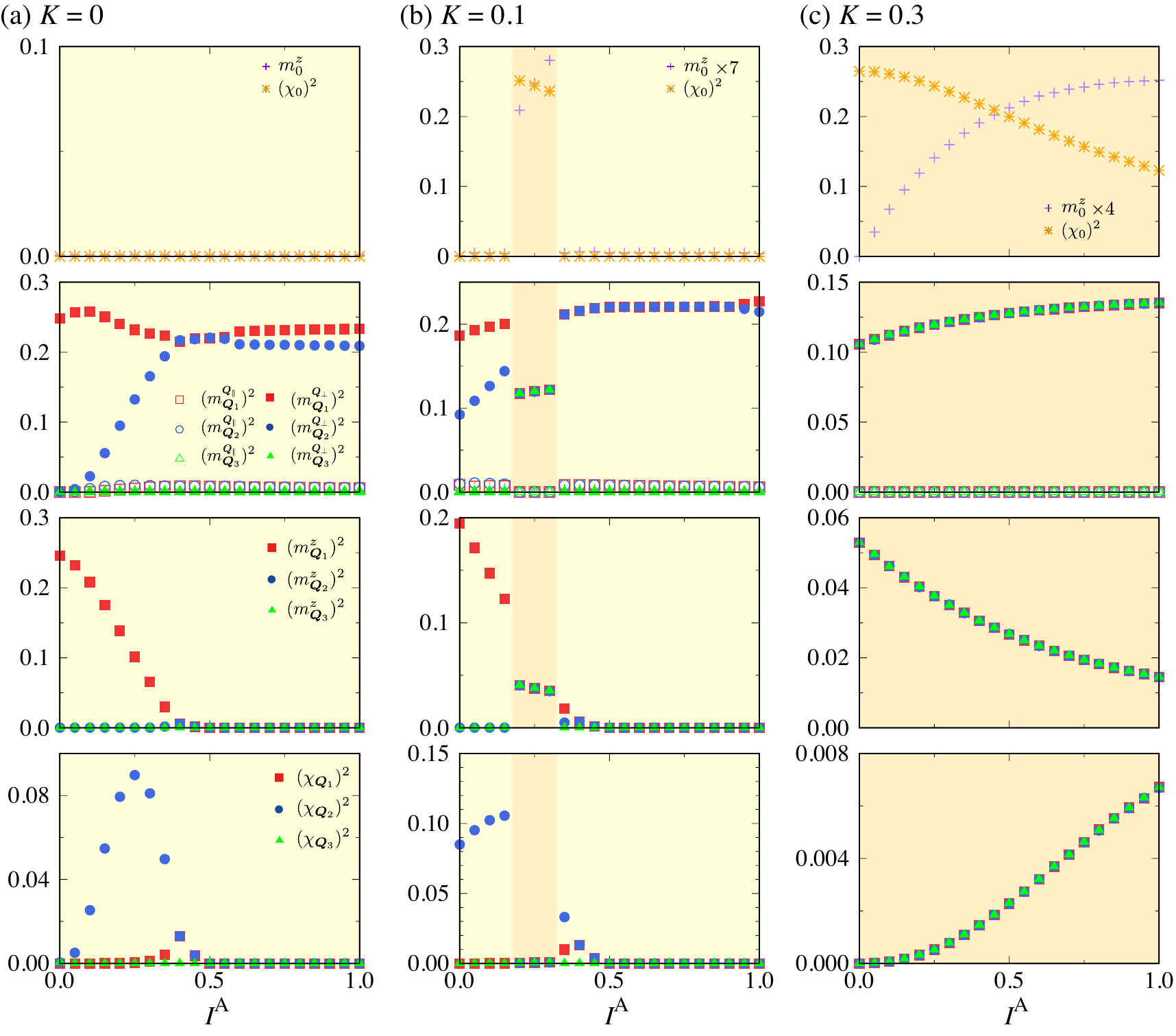} 
\caption{
\label{fig:bonddep_H=0}
$I^{\rm A}$ dependence of (first row) $m^z_0$ and $(\chi_0)^2$, (second row) $(m^{\bm{Q}_\perp}_{\bm{Q}_\nu})^2$ and $(m^{\bm{Q}_\parallel}_{\bm{Q}_\nu})^2$, (third row) $(m^{z}_{\bm{Q}_\nu})^2$, and (fourth row) $ (\chi_{\bm{Q}_\nu})^2$ for (a) $K=0$, (b) $K=0.1$, and (c) $K=0.3$ in the absence of the magnetic field. 
}
\end{center}
\end{figure*}

\begin{figure}[htb!]
\begin{center}
\includegraphics[width=1.0 \hsize]{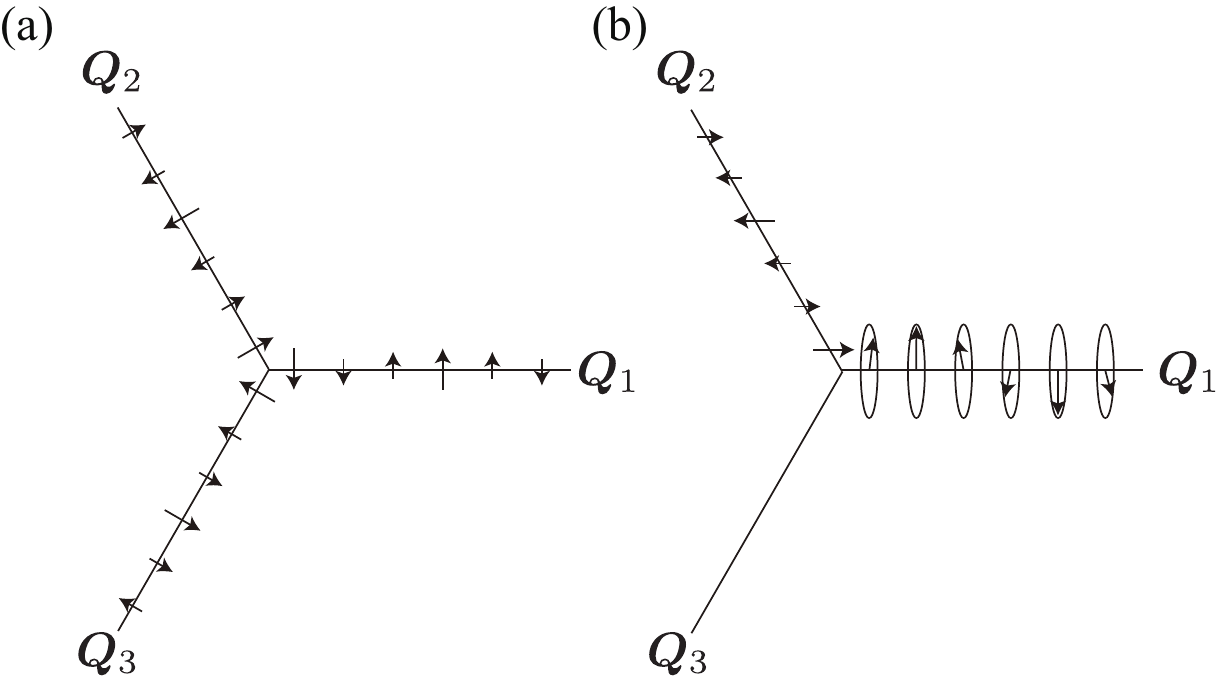} 
\caption{
\label{fig:ponti}
Schematic pictures of (a) three sinusoidal waves consisting the $n_{\rm sk}=2$ skyrmion crystal and (b) the single-$Q$ helical and single-$Q$ sinusoidal waves consisting the double-$Q$ chiral stripe state in Fig.~\ref{fig:souzu_bond-K} for $I^{\rm A}>0$ at zero field. 
See also Figs.~\ref{fig:Spin_bond_K=03_IA=05}(a), \ref{fig:Spin_bond_K=03_IA=05}(b), \ref{fig:Spin_bond_K=03_IA=05}(e), and \ref{fig:Spin_bond_K=03_IA=05}(f)
. 
}
\end{center}
\end{figure}

First, we present the magnetic phase diagram for the model in Eq.~(\ref{eq:Ham}) with $\mathcal{H}^{\rm SIA} = \mathcal{H}^{\rm Z} = 0$ obtained by the simulated annealing in Fig.~\ref{fig:souzu_bond-K}. 
There are six phases including the single-$Q$ spiral state at $K=0$ and $I^{\rm A}=0$, whose spin and chirality configurations are exemplified in Figs.~\ref{fig:Spin_bond_K=03_IA=05} and \ref{fig:Spin_bond_cop}. 
The spin and chirality related quantities are plotted in Fig.~\ref{fig:bonddep_H=0} as functions of $I^{\rm A}$ for $K=0$, $0.1$, and $0.3$. 

At $K=0$, the introduction of $I^{\rm A}$ stabilizes a double-$Q$ state with different intensities at $\bm{Q}_1$ and $\bm{Q}_2$; the dominant component is characterized by $\bm{Q}_1$ whose spiral plane lies on the $yz$ plane, i.e., $(m^{\bm{Q}_\perp}_{\bm{Q}_1})^2$ and $(m^z_{\bm{Q}_1})^2$, whereas the subdominant component is induced along the direction perpendicular to $\bm{Q}_2$, i.e., $(m^{\bm{Q}_\perp}_{\bm{Q}_2})^2$, as shown in the middle two panels of Fig.~\ref{fig:bonddep_H=0}(a). 
The $\bm{Q}_2$ component is increased by $I^{\rm A}$. 
We note that $(m^{\bm{Q}_\parallel}_{\bm{Q}_1})^2$ and $(m^{\bm{Q}_\parallel}_{\bm{Q}_2})^2$ are also induced by $I^{\rm A}$, as shown in the second panel of Fig.~\ref{fig:bonddep_H=0}(a). 
The real-space spin and chirality configurations in this phase obtained by the simulated annealing are exemplified in Figs.~\ref{fig:Spin_bond_K=03_IA=05}(a) and \ref{fig:Spin_bond_K=03_IA=05}(b), respectively. 
The chirality has a stripe pattern with the $\bm{Q}_2$ component, as indicated by nonzero $(\chi_{\bm{Q}_2})^2$ in the lowest panel of Fig.~\ref{fig:bonddep_H=0}(a).
The result indicates that the single-$Q$ spiral state at $I^{\rm A}=0$ turns into the double-$Q$ chiral stripe state even in the case of $K=0$. 

While increasing $I^{\rm A}$, the double-$Q$ chiral stripe state changes into the double-$Q$ helical state for $0.37 \lesssim I^{\rm A} \lesssim 0.47$. 
In this state, the spin pattern is characterized by two dominant contributions from $(m^{\bm{Q}_\perp}_{\bm{Q}_1})^2$ and $(m^{\bm{Q}_\perp}_{\bm{Q}_2})^2$ and subdominant contributions from $(m^{z}_{\bm{Q}_1})^2$, $(m^{z}_{\bm{Q}_2})^2$, $(m^{\bm{Q}_\parallel}_{\bm{Q}_1})^2$, and $(m^{\bm{Q}_\parallel}_{\bm{Q}_2})^2$, as shown in the middle two panels of Fig.~\ref{fig:bonddep_H=0}(a). 
Due to the small contributions from $(m^{z}_{\bm{Q}_1})^2$ and $(m^{z}_{\bm{Q}_2})^2$, this spin state is noncoplanar, which is also indicated from nonzero $(\chi_{\bm{Q}_1})^2$ and $(\chi_{\bm{Q}_2})^2$ shown in the lowest panel of Fig.~\ref{fig:bonddep_H=0}(a). 
The real-space spin and chirality configurations in this phase are shown in Figs.~\ref{fig:Spin_bond_K=03_IA=05}(c) and \ref{fig:Spin_bond_K=03_IA=05}(d), respectively. 

While further increasing $I^{\rm A}$, $(\chi_{\bm{Q}_1})^2$ and $(\chi_{\bm{Q}_2})^2$ vanish continuously at $I^{\rm A} \simeq 0.47$, whereas the $xy$ components of $(\bm{m}_{\bm{Q}_1})^2$ and $(\bm{m}_{\bm{Q}_2})^2$ are almost unchanged. 
This means that a double-$Q$ coplanar state is realized for $I^{\rm A} \gtrsim 0.47$. 
There are two types of the double-$Q$ coplanar states: the isotropic one with $(\bm{m}_{\bm{Q}_1})^2 = (\bm{m}_{\bm{Q}_2})^2$ for $0.47 \lesssim I^{\rm A} \lesssim 0.58$ (denoted as $2Q$ coplanar in Fig.~\ref{fig:souzu_bond-K}) and the anisotropic one with $(\bm{m}_{\bm{Q}_1})^2 > (\bm{m}_{\bm{Q}_2})^2$ for $I^{\rm A} \gtrsim 0.58$ (denoted as $2Q'$ coplanar in Fig.~\ref{fig:souzu_bond-K}). 
The spin configurations of these two states are shown in Figs.~\ref{fig:Spin_bond_cop}(a) and \ref{fig:Spin_bond_cop}(b). 

Thus, the results indicate that the anisotropic bond-dependent interaction $I^{\rm A}$ induces various double-$Q$ states even for $K=0$ and $H^z=0$. 
This is in contrast to the result under the single-ion anisotropy in Sec.~\ref{sec:At zero field_SIA} where no multiple-$Q$ states appear for $K=0$ and $H^z=0$. 

These double-$Q$ states remain robust against the introduction of $K$, as shown in Fig.~\ref{fig:souzu_bond-K}. 
At $I^{\rm A}=0$, the system undergoes the phase transitions from the single-$Q$ spiral state at $K=0$, to the double-$Q$ chiral stripe state for $0<K\lesssim 0.19$, and to the $n_{\rm sk}=2$ skyrmion crystal for $K\gtrsim 0.19$. 
The result is consistent with that obtained in Ref.~\onlinecite{Hayami_PhysRevB.95.224424}. 
The phase boundary between the double-$Q$ chiral stripe and the $n_{\rm sk}=2$ skyrmion crystal shifts downward while increasing $I^{\rm A}$, as shown in Fig.~\ref{fig:souzu_bond-K}; namely, $I^{\rm A}$ stabilizes the $n_{\rm sk}=2$ skyrmion crystal against the double-$Q$ chiral stripe state. 
This is qualitatively understood from their spin configurations as follows. 
For $I^{\rm A}>0$, the spin pattern in the $n_{\rm sk}=2$ skyrmion crystal is modulated so that all the parallel components of the magnetic moments with $\bm{Q}_\nu$, $(m_{\bm{Q}_\nu}^{\bm{Q}_\parallel})^2$, become zero, as shown in the second  panel of Fig.~\ref{fig:bonddep_H=0}(c); namely, the spin texture for $I^{\rm A}>0$ is characterized by a superposition of three sinusoidal waves perpendicular to $\bm{Q}_\nu$, as schematically shown in Fig.~\ref{fig:ponti}(a). 
Each sinusoidal component is composed of a linear combination of $m^{\bm{Q}_\perp}_{\bm{Q}_\eta}$ and $m^{\bm{Q}_z}_{\bm{Q}_\eta}$. 
On the other hand, the spin pattern in the double-$Q$ chiral stripe state is given by a superposition of the single-$Q$ helical and single-$Q$ sinusoidal waves. 
As the sinusoidal direction is perpendicular to the helical plane in spin space, the second-$Q$ ($\bm{Q}_2$) component is represented by a linear combination of $(m^{\bm{Q}_\perp}_{\bm{Q}_2})^2$ and $(m^{\bm{Q}_\parallel}_{\bm{Q}_2})^2$, as schematically shown in Fig.~\ref{fig:ponti}(b). 
Thus, the double-$Q$ chiral stripe state has both $(m^{\bm{Q}_\perp}_{\bm{Q}_\eta})^2$ and $(m^{\bm{Q}_\parallel}_{\bm{Q}_\eta})^2$ components for $\eta=1$ and $2$, as shown in the middle two panels of Fig.~\ref{fig:bonddep_H=0}(b). 
Since the bond-dependent interaction $I^{\rm A}$ prefers a proper screw with the spiral plane perpendicular to the helical direction, the above argument suggests that the energy gain by the introduction of $I^{\rm A}$ becomes larger for the $n_{\rm sk}=2$ skyrmion crystal than the double-$Q$ chiral stripe state. 
This is consistent with our result in Fig.~\ref{fig:souzu_bond-K} where the phase boundary between the two states is shifted to lower $K$ while increasing $I^{\rm A}$ in the small $I^{\rm A}$ region. 

In the $n_{\rm sk}=2$ skyrmion crystal in the large $K$ region, the uniform ferromagnetic moment along the $z$ direction, $m_0^z$, is induced by the introduction of $I^{\rm A}$, as shown in the top panel of Fig.~\ref{fig:bonddep_H=0}(c). 
The real-space spin and chirality configurations obtained by the simulated annealing are shown in Figs.~\ref{fig:Spin_bond_K=03_IA=05}(e) and \ref{fig:Spin_bond_K=03_IA=05}(f), respectively; they show a positive out-of-plane magnetization ($m^{\rm total}=\sum_i S_i^z>0$) and a negative scalar chirality ($\chi^{\rm total}=\sum_{\bm{R}} \chi_{\bm{R} }<0$). 
We note that the state is energetically degenerate with the one with $m^{\rm total}<0$ and $\chi^{\rm total}>0$. 
This is in contrast to the situation in the absence of $I^{\rm A}$ where $m^{\rm total}=0$ and $\chi^{\rm total}$ takes either a positive or negative value. 
The nonzero $m^{\rm total}$ indicates that the remaining degeneracy for $I^{\rm A}>0$ can be lifted by a magnetic field, as indeed shown in Sec.~\ref{sec:Field along the $z$ direction}. 

In the region for $0.37 \lesssim I^{\rm A} \lesssim 0.47$, the double-$Q$ helical state 
changes into the $n_{\rm sk}=2$ skyrmion crystal in the range of $0.1 \lesssim K \lesssim  0.15$, as shown in Fig.~\ref{fig:souzu_bond-K}. 
The phase boundary moves upward while increasing $I^{\rm A}$, which indicates that the energy gain by $I^{\rm A}$ is larger for the double-$Q$ helical state than the $n_{\rm sk}=2$ skyrmion crystal, in contrast to the case for the double-$Q$ chiral stripe state discussed above.
The isotropic double-$Q$ state for $0.47 \lesssim I^{\rm A} \lesssim 0.58$ shows a similar behavior; it changes into the $n_{\rm sk}=2$ skyrmion crystal in the range of $0.15 \lesssim K \lesssim  0.2$, where the critical value of $K$ increases while increasing $I^{\rm A}$. 
For $I^{\rm A} \gtrsim  0.58  $, the anisotropic double-$Q$ state turns into the isotropic double-$Q$ state, and then, into the $n_{\rm sk}=2$ skyrmion crystal while increasing $K$. 
In this region, the critical value of $K$ between the isotropic double-$Q$ state and the $n_{\rm sk}=2$ skyrmion crystal is almost unchanged against $I^{\rm A}$, indicating the energy gain from $I^{\rm A}$ is almost the same for these two states in the large $I^{\rm A}$ region. 

Meanwhile, the phase boundaries between the four different double-$Q$ states show distinct behavior in the $I^{\rm A}$-$K$ plane, as shown in Fig.~\ref{fig:souzu_bond-K}. 
This is qualitatively understood as follows. 
In the double-$Q$ chiral stripe state, the $\bm{Q}_2$ component becomes more dominant and $m_{\bm{Q}_1}^z$ becomes smaller for larger $K$, namely, the state is gradually modulated to approach the adjacent double-$Q$ helical one. 
This suggests that the phase boundary between the two states shifts to a smaller $I_{\rm A}$ region while increasing $K$ as shown in Fig.~\ref{fig:souzu_bond-K}, although the boundary looks almost independent of $I^{\rm A}$ in the small $K$ region. 
With regard to the boundary between the double-$Q$ helical and isotropic double-$Q$ states, both states are isotropic with respect to the two components, and hence, the energy gain from $K$ is almost the same and the boundary is almost independent of $I_{\rm A}$. 
On the other hand, the boundary between the isotropic and anisotropic double-$Q$ states shifts to a larger $I_{\rm A}$ region while increasing $K$, as $K$ favors the isotropic multiple-$Q$ state. 

\subsection{Field along the $z$ direction}
\label{sec:Field along the $z$ direction}

\begin{figure*}[htb!]
\begin{center}
\includegraphics[width=1.0 \hsize]{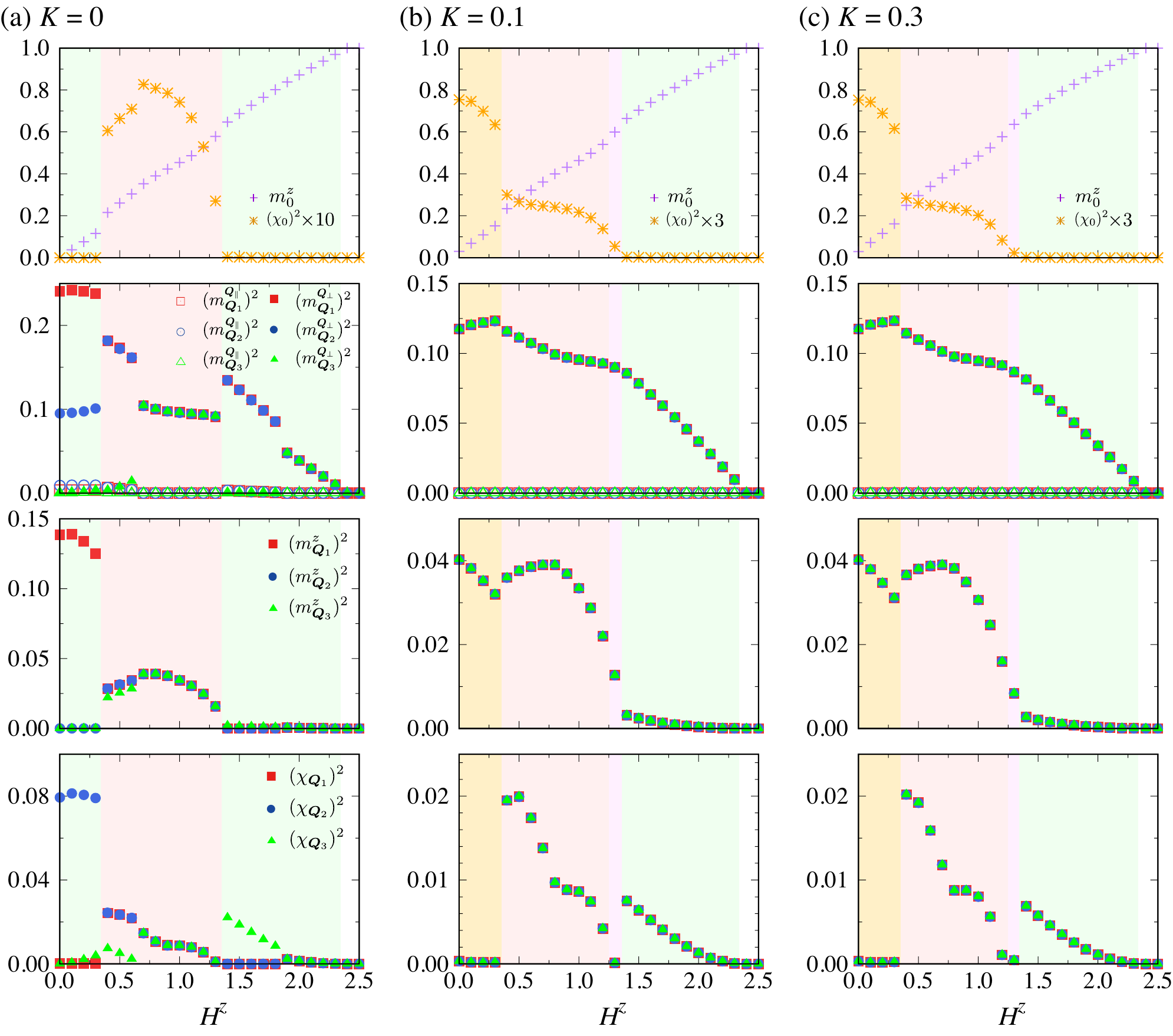} 
\caption{
\label{fig:bond_Hz_I=02}
$H^z$ dependence of (first row) $m^z_0$ and $(\chi_0)^2$, (second row) $(m^{\bm{Q}_\perp}_{\bm{Q}_\nu})^2$ and $(m^{\bm{Q}_\parallel}_{\bm{Q}_\nu})^2$, (third row) $(m^{z}_{\bm{Q}_\nu})^2$, and (fourth row) $ (\chi_{\bm{Q}_\nu})^2$ for $K=0$ for (a) $K=0$, (b) $K=0.1$, and (c) $K=0.3$ at $I^{\rm A}=0.2$. 
}
\end{center}
\end{figure*}

\begin{figure}[htb!]
\begin{center}
\includegraphics[width=1.0 \hsize]{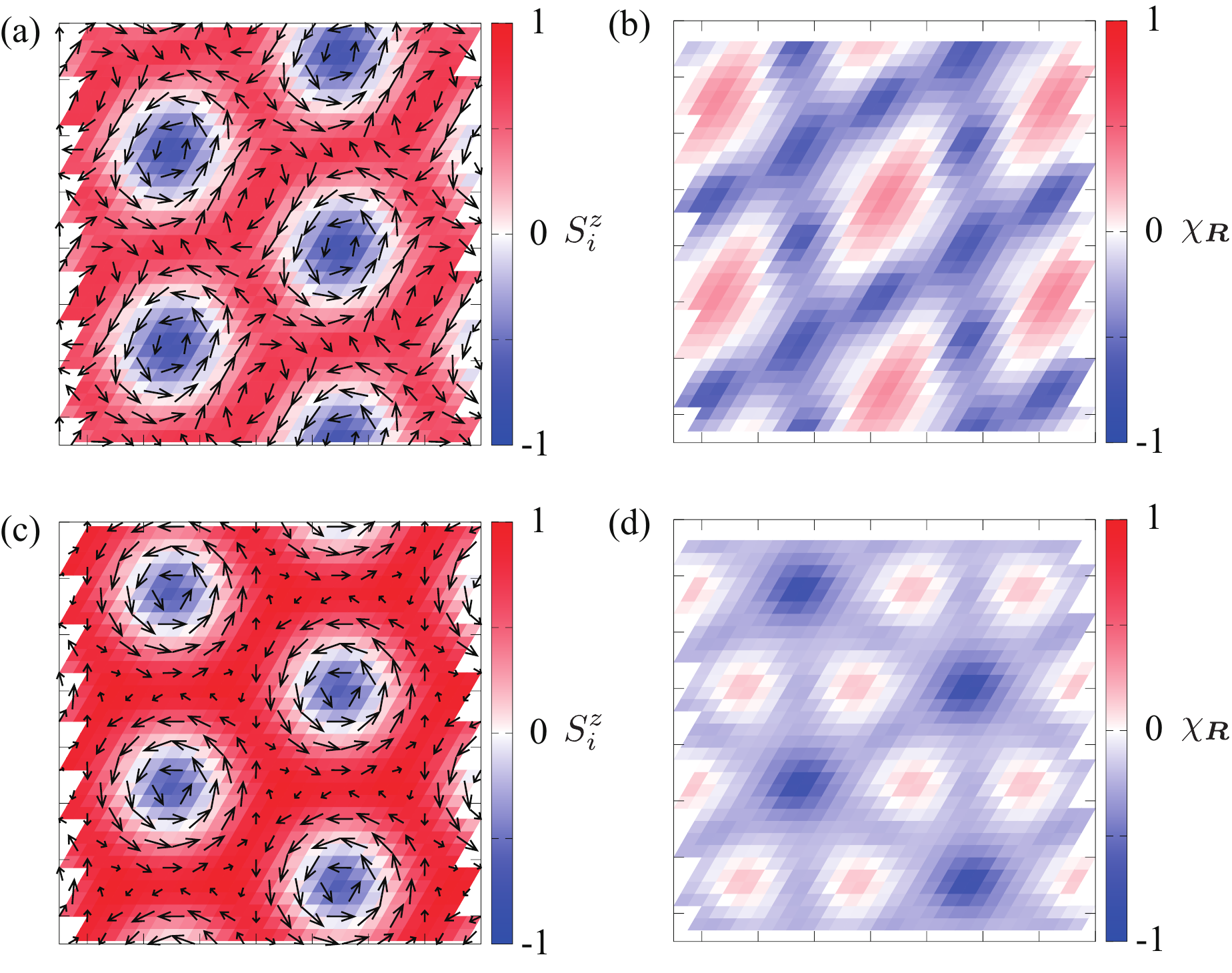} 
\caption{
\label{fig:Spin_Hz_K=0_IA=02}
Real-space spin and chirality configurations of the $n_{\rm sk}=1$ skyrmion crystals at $K=0$ and  $I^{\rm A}=0.2$. 
The magnetic field is taken at $H^z=0.5$ for (a) and (b), and at $H^z=1$ for (c) and (d).
In (a) and (c), the contour shows the $z$ component of the spin moment, and the arrows represent the $xy$ components. 
In (b) and (d), the contour shows the scalar chirality. 
}
\end{center}
\end{figure}

\begin{figure}[htb!]
\begin{center}
\includegraphics[width=1.0 \hsize]{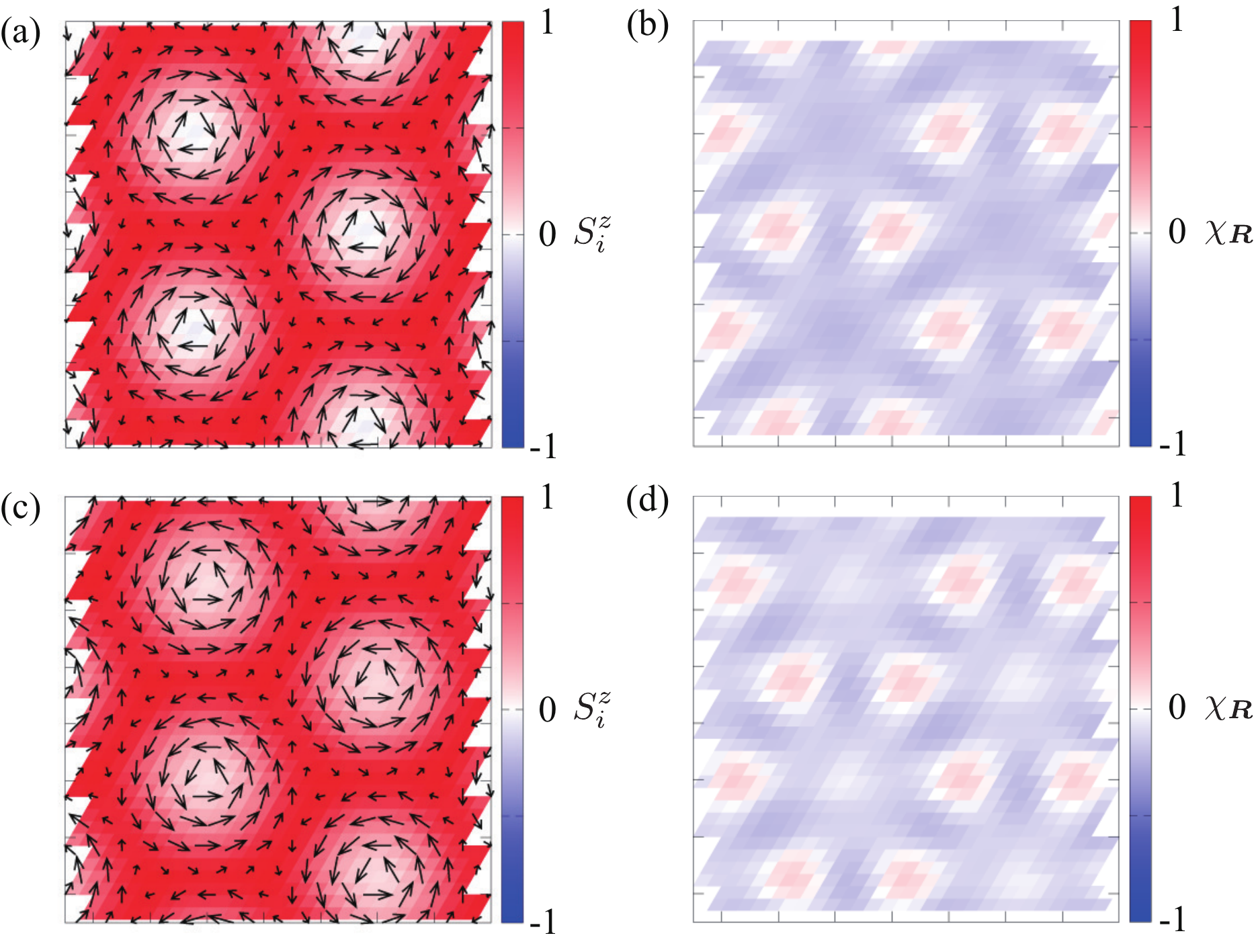} 
\caption{
\label{fig:Spin_Hz_K=0_IA=02_SkX2}
Real-space spin and chirality configurations of the (a), (b) $n_{\rm sk}=1$ skyrmion crystal at $K=0$ and (c), (d) the triple-$Q$ crystal at $K=0.1$ for $I^{\rm A}=0.2$ and $H^z=1.3$.
In (a) and (c), the contour shows the $z$ component of the spin moment, and the arrows represent the $xy$ components. 
In (b) and (d), the contour shows the scalar chirality. 
}
\end{center}
\end{figure}

\begin{figure}[htb!]
\begin{center}
\includegraphics[width=1.0 \hsize]{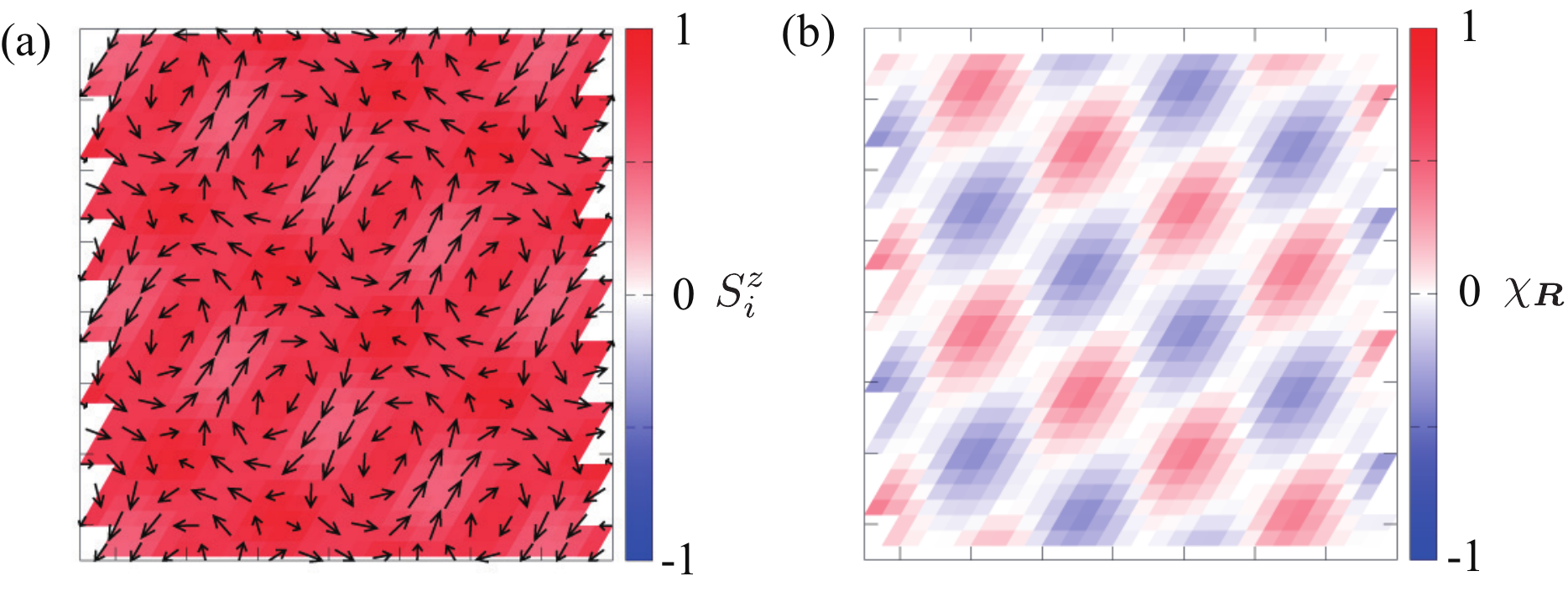} 
\caption{
\label{fig:Spin_Hz_K=0_IA=02_3Q}
Real-space spin and chirality configurations of the anisotropic triple-$Q$ state at $K=0$, $I^{\rm A}=0.2$, and $H^z=1.5$. 
In (a), the contour shows the $z$ component of the spin moment, and the arrows represent the $xy$ components. 
In (b), the contour shows the scalar chirality. 
}
\end{center}
\end{figure}

\begin{figure}[htb!]
\begin{center}
\includegraphics[width=1.0 \hsize]{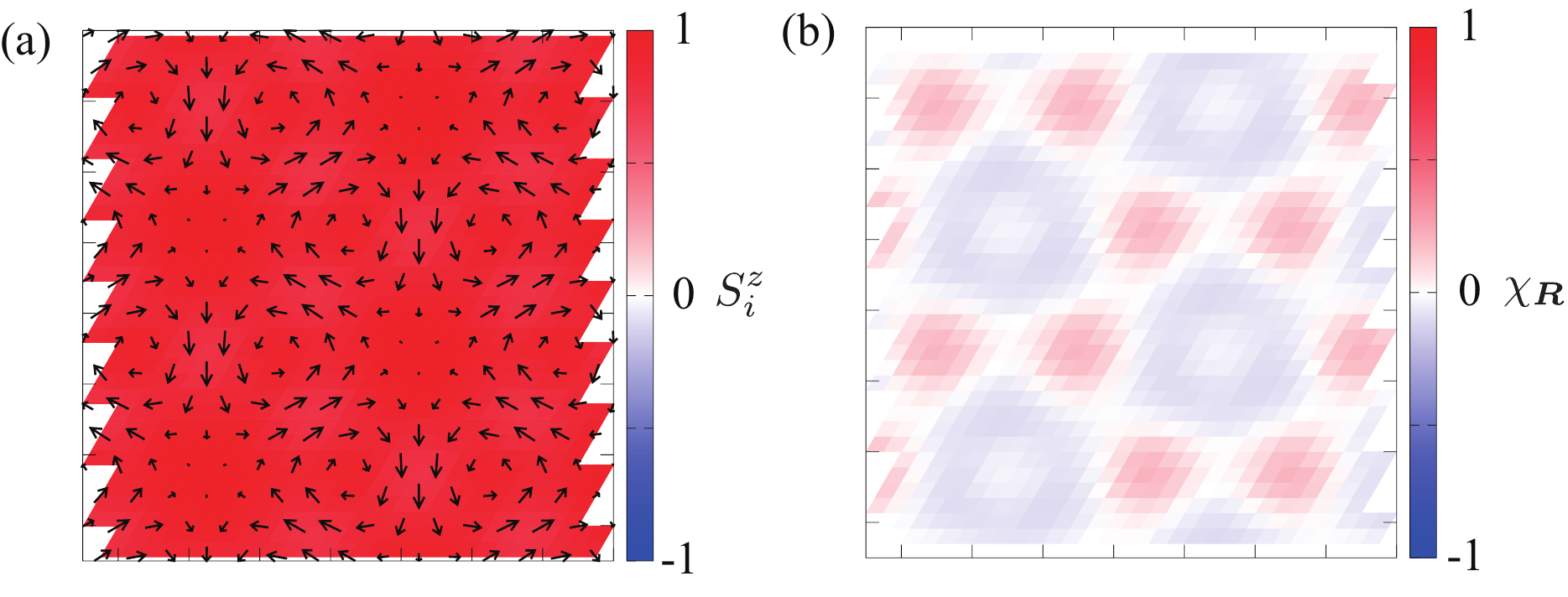} 
\caption{
\label{fig:Spin_Hz_K=0_IA=02_highfield}
Real-space spin and chirality configurations of the isotropic triple-$Q$ state at $K=0$, $I^{\rm A}=0.2$, and $H^z=2$. 
In (a), the contour shows the $z$ component of the spin moment, and the arrows represent the $xy$ components. 
In (b), the contour shows the scalar chirality. 
}
\end{center}
\end{figure}

Next, we examine the effect of the magnetic field along the $z$ direction, $H^z$, on each magnetic phase obtained in Fig.~\ref{fig:souzu_bond-K}. 
In the following, we present the results for the $3\times 3$ parameter sets with $K=(0,  0.1, 0.3)$ and $I^{\rm A}=(0.2, 0.4, 0.6)$ to show the systematic evolution with $H^z$ of the five multiple-$Q$ phases in Fig.~\ref{fig:souzu_bond-K}.

Figure~\ref{fig:bond_Hz_I=02} shows the result at $I^{\rm A}=0.2$ for $K=0$, 0.1, and 0.3.
For $K=0$ in Fig.~\ref{fig:bond_Hz_I=02}(a), the introduction of $H^z$ induces small $\bm{Q}_3$ components, e.g., $(m^{\parallel}_{\bm{Q}_3})^2 \simeq 0.001$ and $(m^{\perp}_{\bm{Q}_3})^2 \simeq 0.004$ at $H^z=0.3$. 
This means that nonzero $H^z$ changes the double-$Q$ chiral stripe state into a triple-$Q$ state.  
The triple-$Q$ state turns into the $n_{\rm sk}=1$ skyrmion crystal at $H^z \simeq 0.4$. 
It is noteworthy that $I^{\rm A}$ can result in the $n_{\rm sk}=1$ skyrmion crystal even without $K$. 
There are two types of the $n_{\rm sk}=1$ skyrmion crystals, which are separated at $H^z \simeq 0.65$ where $(\chi_0)^2$ exhibits a clear jump, as shown in the top panel of Fig.~\ref{fig:bond_Hz_I=02}(a). 
The spin structure for $0.4\lesssim  H^z \lesssim 0.65$ is characterized by the dominant double-$Q$ peak at $\bm{Q}_1$ and $\bm{Q}_2$ and the subdominant single-$Q$ peak at $\bm{Q}_3$, while that for $0.65\lesssim H^z \lesssim 1.3$ is by the triple-$Q$ peak with equal intensities, as shown in the middle two panels of Fig.~\ref{fig:bond_Hz_I=02}(a).
Accordingly, the chirality structure is characterized by $(\chi_{\bm{Q}_1})^2=(\chi_{\bm{Q}_2})^2 > (\chi_{\bm{Q}_3})^2$ in the lower-field state, whereas $(\chi_{\bm{Q}_1})^2=(\chi_{\bm{Q}_2})^2=(\chi_{\bm{Q}_3})^2$ in the higher-field state, as shown in the lowest panel of Fig.~\ref{fig:bond_Hz_I=02}(a). 
Thus, the threefold rotational symmetry is broken in the former, while it is recovered in the latter. 
The symmetry difference is clearly seen in the real-space spin and chirality configurations as well, as shown in Fig.~\ref{fig:Spin_Hz_K=0_IA=02}: 
The spin and chirality distributions around the skyrmion cores are elongated along the $\bm{Q}_3$ direction in the lower-field state as shown in Figs.~\ref{fig:Spin_Hz_K=0_IA=02}(a) and \ref{fig:Spin_Hz_K=0_IA=02}(b), while they are isotropic with respect to $\bm{Q}_1$, $\bm{Q}_2$, and $\bm{Q}_3$, and form a hexagonal lattice in the higher-field one as shown in Figs.~\ref{fig:Spin_Hz_K=0_IA=02}(c) and \ref{fig:Spin_Hz_K=0_IA=02}(d). 

For these $n_{\rm sk}=1$ skyrmion crystals, the application of $H^z$ in the presence of $I^{\rm A}$ chooses the state with fixed signs of $m^{\rm total}>0$ and $\chi^{\rm total}<0$, as deduced in the end of in Sec.~\ref{sec:At zero field_bond}. 
In terms of the helicity and vorticity, the obtained skyrmion crystals are categorized into the Bloch-type ones with the helicity $\pm \pi/2$ and the vorticity $1$, where the states with the helicity $ \pi/2$ or $- \pi/2$ are energetically degenerate in contrast to the skyrmion crystals stabilized in the chiral lattice structures by the DM interaction~\cite{nagaosa2013topological}. 
When the sign of $I^{\rm A}$ is reversed, the N\'eel-type skyrmions with the helicity $0$ or $\pi$ and the vorticity $1$ are realized. 
The antiskyrmions with the vorticity $-1$, however, are not stabilized in the present system.

While increasing $H^z$, the $z$-spin component at the skyrmion core takes almost zero and the chirality reduces, as shown in Figs.~\ref{fig:Spin_Hz_K=0_IA=02_SkX2}(a) and \ref{fig:Spin_Hz_K=0_IA=02_SkX2}(b), while the skyrmion number remains one. 
It turns into another triple-$Q$ state at $H^z \simeq 1.4$, as shown in Fig.~\ref{fig:bond_Hz_I=02}(a). 
This state is characterized by the dominant double-$Q$ structure with $(m^{\bm{Q}_\perp}_{\bm{Q}_1})^2$ and $(m^{\bm{Q}_\perp}_{\bm{Q}_2})^2$, accompanied by a small $(m^{z}_{\bm{Q}_3})^2$, which results in the chirality density wave with $(\chi_{\bm{Q}_3})^2$, as shown in the lowest panel of Fig.~\ref{fig:bond_Hz_I=02}(a). 
The uniform component $(\chi_0)^2$ vanishes in this state, as shown in the top panel of Fig.~\ref{fig:bond_Hz_I=02}(a). 
The real-space spin and chirality configurations in this state are shown in Fig.~\ref{fig:Spin_Hz_K=0_IA=02_3Q}. 
It is noted that this state has an additional component at $\bm{Q}_1 -\bm{Q}_2$ in the chirality in addition to that at $\bm{Q}_3$ (not shown), leading to the checkerboard-like pattern shown in Fig.~\ref{fig:Spin_Hz_K=0_IA=02_3Q}(b). 
While further increase of $H^z$, the triple-$Q$ state changes its spin and chirality structures to have the same intensities at $\bm{Q}_1$, $\bm{Q}_2$, and $\bm{Q}_3$ for $1.9\lesssim H^z \lesssim 2.4$, as shown in the middle two panels of Fig.~\ref{fig:bond_Hz_I=02}(a). 
The real-space spin structure changes into a periodic array of two types of vortices with the vorticity $1$ and $-2$, as shown in Fig.~\ref{fig:Spin_Hz_K=0_IA=02_highfield}(a). 
The opposite sign of the vorticity leads to the opposite sign of the scalar chirality, as shown in Fig.~\ref{fig:Spin_Hz_K=0_IA=02_highfield}(b). 
The number of vortices with the vorticity 1 is twice as that of vortices with the vorticity $-2$, and $(\chi_0)^2$ cancels out between the two types of the vortices, as plotted in the top panel of Fig.~\ref{fig:bond_Hz_I=02}(a).

\begin{figure*}[htb!]
\begin{center}
\includegraphics[width=1.0 \hsize]{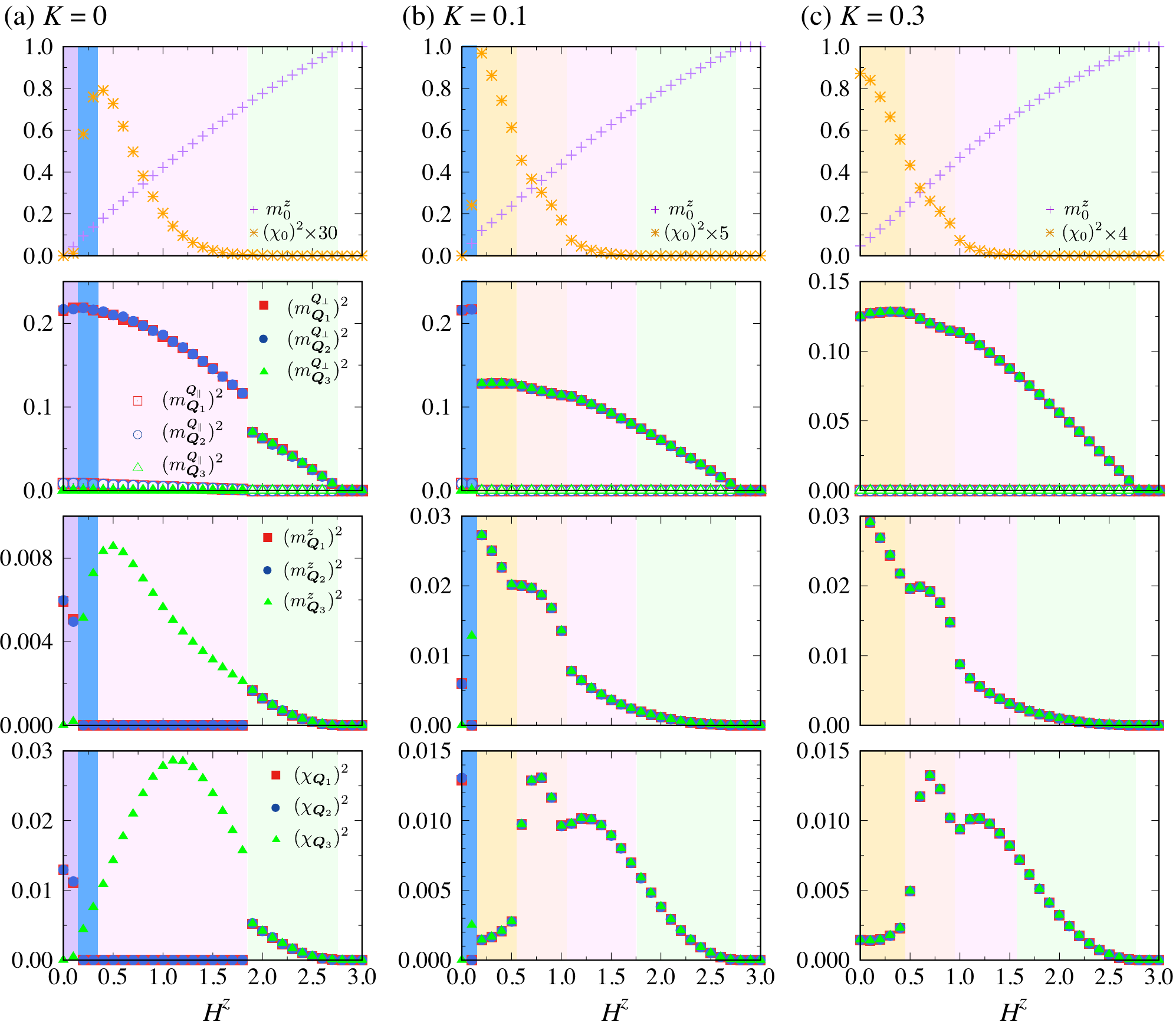} 
\caption{
\label{fig:bond_Hz_I=04}
The same plots as in Fig.~\ref{fig:bond_Hz_I=02} for $I^{\rm A}=0.4$.
}
\end{center}
\end{figure*}

\begin{figure}[htb!]
\begin{center}
\includegraphics[width=1.0 \hsize]{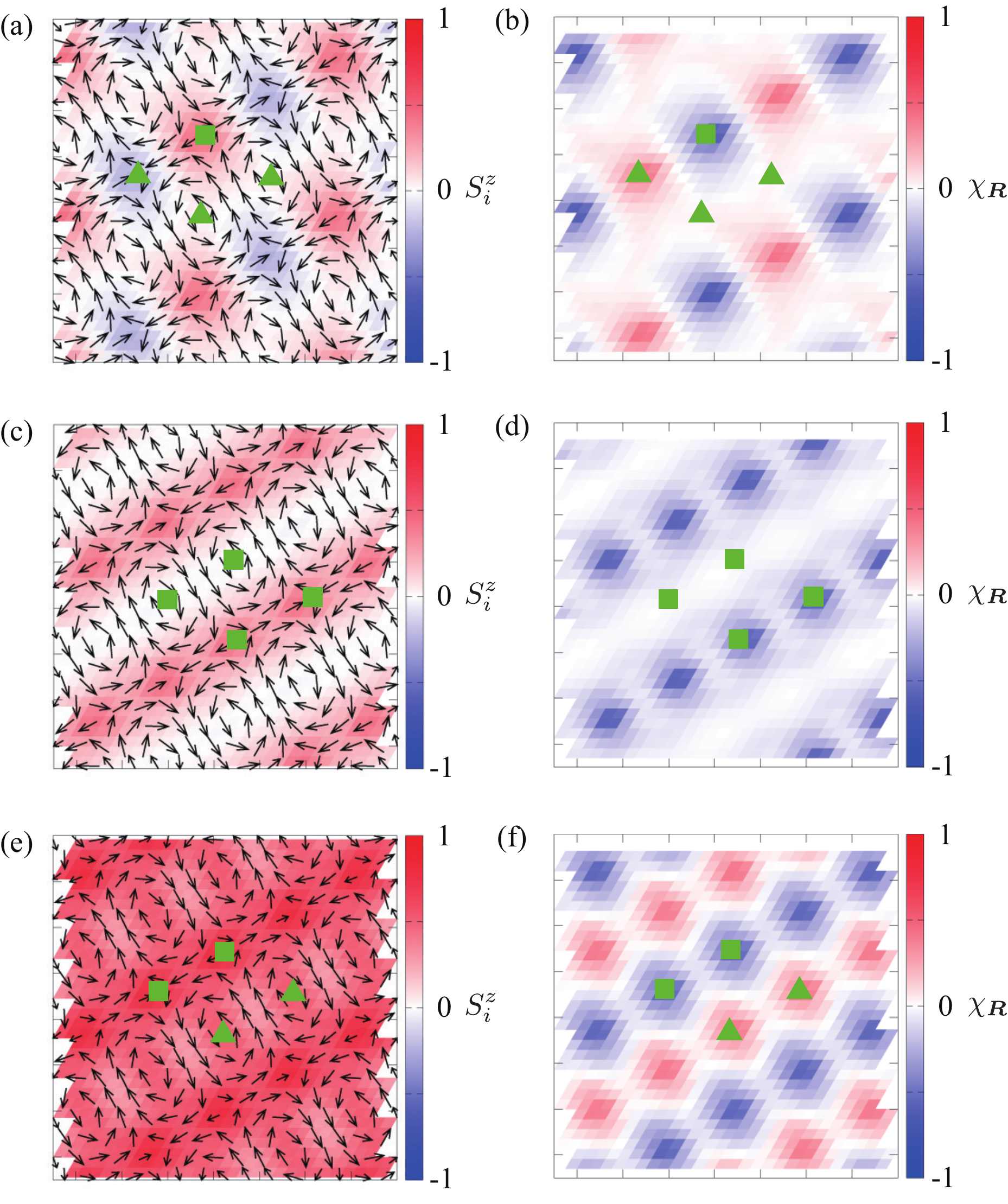} 
\caption{
\label{fig:Spin_Hz_K=0_IA=04}
Real-space spin and chirality configurations at $K=0$ and $I^{\rm A}=0.4$. 
The magnetic field is taken at $H^z=0.1$ for (a) and (b) the $n_{\rm sk}=1$ meron crystal, $H^z=0.3$ for (c) and (d) the $n_{\rm sk}=2$ meron crystal, and $H^z=1.2$ for (e) and (f) the anisotropic triple-$Q$ state.
In (a), (c), and (e), the contour shows the $z$ component of the spin moment, and the arrows represent the $xy$ components. 
In (b), (d), and (f), the contour shows the scalar chirality. 
The green triangles and squares represent the cores with the positive and negative skyrmion numbers, respectively.
}
\end{center}
\end{figure}

\begin{figure}[htb!]
\begin{center}
\includegraphics[width=1.0 \hsize]{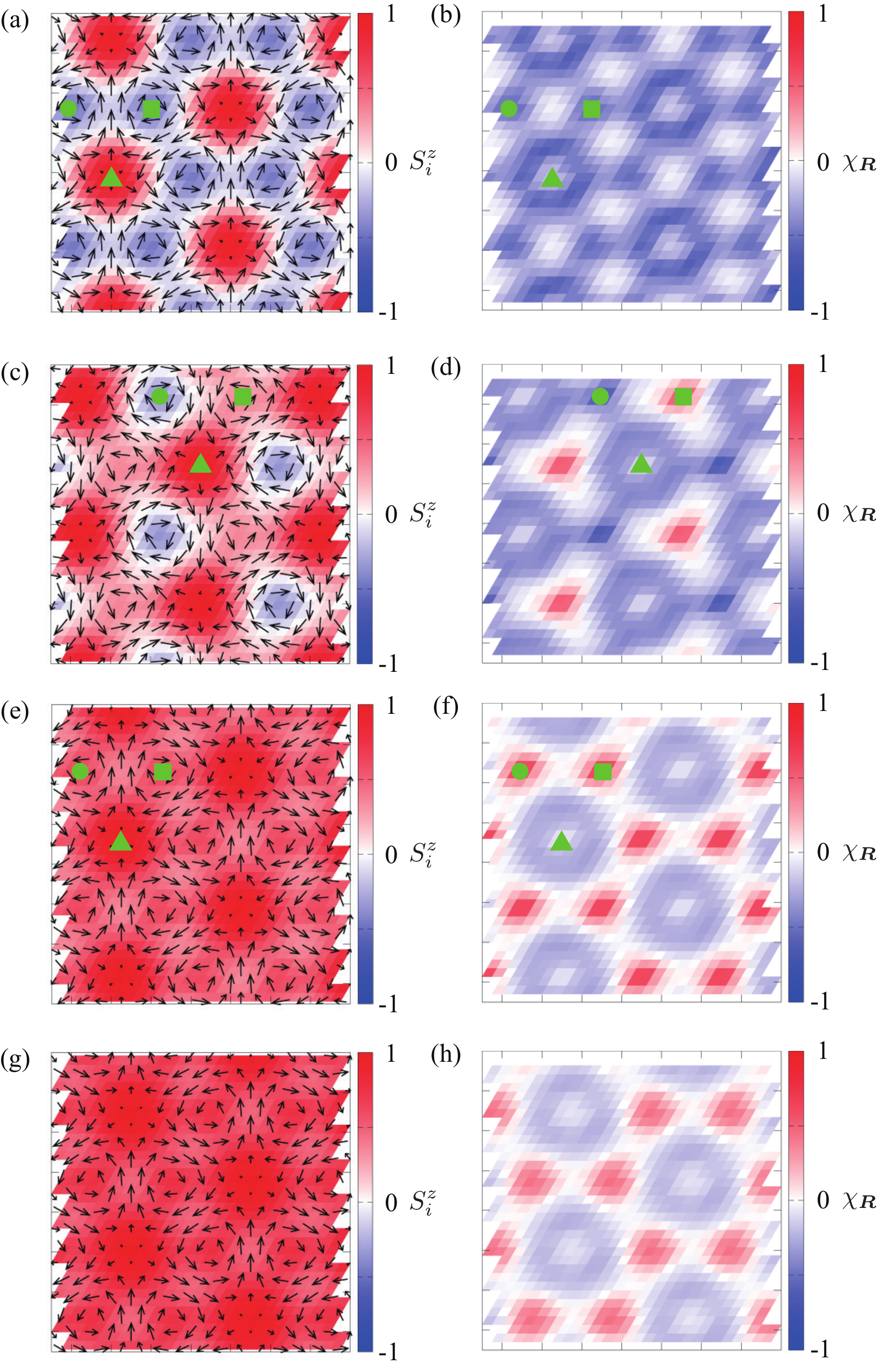} 
\caption{
\label{fig:Spin_Hz_K=03_IA=04}
Real-space spin and chirality configurations at $K=0.3$ and  $I^{\rm A}=0.4$. 
The magnetic field is taken at $H^z=0.1$ for (a) and (b) the $n_{\rm sk}=2$ skyrmion crystal, $H^z=0.6$ for (c) and (d) the $n_{\rm sk}=1$ skyrmion crystal, $H^z=1.3$ for (e) and (f) the triple-$Q$ state with nonzero $(\chi_0)^2$, and $H^z=1.6$ for (g) and (h) the triple-$Q$ state without $(\chi_0)^2$.
In (a), (c), (e), and (g), the contour shows the $z$ component of the spin moment, and the arrows represent the $xy$ components. 
In (b), (d), (f), and (h), the contour shows the scalar chirality. 
In (a)-(f), the green squares and circles represent the cores of the type-I and II vortices with vorticity $+1$, respectively, whereas the green triangles represent the cores of the vortices with vorticity $-2$. See the main text in the details. 
}
\end{center}
\end{figure}

For $K=0.1$ and $0.3$, the results are qualitatively the same as those for $K=0$, except for the low-field region for $0 < H^z \lesssim 0.4$ and the intermediate-field region for $1.3 \lesssim H^z \lesssim 1.4$, as shown in Figs.~\ref{fig:bond_Hz_I=02}(b) and \ref{fig:bond_Hz_I=02}(c), respectively. 
For both values of $K$, there are four phases in addition to the fully-polarized state for $H^z \gtrsim 2.4$. 
The low-field phase for $0 < H^z \lesssim 0.4$ corresponds to the $n_{\rm sk}=2$ skyrmion crystal with $m^{\rm total}>0$ and $\chi^{\rm total}<0$ similar to the case with $I^{\rm A}=0$. 
Meanwhile, in the region for $0.4\lesssim H^z \lesssim 1.2$, we obtain only one type of the $n_{\rm sk}=1$ skyrmion crystal, which has threefold rotational symmetry similar to the one found for $0.65\lesssim H^z\lesssim 1.4$ at $K=0$. 
This is presumably owing to the tendency that $K$ favors isotropic multiple-$Q$ states rather than anisotropic ones. Such a tendency is also found in the case of $I^{\rm A}=0$ where the anisotropic double-$Q$ chiral stripe is replaced by the isotropic $n_{\rm sk}=2$ skyrmion crystal, as discussed in Sec.~\ref{sec:At zero field_bond}. 
In the higher-field region, there are two states; the triple-$Q$ crystal with nonzero $(\chi_0)^2$ for $1.3 \lesssim H^z \lesssim 1.4$ and the other triple-$Q$ crystal for $1.4 \lesssim H^z \lesssim 2.4$, both of which have the same intensities at $\bm{Q}_1$, $\bm{Q}_2$, and $\bm{Q}_3$. 
The latter triple-$Q$ crystal corresponds to the state for $1.9\lesssim H^z \lesssim 2.4$ at $K=0$. 
The change of the skyrmion number at $H^z \simeq 1.3$ is owing to the positive $z$-spin component at the vortex core in Fig.~\ref{fig:Spin_Hz_K=0_IA=02_SkX2}(c), which is in contrast to the small negative $z$-spin component at the vortex core in the $n_{\rm sk}=1$ skyrmion crystal [for comparison, see Fig.~\ref{fig:Spin_Hz_K=0_IA=02_SkX2}(a) as an example].
Meanwhile, the scalar chirality distributions for these states are similar with each other as shown in Figs.~\ref{fig:Spin_Hz_K=0_IA=02_SkX2}(d) and \ref{fig:Spin_Hz_K=0_IA=02_SkX2}(b). 
Compared to the results at $K=0$, these isotropic states appear in wider field ranges, overcoming the anisotropic triple-$Q$ state for $1.4\lesssim H^z \lesssim 1.9$ at $K=0$, from the same reason stated above.

Next, we discuss the results for $I^{\rm A}=0.4$ shown in Fig.~\ref{fig:bond_Hz_I=04}.  
For $K=0$, a small but nonzero $(\chi_0)^2$ is induced by applying the magnetic field to the double-$Q$ helical state, as shown in the top panel of Fig.~\ref{fig:bond_Hz_I=04}(a). 
This is in contrast to the result for $I^{\rm A}=0.2$ where no $(\chi_0)^2$ is induced from the double-$Q$ chiral stripe state by the magnetic field [see the top panel of Fig.~\ref{fig:bond_Hz_I=02}(a)]. 
The spin and chirality patterns at $H^z= 0.1$ are shown in Figs.~\ref{fig:Spin_Hz_K=0_IA=04}(a) and \ref{fig:Spin_Hz_K=0_IA=04}(b), respectively. 
The in-plane magnetic moments form a vortex crystal and the out-of-plane ones $S_i^z$ show a checkerboard modulation, both of which are represented by the dominant double-$Q$ structure with $\bm{Q}_1$ and $\bm{Q}_2$ shown in the middle two panels of Fig.~\ref{fig:bond_Hz_I=04}(a). 
The real-space distribution of the chirality $\chi_{\bm{R}}$ also has a checkerboard modulation, which is described by the anisotropic triple-$Q$ structure in $\chi_{\bm{Q}_\nu}$ shown in the lowest panel of Fig.~\ref{fig:bond_Hz_I=04}(a). 
In Figs.~\ref{fig:Spin_Hz_K=0_IA=04}(a) and \ref{fig:Spin_Hz_K=0_IA=04}(b), the magnitude of $S^z_i$ ($\chi_{\bm{R}}$) in the red (blue) regions is larger than that in the blue (red) regions, resulting in nonzero $m^z_0$ [$(\chi_0)^2$] in the top panel of Fig.~\ref{fig:bond_Hz_I=04}(a). 
By calculating the skyrmion number, we find that this state for $0 < H^z\lesssim 0.1$ has $n_{\rm sk}=1$ consisting of three vortices with a positive topological charge around $1/2$ denoted as the green triangle in Fig.~\ref{fig:Spin_Hz_K=0_IA=04}(a) and one vortex with a negative topological charge around $-1/2$ denoted as the green square in Fig.~\ref{fig:Spin_Hz_K=0_IA=04}(a) in the magnetic unit cell.
Since these vortices have meron-like spin textures and the skyrmion number becomes $+1$ by summing up the skyrmion number in the magnetic unit cell, we call this state the $n_{\rm sk}=1$ meron crystal. 

While increasing $H^z$, another topological spin texture appears for $0.1 \lesssim H^z \lesssim 0.4$. 
In this state, $(m^{\bm{Q}_\perp}_{\bm{Q}_\nu})^2$ and $(m^{\bm{Q}_\parallel}_{\bm{Q}_\nu})^2$ are similar to those in the lower-field meron crystal, while $(m^{z}_{\bm{Q}_\nu})^2$ and $(\chi_{\bm{Q}_\nu})^2$ show distinct features with a single-$Q$ structure, as shown in the lower three panels of Fig.~\ref{fig:bond_Hz_I=04}(a). 
The real-space spin and chirality configurations are shown in Figs.~\ref{fig:Spin_Hz_K=0_IA=04}(c) and \ref{fig:Spin_Hz_K=0_IA=04}(d), respectively. 
Interestingly, we find that this state has $n_{\rm sk}=2$ in the magnetic unit cell, although the spin texture looks very different from that in the $n_{\rm sk}=2$ skyrmion crystal exemplified in Fig.~\ref{fig:Spin_bond_K=03_IA=05}(e). 
In fact, the real-space spin texture is charactered by the periodic array of the clockwise and counterclockwise vortices, as shown in Fig.~\ref{fig:Spin_Hz_K=0_IA=04}(c). 
In other words, the spin structure includes four different vortices in the magnetic unit cell, all of which have negative topological charges. Since this is regarded as four meron-like structures in each magnetic unit cell in the real-space picture, we call this state the $n_{\rm sk}=2$ meron crystal.

With a further increase of $H^z$, there is a topological phase transition from the $n_{\rm sk}=2$ meron crystal to another triple-$Q$ state with $n_{\rm sk}=0$ at $H^z \simeq 0.4$. 
Despite the change in $n_{\rm sk}$, the spin and chirality related quantities are continuous through this transition, as shown in Fig.~\ref{fig:bond_Hz_I=04}(a). 
The spin texture looks similar to that in the lower-field $n_{\rm sk}=2$ meron crystal, as shown in Fig.~\ref{fig:Spin_Hz_K=0_IA=04}(e). 
By closely looking into the spin configurations in Figs.~\ref{fig:Spin_Hz_K=0_IA=04}(c) and \ref{fig:Spin_Hz_K=0_IA=04}(e), however, we notice that two of four vortices in the $n_{\rm sk}=2$ meron crystal have a negative $z$-spin component at the cores, while all the vortices for the higher-field triple-$Q$ state have a positive $z$-spin component at the cores.
Thus, the skyrmion number is canceled out for the higher-field triple-$Q$ state and becomes zero. 
The corresponding chirality pattern is displayed in Fig.~\ref{fig:Spin_Hz_K=0_IA=04}(f); the regions with positive and negative chirality form a stripy pattern, but the cancellation between them is not perfect and results in the nonzero $(\chi_0)^2$, as plotted in the top panel of Fig.~\ref{fig:bond_Hz_I=04}(a). 
As $H^z$ increases, the cancellation approaches perfect, and $(\chi_0)^2$ decreases with the suppression of $(m^{z}_{\bm{Q}_3})^2$ plotted in the third panel of Fig.~\ref{fig:bond_Hz_I=04}(a). 
$(\chi_0)^2$ vanishes at $H^z \simeq 1.9$, where the system undergoes a transition to the triple-$Q$ state whose spin and chirality configurations are similar to those obtained at $I^{\rm A}=0.2$ in Figs.~\ref{fig:Spin_Hz_K=0_IA=02_highfield}(a) and \ref{fig:Spin_Hz_K=0_IA=02_highfield}(b), respectively.

For $K=0.1$ and $0.3$, the $H^z$ dependences of $(\bm{m}_{\bm{Q}_\nu})^2$ and $(\chi_{\bm{Q}_\nu})^2$ are similar to each other, except for the low-field region for $H^z \lesssim 0.2$, as shown in Figs.~\ref{fig:bond_Hz_I=04}(b) and \ref{fig:bond_Hz_I=04}(c).
In the case with $K=0.1$, the $n_{\rm sk}=2$ meron crystal is obtained for $0<H^z\lesssim 0.2$, and the $n_{\rm sk}=2$ skyrmion crystal is realized for $0.2\lesssim H^z\lesssim 0.6$. 
The spin and chirality configurations are similar to those in Figs.~\ref{fig:Spin_Hz_K=0_IA=04}(c) and \ref{fig:Spin_Hz_K=0_IA=04}(d), Figs.~\ref{fig:Spin_bond_K=03_IA=05}(e) and \ref{fig:Spin_bond_K=03_IA=05}(f). 
Meanwhile, for $K=0.3$, the $n_{\rm sk}=2$ skyrmion crystal is stabilized for $0<H^z\lesssim 0.5$, and the $n_{\rm sk}=2$ meron crystal does not appear. 
For both $K=0.1$ and $0.3$, there are three triple-$Q$ states in the larger $H^z$ region, and all of them retain the threefold rotational symmetry with equal intensities at the three wave numbers, as shown in the lower three panels of Figs.~\ref{fig:bond_Hz_I=04}(b) and \ref{fig:bond_Hz_I=04}(c). 
In this field region, the uniform $(\chi_0)^2$ decreases monotonically as increasing $H^z$ as shown in the top panels of Figs.~\ref{fig:bond_Hz_I=04}(b) and \ref{fig:bond_Hz_I=04}(c). 
There are two topological phase transitions with changes in $n_{\rm sk}$: One is from the $n_{\rm sk}=2$ skyrmion crystal to the $n_{\rm sk}=1$ skyrmion crystal at $H^z \simeq 0.6$ for $K=0.1$ and at $H^z \simeq 0.5$ for $K=0.3$ and the other is from the $n_{\rm sk}=1$ skyrmion crystal to another chiral magnetic state with $n_{\rm sk}=0$ at $H^z \simeq 1.1$ for $K=0.1$ and at $H^z \simeq 1$ for $K=0.3$. 
While further increasing $H^z$, $(\chi_0)^2$ vanishes at $H^z \simeq 1.8$ for $K=0.1$ and at $H^z \simeq 1.6$ for $K=0.3$, where the system undergoes a phase transition to a nonchiral triple-$Q$ state. 
We note that similar changes with monotonous decrease of $(\chi_0)^2$ while keeping equal intensities $(\bm{m}_{\bm{Q}_1})^2=(\bm{m}_{\bm{Q}_2})^2=(\bm{m}_{\bm{Q}_3})^2$ against the magnetic field have also been found in itinerant magnets with an anisotropic bond interaction on a square lattice~\cite{Hayami_PhysRevLett.121.137202, Hayami_PhysRevB.103.024439}.

The spin and chirality configurations for the three triple-$Q$ states as well as the $n_{\rm sk}=2$ skyrmion crystal are displayed in Fig.~\ref{fig:Spin_Hz_K=03_IA=04} for $K=0.3$. 
In the $n_{\rm sk}=2$ skyrmion crystal for $0<H^z\lesssim 0.5$, the spin and chirality patterns in Figs.~\ref{fig:Spin_Hz_K=03_IA=04}(a) and \ref{fig:Spin_Hz_K=03_IA=04}(b), respectively, look similar to those obtained at zero field in Figs.~\ref{fig:Spin_bond_K=03_IA=05}(e) and \ref{fig:Spin_bond_K=03_IA=05}(f). 
By closely looking into the real-space spin structure in Fig.~\ref{fig:Spin_Hz_K=03_IA=04}(a), the spin texture consists of two types of vortices: one with vorticity $-2$ around $S^z \simeq +1$ (denoted as the green triangle) and the other with vorticity $+1$ around $S^z \simeq -1$ (denoted as the green square and circle). 
The number of the former is half of the latter.
It is noted that the latter vortices are equivalent between the green square and circle ones in this state, although they show different behaviors in the states for larger $H^z$, as discussed below.
In the following, we call the green square ones type-I vortices, 
while the green circle ones types-II vortices. 
In this state, all the vortices give a negative chirality as shown in Fig.~\ref{fig:Spin_Hz_K=03_IA=04}(b).

When the system enters into the $n_{\rm sk}=1$ skyrmion crystal by increasing $H^z$, the type-I and type-II vortices with vorticity $+1$ becomes inequivalent; the $z$-spin component near the type-I vortex core changes gradually from negative to positive, while that near the type-II vortex remains $S_i^z<0$, as shown in Fig.~\ref{fig:Spin_Hz_K=03_IA=04}(c).
Accordingly, the scalar chirality around the type-I vortex is reversed, as shown in Fig.~\ref{fig:Spin_Hz_K=03_IA=04}(d). 
In spite of the continuous changes of the spin and chirality configurations, we find that the skyrmion number remains one in the entire region of $0.5 \lesssim H^z \lesssim 1$. 

While further increasing $H^z$ to the state for $1 \lesssim H^z \lesssim 1.6$ appearing after the $n_{\rm sk}=1$ skyrmion crystal, the $z$-spin components in the type-I and II vortices become equivalent as shown in Figs.~\ref{fig:Spin_Hz_K=03_IA=04}(e) and \ref{fig:Spin_Hz_K=03_IA=04}(f), each of which retains the same skyrmion number. 
Consequently, these contributions cancel out that from the vortex with vorticity $-2$, 
resulting in the skyrmion number of zero, although $(\chi_0)^2$ retains a nonzero small value as shown in the top panel of Fig.~\ref{fig:bond_Hz_I=04}(c). 
While further increasing $H^z$, $(\chi_0)^2$ vanishes continuously in the triple-$Q$ state for $H^z \gtrsim 1.8$, whose spin and chirality configurations remain similar, as shown in Figs.~\ref{fig:Spin_Hz_K=03_IA=04}(g) and \ref{fig:Spin_Hz_K=03_IA=04}(h).

\begin{figure*}[htb!]
\begin{center}
\includegraphics[width=1.0 \hsize]{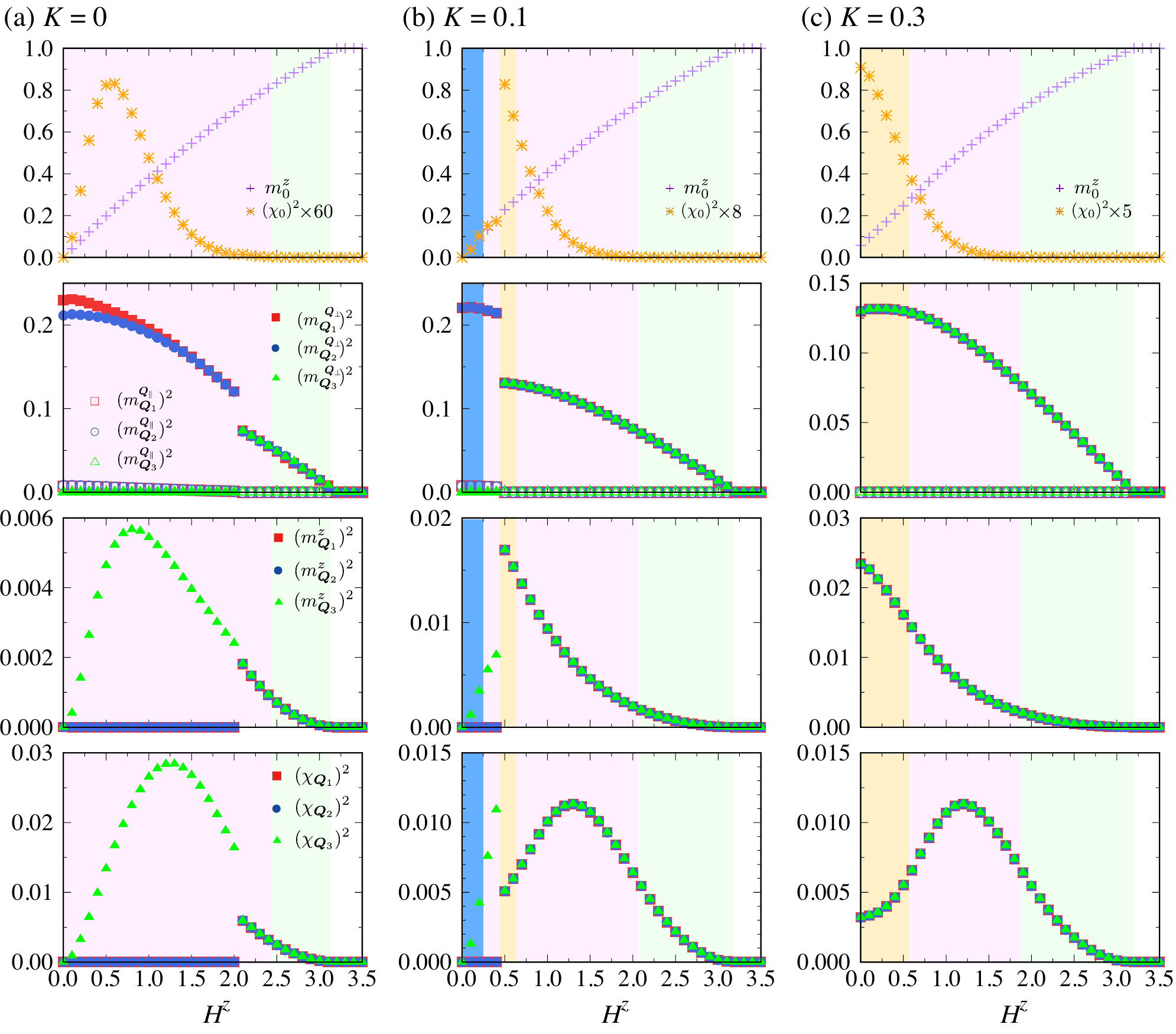} 
\caption{
\label{fig:bond_Hz_I=06}
The same plots as in Fig.~\ref{fig:bond_Hz_I=02} for $I^{\rm A}=0.6$.
}
\end{center}
\end{figure*}

\begin{figure}[htb!]
\begin{center}
\includegraphics[width=1.0 \hsize]{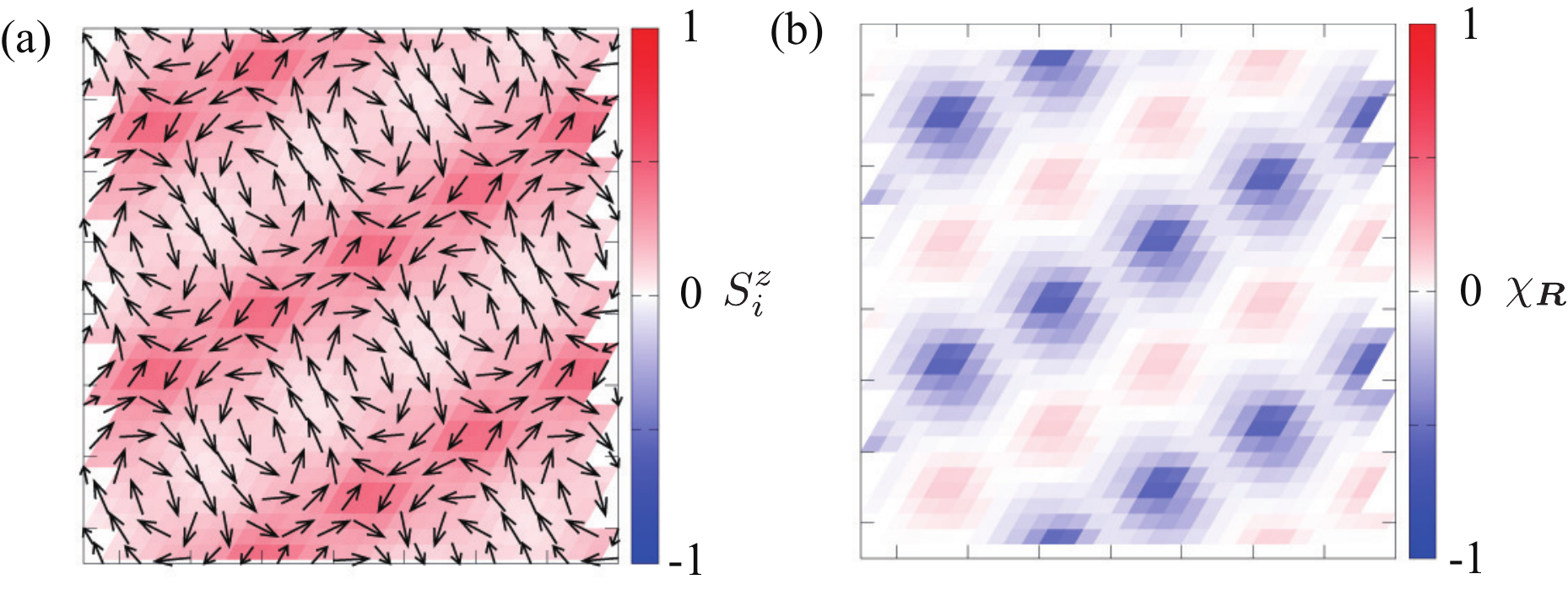} 
\caption{
\label{fig:Spin_Hz_K=0_IA=06}
Real-space spin and chirality configurations of the anisotropic triple-$Q$ state at $K=0$ and  $I^{\rm A}=0.6$. 
The magnetic field is taken at $H^z=0.5$. 
In (a), the contour shows the $z$ component of the spin moment, and the arrows represent the $xy$ components. 
In (b), the contour shows the scalar chirality. 
}
\end{center}
\end{figure}

\begin{figure}[htb!]
\begin{center}
\includegraphics[width=1.0 \hsize]{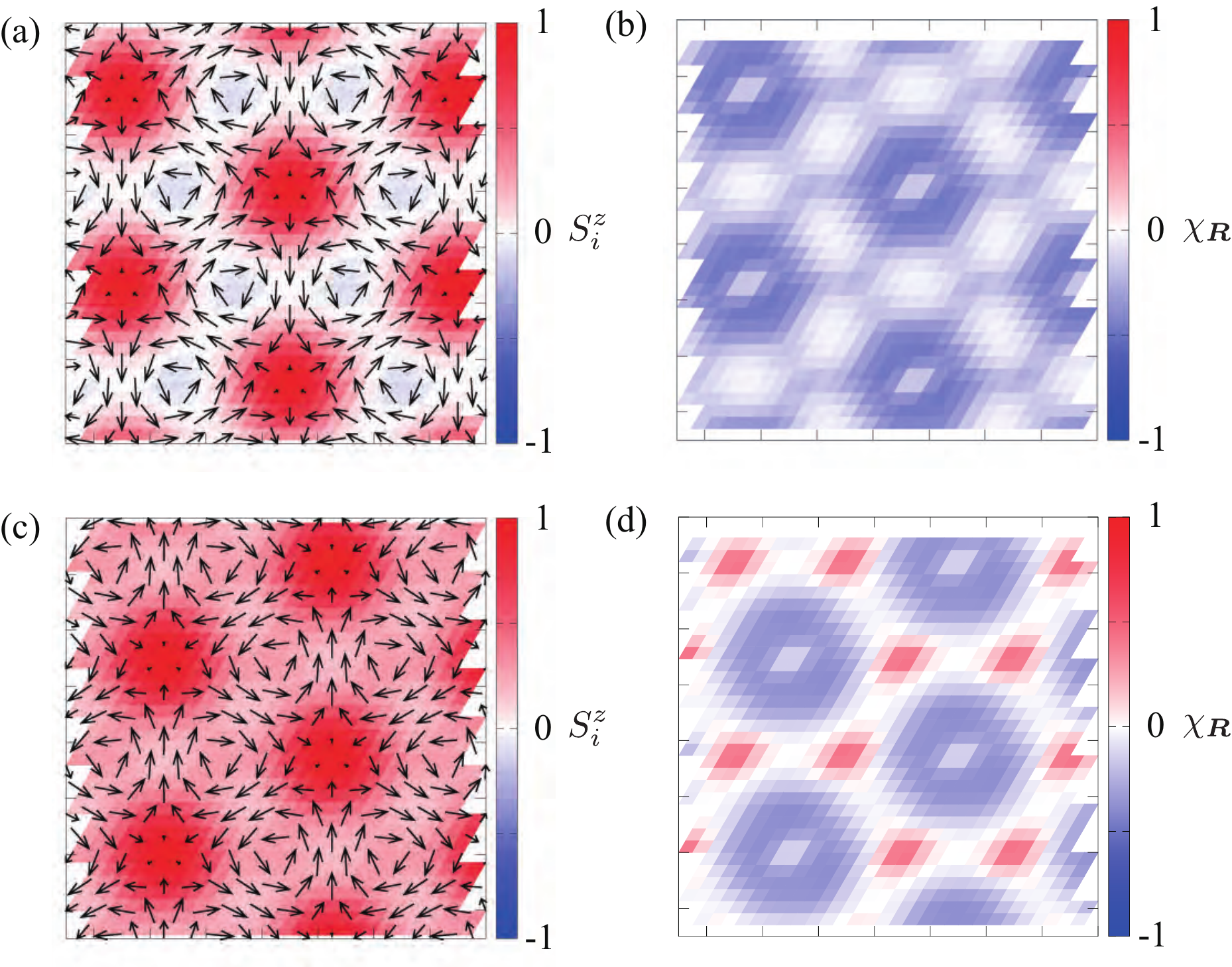} 
\caption{
\label{fig:Spin_Hz_K=01_IA=06}
Real-space spin and chirality configurations of the $n_{\rm sk}=1$ skyrmion crystals at $K=0.1$ and  $I^{\rm A}=0.6$. 
The magnetic field is taken at $H^z=0.5$ for (a) and (b) and $H^z=1$ for (c) and (d). 
In (a) and (c), the contour shows the $z$ component of the spin moment, and the arrows represent the $xy$ components. 
In (b) and (d), the contour shows the scalar chirality. 
}
\end{center}
\end{figure}

Figure~\ref{fig:bond_Hz_I=06} shows the results at $I^{\rm A}=0.6$. 
For $K=0$, the anisotropic double-$Q$ coplanar state stabilized at $H^z=0$ is deformed to show nonzero $(m_{\bm{Q}_3}^z)^2$, as shown in the third panel of Fig.~\ref{fig:bond_Hz_I=06}(a). 
Accordingly, $(\chi_0)^2$ is induced, as shown in the top panel of Fig.~\ref{fig:bond_Hz_I=06}(a). 
The resultant spin and chirality textures are similar to those realized in the region for $0.4 \lesssim  H^z \lesssim 1.9$ at $K=0$ and $I^{\rm A}=0.4$ shown in Figs.~\ref{fig:Spin_Hz_K=0_IA=04}(c) and \ref{fig:Spin_Hz_K=0_IA=04}(d). 
The spin and chirality structures are shown in Figs.~\ref{fig:Spin_Hz_K=0_IA=06}(a) and \ref{fig:Spin_Hz_K=0_IA=06}(b), respectively, which is similar to those in Figs.~\ref{fig:Spin_Hz_K=0_IA=04}(e) and \ref{fig:Spin_Hz_K=0_IA=04}(f). 
While further increasing $H^z$, this triple-$Q$ state changes into an isotropic one with $(\bm{m}_{\bm{Q}_1})^2=(\bm{m}_{\bm{Q}_2})^2=(\bm{m}_{\bm{Q}_3})^2$ and $(\chi_{\bm{Q}_1})^2=(\chi_{\bm{Q}_2})^2=(\chi_{\bm{Q}_3})^2$ 
for $2.1 \lesssim H^z \lesssim 2.5$, whose spin and chirality textures are similar to those shown in Figs.~\ref{fig:Spin_Hz_K=03_IA=04}(i) and \ref{fig:Spin_Hz_K=03_IA=04}(j) for $I^{\rm A}=0.4$. 
The skyrmion number is zero also in this state despite nonzero $(\chi_0)^2$. 
Finally, $(\chi_0)^2$ vanishes at $H^z\simeq 2.5$, and the system turns into the nonchiral triple-$Q$ state.

In the case of $K=0.1$ where the double-$Q$ coplanar state with equal intensities at $\bm{Q}_1$ and $\bm{Q}_2$ is stabilized at zero field, the $n_{\rm sk}=2$ meron crystal with nonzero $(\chi_0)^2$ appears for $0 < H^z \lesssim 0.3$, as shown in Fig.~\ref{fig:bond_Hz_I=06}(b).  
It turns into the other triple-$Q$ state at $H^z\simeq 0.3$.  
This is a triple-$Q$ state with a small contribution from $(m^z_{\bm{Q}_3})^2$ as shown in the third panel of Fig.~\ref{fig:bond_Hz_I=06}(b), leading to the nonzero $(\chi_0)^2$. 
While increasing $H^z$, the system undergoes a phase transition at $H^z \simeq 0.5$ by showing a jump of $(\chi_0)^2$ as shown in the top panel of Fig.~\ref{fig:bond_Hz_I=06}(b); the spin texture changes into the isotropic triple-$Q$ structure and the skyrmion number changes from 0 to 2. 
In other words, the anisotropic triple-$Q$ state changes into the $n_{\rm sk}=2$ skyrmion crystal at this transition. 
The spin and chirality structures in this $n_{\rm sk}=2$ state are shown in Figs.~\ref{fig:Spin_Hz_K=01_IA=06}(a) and \ref{fig:Spin_Hz_K=01_IA=06}(b), which is similar to those in Figs.~\ref{fig:Spin_Hz_K=03_IA=04}(a) and \ref{fig:Spin_Hz_K=03_IA=04}(b).
While increasing $H^z$, the $n_{\rm sk}=2$ skyrmion crystal changes into the triple-$Q$ state with $n_{\rm sk}=0$ at $H^z \simeq 0.7$.
The spin and chirality configurations in this state are shown in Figs.~\ref{fig:Spin_Hz_K=01_IA=06}(c) and \ref{fig:Spin_Hz_K=01_IA=06}(d), which is similar to the triple-$Q$ state in Figs.~\ref{fig:Spin_Hz_K=03_IA=04}(e) and \ref{fig:Spin_Hz_K=03_IA=04}(f). 
While further increasing $H^z$, the system undergoes a phase transition to the state with vanishing $(\chi_{0})^2$ at $H^z \simeq 2.1$. 
The spin and chirality textures are similar to those obtained at $K=0.1$; see Figs.~\ref{fig:Spin_Hz_K=03_IA=04}(g) and \ref{fig:Spin_Hz_K=03_IA=04}(h). 

The result at $K=0.3$ and $I^{\rm A}=0.6$ shown in Fig.~\ref{fig:bond_Hz_I=06}(c) is similar to that at $K=0.3$ and $I^{\rm A}=0.4$ shown in Fig.~\ref{fig:bond_Hz_I=04}(c), except for the $n_{\rm sk}=1$ skyrmion crystal for $I^{\rm A}=0.4$; in the case with $I^{\rm A}=0.6$, the $n_{\rm sk}=2$ skyrmion crystal directly turns into the chiral triple-$Q$ state with $n_{\rm sk}=0$ at $H^z \simeq 0.6$. 
While increasing $H^z$, the chiral triple-$Q$ state turns into the triple-$Q$ state with vanishing $(\chi_0)^2$ at $H^z \simeq 1.9$.

\subsection{Discussion}
\label{sec:Summary of this section_4}

The results obtained in this section are summarized in Fig.~\ref{fig:Summary}(d). 
While the bond-dependent anisotropy $I^{\rm A}$ and the single-ion anisotropy $A$ are both rooted in the spin-orbit coupling, we obtained a further variety of the multiple-$Q$ instabilities by $I^{\rm A}$, especially toward chiral magnetic spin textures different from the $n_{\rm sk}=1$ and $n_{\rm sk}=2$ skyrmion crystals. 
In the following, we discuss the main results obtained in this section. 

In the absence of the magnetic field, we obtained the $n_{\rm sk}=2$ skyrmion crystal in the wide parameter range of $I^{\rm A}$ and $K$.  
We showed that $I^{\rm A}$ induces nonzero out-of-plane magnetization in the $n_{\rm sk}=2$ skyrmion crystal. 
The sign of the scalar chirality is set to be opposite to that of the magnetization. 
This is in contrast to the situation in the absence of $I^{\rm A}$ where the magnetization is zero and the sign of the chirality is free due to the in-plane spin rotational symmetry. 
We also showed that $I^{\rm A}$ brings about multiple-$Q$ instabilities even without $K$ and $H^z$, which is also in contrast to the case with the single-ion anisotropy $A$. 

When the magnetic field is applied along the $z$ direction, we found further intriguing chiral phases including the skyrmion crystals. 
Similar to the cases with nonzero $A$, we obtained the $n_{\rm sk}=1$ skyrmion crystal for nonzero $I^{\rm A}$ even without $K$, as shown in Fig.~\ref{fig:bond_Hz_I=02}(a). 
The difference from the result for nonzero $A$ is found in the degeneracy lifting between the states with different vorticity; the Bloch(N\'eel)-type skyrmion is stabilized for $I^{\rm A}>0$ ($I^{\rm A}<0$), while in the absence of $I^{\rm A}$, the energy for different types of the skyrmion crystals is degenerate for $A\neq 0$. 

Besides the skyrmion crystals, we obtained a variety of chiral magnetic states with nonzero scalar chirality, which have not been obtained in the case with the single-ion anisotropy. 
The double-$Q$ helical state is modulated to exhibit nonzero scalar chirality by applying the magnetic field, being the $n_{\rm sk}=1$ meron crystal composed of one meron- and three antimerion-like spin textures in the magnetic unit cell for $0< H^z \lesssim 0.1$ at $K=0$ and $I^{\rm A}=0.4$ [Fig.~\ref{fig:bond_Hz_I=04}(a)]. 
We also obtained the $n_{\rm sk}=2$ meron crystal composed of four meron-like spin textures in the magnetic unit cell [Figs.~\ref{fig:bond_Hz_I=04}(a), \ref{fig:bond_Hz_I=04}(b), and \ref{fig:bond_Hz_I=06}(b)].  
Moreover, we found multiple-$Q$ states with nonzero scalar chirality in the wide range of $H^z$ (Figs.~\ref{fig:bond_Hz_I=02}, \ref{fig:bond_Hz_I=04} and \ref{fig:bond_Hz_I=06}). 
The competition between these multiple-$Q$ states leads to a plethora of topological phase transitions accompanied by changes in the skyrmion number. 

The present results are useful to narrow down the origin of the multiple-$Q$ magnetic states found in experiments. 
The conditions for the emergence of the $n_{\rm sk}=1$ and $n_{\rm sk}=2$ skyrmion crystals are similar to those in the case of the single-ion anisotropy. 
The $n_{\rm sk}=2$ skyrmion crystal is realized only for nonzero $K$, while the $n_{\rm sk}=1$ one is stabilized even without $K$. 
Meanwhile, the stability of the other multiple-$Q$ states except for the skyrmion crystals are strongly dependent of the type of anisotropy and $K$, as shown in Fig.~\ref{fig:Summary}.  
Thus, the systematic study of the phase diagram in the magnetic field in experiments provides which interactions play an important role in the target materials.

\section{Concluding remarks}
\label{sec:Summary}

We have theoretically investigated the instabilities toward multiple-$Q$ states in centrosymmetric itinerant magnets, focusing on the effects of single-ion anisotropy and bond-dependent anisotropy. 
By performing the simulated annealing for the effective spin model on a triangular lattice, we found a plethora of multiple-$Q$ states with and without the scalar chirality in the wide range of the model parameters. 
As we have already shown the brief summary of the results in Sec.~\ref{sec:Brief Summary of main results} and the discussions in Secs.~\ref{sec:Summary of this section_1}, \ref{sec:Summary of this section_2}, \ref{sec:Summary of this section_3}, and \ref{sec:Summary of this section_4}, we here make some remarks on the relevant parameters to the emergence of topological spin textures, which would be useful for experimental identification of the microscopic mechanism. 

On the whole, we obtained four types of topological spin textures with nonzero skyrmion numbers: the $n_{\rm sk}=1$ skyrmion crystal, the $n_{\rm sk}=2$ skyrmion crystal, the $n_{\rm sk}=1$ meron crystal, and the $n_{\rm sk}=2$ meron crystal. 
Among them, we showed that there are several mechanisms for stabilizing the $n_{\rm sk}=1$ skyrmion crystal in a magnetic field; either the biquadratic interaction, single-ion anisotropy, or bond-dependent anisotropy can stabilize it. 
Thus, one can expect that the $n_{\rm sk}=1$ skyrmion crystal prevails in a wider range of materials compared to the other topological spin textures in centrosymmetric itinerant magnets. 
In fact, the $n_{\rm sk}=1$ skyrmion crystal has been recently observed in several centrosymmetric compounds, such as Gd$_2$PdSi$_3$~\cite{kurumaji2019skyrmion,Hirschberger_PhysRevLett.125.076602,Hirschberger_PhysRevB.101.220401,Nomoto_PhysRevLett.125.117204,moody2020charge}, Gd$_3$Ru$_4$Al$_{12}$~\cite{hirschberger2019skyrmion}, and GdRu$_2$Si$_2$~\cite{khanh2020nanometric,Yasui2020imaging}. 

Meanwhile, the various stabilization mechanisms for the $n_{\rm sk}=1$ skyrmion crystal make it difficult to identify its microscopic origin. 
To narrow down the origin of the $n_{\rm sk}=1$ skyrmion crystal, it is useful to investigate the magnetic phases around it, especially (i) in the lower- and higher-field regions and (ii) in the different field directions. 
With respect to (i), our results indicate that the lower-field state becomes the single-$Q$ spiral state when the single-ion anisotropy is the key parameter for the $n_{\rm sk}=1$ skyrmion crystal. 
When the itinerant nature of electrons becomes important (i.e., the biquadratic interaction becomes large in our model), the lower-field state of the $n_{\rm sk}=1$ skyrmion crystal becomes the anisotropic triple-$Q$ state or the $n_{\rm sk}=2$ skyrmion crystal. 
Meanwhile, the $n_{\rm sk}=1$ and $n_{\rm sk}=2$ meron crystals will be observed when the bond-dependent interaction has a significant contribution. 
On the other hand, in the higher-field region, the anisotropic triple-$Q$ state appears under the biquadratic interaction and the single-ion anisotropy, whereas the isotropic triple-$Q$ state is stabilized under the bond-dependent anisotropy. 
With respect to (ii), the $n_{\rm sk}=1$ skyrmion crystal remains stable against the field rotation when the biquadratic interaction is predominant owing to spin-rotational symmetry, while it is unstable when the single-ion or bond-dependent anisotropy is relevant. 

In this way, the systematic investigation of the phase diagram by changing the magnitude and direction of the magnetic field in experiments will provide which interaction plays an important role in stabilizing the skyrmion crystals. 
Our study gives a good starting reference to understand the origin of topological magnetism and a guiding principle to explore further exotic magnetic textures in centrosymmetric itinerant electrons.

\begin{acknowledgments}
We thank for T. Kurumaji, M. Hirschberger, S. Seki, R. Takagi, and S. Ishiwata for fruitful discussions. 
This research was supported by JSPS KAKENHI Grants Numbers JP18K13488, JP19K03752, JP19H01834, JP19H05825, and by JST PREST (JPMJPR20L8) and JST CREST (JP-MJCR18T2). 
This work was also supported by the Toyota Riken Scholarship. 
Parts of the numerical calculations were performed in the supercomputing systems in ISSP, the University of Tokyo.
\end{acknowledgments}

\bibliography{ref}

\begin{thebibliography}{147}
\bibitem [{com()}]{comment_contour}%
  \BibitemOpen
  \href@noop {} {}\bibinfo {note} {We plot the color in each pixel by taking
  the average of $S_i^z$ for four neighboring spins located at $\bm{r}_i$,
  $\bm{r}_i+(1,0,0)$, $\bm{r}_i+(1/2,\sqrt{3}/2,0)$, and
  $\bm{r}_i+(3/2,\sqrt{3}/2,0)$. The reason why the white region appears in the
  collinear state in (c) is owing to the averaging. We also use the
  interpolation for a better readability. The colored region in spite of the
  in-plane coplanar states in (b) and (d) is owing to thermal
  fluctuations.}\BibitemShut {Stop}%
\end{thebibliography}

\end{document}